\documentclass[aps,pra,reprint,superscriptaddress,twocolumn,showkeys,amsmath,amssymb,longbibliography]{revtex4-1}
\usepackage[english]{babel}
\usepackage{amsmath,amssymb,bbm,mathrsfs,bm,braket,color,graphicx,comment,amsfonts,dsfont}
\usepackage[colorlinks,citecolor=blue,urlcolor=blue]{hyperref}
\usepackage[normalem]{ulem}
\usepackage{comment}
\newcommand{\pd}{{\phantom{\dag}}}

\usepackage[normalem]{ulem}
\newcommand\delt{\bgroup\markoverwith{\textcolor{red}{\rule[0.5ex]{2pt}{0.4pt}}}\ULon}
\definecolor{purple}{rgb}{0.6, 0., 0.8}

\begin{document}
\title{Network model for higher-order topological phases}

\author{Hui Liu}
\affiliation{IFW Dresden and W{\"u}rzburg-Dresden Cluster of Excellence ct.qmat, Helmholtzstrasse 20, 01069 Dresden, Germany}

\author{Selma Franca}
\affiliation{IFW Dresden and W{\"u}rzburg-Dresden Cluster of Excellence ct.qmat, Helmholtzstrasse 20, 01069 Dresden, Germany}

\author{Ali G. Moghaddam}
\affiliation{Department of Physics, Institute for Advanced Studies in Basic Sciences (IASBS), Zanjan 45137-66731, Iran}
\affiliation{Research Center for Basic Sciences $\&$ Modern Technologies (RBST), Institute for Advanced Studies in Basic Science (IASBS), Zanjan 45137-66731, Iran}
\affiliation{IFW Dresden and W{\"u}rzburg-Dresden Cluster of Excellence ct.qmat, Helmholtzstrasse 20, 01069 Dresden, Germany}

\author{Fabian Hassler}
\affiliation{JARA-Institute for Quantum Information, RWTH Aachen University, 52056 Aachen, Germany}

\author{Ion Cosma Fulga}
\affiliation{IFW Dresden and W{\"u}rzburg-Dresden Cluster of Excellence ct.qmat, Helmholtzstrasse 20, 01069 Dresden, Germany}

\begin{abstract}
We introduce a two-dimensional network model that realizes a higher-order topological phase (HOTP). 
We find that in the HOTP the bulk and boundaries of the system are gapped, and a total of 16 corner states are protected by the combination of a four-fold
rotation, a phase-rotation, and a particle-hole symmetry. 
In addition, the model exhibits a strong topological phase at a point of
maximal coupling. This behavior is in opposition to  conventional network
models, which are gapless at this point. 
By introducing the appropriate topological invariants, we show how a point group symmetry can protect a topological phase in a network.
Our work provides the basis for the realization of HOTP in alternative experimental platforms implementing the network model.
\end{abstract}

\maketitle

\section{Introduction}
\label{sec:intro}

Topological phases of matter are known to support boundary modes which are protected by bulk topological invariants through the so-called bulk-boundary correspondence. 
The topological protection is typically associated with fundamental symmetries (time-reversal, particle-hole, and chiral symmetry) which have led to the tenfold classification of (gapped) topological states \cite{Schnyder2008, Ryu2010, Chiu2016}. 
By further including discrete spatial symmetries, new phases such as topological crystalline phases \cite{Mong2010, fu2011prl, slager2013natphys, fu2015review, Chiu2013} and, very recently, higher-order topological phases (HOTPs) have been identified
\cite{Benalcazar2017, Benalcazar2017prb, Langbehn2017, Song2017, Schindler2018natphys, Schindler2018sciadv, Geier2018, Khalaf2018, Trifunovic2019, Hughes2018, Ortix2018, You2018, Roy2019, Araki2019, Hatsugai2019, Tiwari2020, Ezawa2018kagome, Ezawa2018phosphorene, loss2018, Wang2018prl, Yan2018prl, sayed2020, sayed2020_1, sayed2019}.

A prototypical HOTP is the Benalcazar-Bernevig-Hughes (BBH) model \cite{Benalcazar2017}, where fundamental and point group symmetries protect zero-energy corner states in an otherwise gapped, two-dimensional (2D) system 
\footnote{Even though the BBH model is probably well known by now, we briefly summarize its main features and compare it with our network model in Appendix \ref{app:BBH}.}. 
HOTPs have already been realized in various platforms ranging from photonic, phononic, and electronic systems to microwave and topoelectric circuits 
\cite{Serra-Garcia2018, Peterson2018, Imhof2018, Serra-Garcia2019prb, zhang2019natphys, Kempkes2019, Yue2019natphys, mittal2019photonic, Xue2018, ni2019observation, elHassan2019corner}.

Ever since the discovery of the quantum Hall effect, a main focus of research has been to study the robustness of topological states against disorder, and the associated localization-delocalization transitions that destroy a topological phase \cite{mirlin2008review}.
Among other approaches, a powerful tool in this study has been the network model description of topological phases \cite{mirlin2008review, chalker1988, kramer2005review}, first introduced by Chalker and Coddington (CC) in the context of the quantum Hall effect \cite{chalker1988}. 
The network model idea has subsequently been adapted to a variety of different topological systems, including the quantum spin-Hall effect, weak topological insulators, Floquet systems, as well as different types of topological superconductors \cite{chalker1999super, chalker2001super, chalker2002qshe, murdy2007qshe, murdy2014TI, fulga2012thermal, chong2014floquet, debeule2020}.
However, despite their versatility, no network models for point group symmetry-protected topological phases have been introduced to date. This includes HOTPs, which are very recent additions to the list of topological systems, but also conventional topological crystalline phases, which are by now a decade old \cite{Mong2010, fu2011prl}.

Here, we construct a network model for a HOTP protected by particle-hole and by fourfold rotation ($C_4$) symmetry, which hosts mid gap corner modes. 
Unlike the $C_4$-symmetric HOTPs in static systems, which have 4 corner states~\cite{Benalcazar2017, Benalcazar2017prb}, or their periodically driven counterparts, where this number can go up to 8~\cite{Rodriguez-Vega2019, Bomentara2019, Huang2020}, our network model shows up to 16 corner modes, due to an additional, phase-rotation symmetry \cite{Delplace2017}.
Upon varying system parameters, we find that the trivial phase and the HOTP are separated by an intermediate, strong topological phase (STP), which surprisingly is present even at the point of maximal mode mixing, \emph{i.e.}, when the reflection and transmission probabilities are equal. 
This behavior is in sharp contrast to the conventional Chalker-Coddington model and its generalizations \cite{chalker1999super, chalker2001super, chalker2002qshe, murdy2007qshe, murdy2014TI, fulga2012thermal, chong2014floquet}, all of which are gapless at this point.

The rest of this work is organized as follows. 
In Sec.~\ref{sec:nm}, we describe the construction of the network model and examine its symmetries. 
In Sec.~\ref{sec:decoupled} we analyze several simple limits of the network model, which realize a HOTP, a trivial, and a STP. 
A detailed study of the phase diagram is performed in Sec.~\ref{sec:pd}, and the topological invariants of the system are calculated in Sec.~\ref{sec:inv}. 
We study the effect of disorder in Sec.~\ref{sec:disorder} and comment on experimental realizations of network models in Sec.~\ref{sec:xp}. We conclude in Sec.~\ref{sec:conc}.

\section{Network model}
\label{sec:nm}

The CC network model is a 2D square lattice consisting of directed links and nodes. 
It models a quantum Hall device where the links correspond to chiral edge states which propagate along equipotential lines, and the nodes correspond to saddle points in the potential where scattering between the edge states takes place \cite{chalker1988}.
Each node of the CC network model connects two incoming to two outgoing states, being described by a $2\times2$ unitary scattering matrix
\begin{equation}
S=\begin{pmatrix}
  \mathfrak{r} & \mathfrak{t}' \\
  \mathfrak{t} & \mathfrak{r}'
  \end{pmatrix},
\end{equation}
where $\mathfrak{r}, \mathfrak{r}'$ and $\mathfrak{t}, \mathfrak{t}'$ are (generally complex valued) reflection and transmission amplitudes, respectively.
By imposing constraints on the form of $S$ while keeping the lattice structure intact, one can include the effect of fundamental symmetries on the network model. 
The CC network model can thus be extended to describe topological phases in a variety of Altland-Zirnbauer (AZ) classes \citep{Altland1997, mirlin2008review, chalker1999super, chalker2001super, fulga2012thermal, murdy2014TI, chalker2002qshe, murdy2007qshe, chong2014floquet, Son2020}.

\begin{figure}[tb]
\centering
\includegraphics[width=\linewidth]{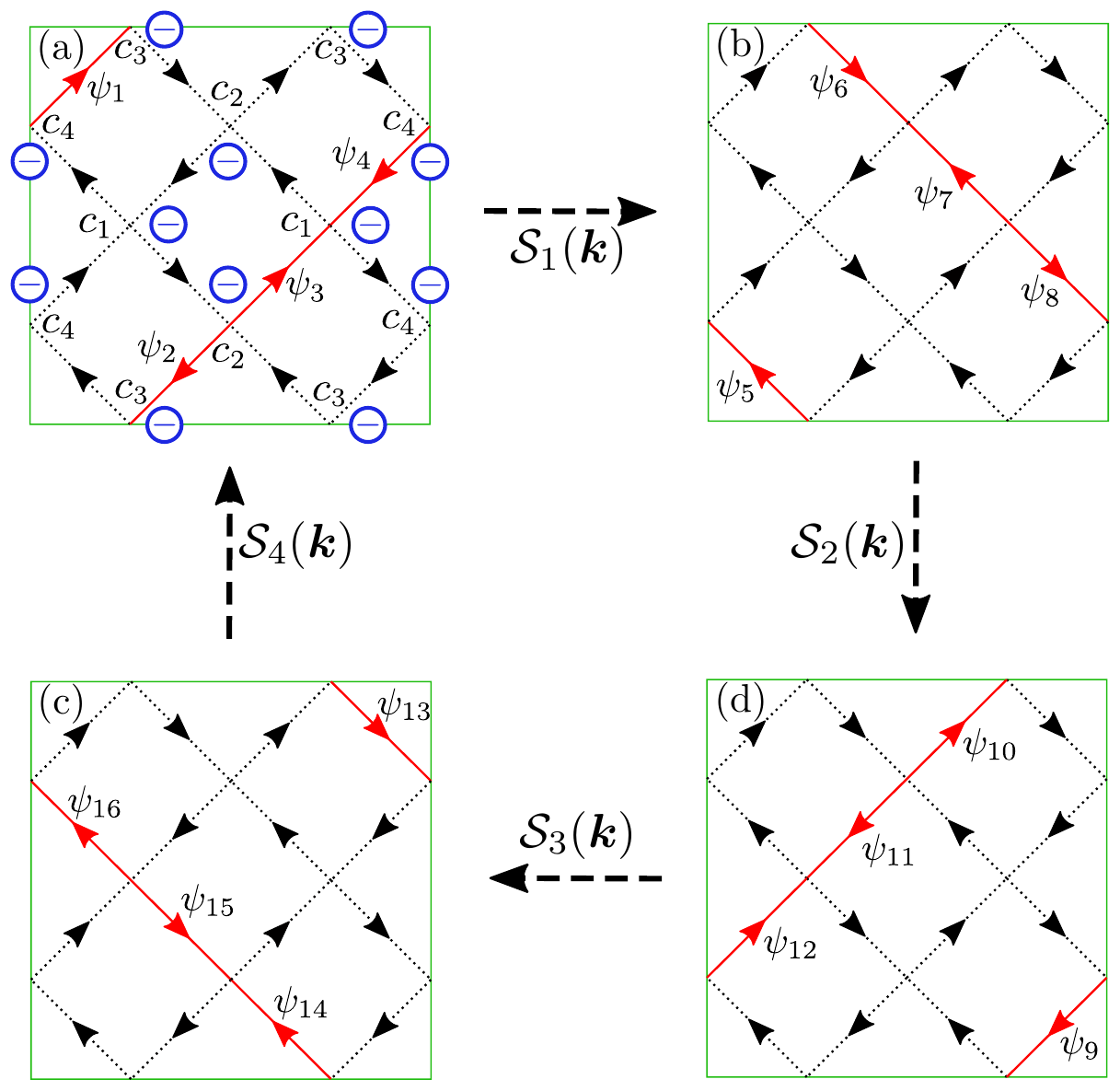}
\caption{Unit cell of the network model. 
The arrows represent chiral Majorana modes with wavefunctions $\psi_i$. 
Their intersections represent scattering nodes. 
The scattering amplitudes at the different nodes are parametrized by the angles $\theta_i$, with $c_i \equiv \cos \theta_i$. 
The symbol $\ominus$ denotes the fact that the scattered wave acquires a $\pi$-phase shift with respect to the incoming amplitude [see Eq.~\eqref{eq:Snode}].
We introduce the notation $\Psi_i=(\psi_{4i-3},\psi_{4i-2},\psi_{4i-1},\psi_{4i})^{T}$ with $\Psi_{i}=\Psi_{i+4}$, $i=1,2,3,4$, see the red arrows. 
The wavefunctions are related to each other via $\Psi_{i+1}=\mathcal{S}_i(\bm{k})\Psi_i$ [see Eqs.~\eqref{eq:Ho-Chalker}--\eqref{eq:HCblock4}].
\label{fig:setup}
}
\end{figure}

To construct a network model for a HOTP, we need both fundamental as well as point group symmetries, meaning that both the form of $S$ as well as the lattice structure of the network model should be modified. 
We choose  a unit cell composed of 8 nodes and 16 links, four times larger than that of the original CC model, see Fig.~\ref{fig:setup}.
We impose particle-hole symmetry, such that the model belongs to the AZ class D. 
The directed links of the network correspond to chiral Majorana modes and the scattering matrices are real, $S^\pd_i=S_i^*$, with $i$ labeling different nodes. 
This allows to parametrize them in terms of an angle $\theta_i$ and a relative sign $\pm$ as 
\begin{equation}\label{eq:Snode}
S_i=\begin{pmatrix}
  \cos{\theta_i} & \mp\sin{\theta_i} \\
  \sin{\theta_i} & \pm\cos{\theta_i}
  \end{pmatrix}.
\end{equation}
In Fig.~\ref{fig:setup}, we have indicated the minus signs connecting an incoming and an outgoing amplitude per node by the symbol $\ominus$. 
In the following, we will examine the behavior of the network model as a function of four different angles $\theta_1, \ldots, \theta_4$ in the unit cell, see Fig.~\ref{fig:setup} for the convention. 
As we will show below, imposing  a $C_4$ symmetry implies setting $\theta_1=\theta_2$ and $\theta_3=\theta_4$.

A network model consisting of $N$ links can be described using the wavefunction $\Phi = (\psi_1, \psi_2, \ldots, \psi_N)^T$, where $\psi_j$ denotes the amplitude on link $j$ of the network. 
As waves propagate through the network, they are scattered into each other at the nodes, a process modeled as a discrete-time evolution \cite{ho1996, zirnbauer1999}
\begin{equation}\label{eq:discrete_time}
 \psi_j(t+1) = \sum_{k} {\cal S}_{jk} \psi_{k}(t),
\end{equation}
with $t\in \mathbb{Z}$ the discrete time unit and ${\cal S}$ a unitary matrix (also known as the Ho-Chalker operator \cite{ho1996}) containing the transmission and reflection amplitudes of the scattering nodes. 
A stationary state of the network model obeys the relation
\begin{equation}\label{eq:Floquet}
 {\cal S} \Phi = e^{-i \varepsilon} \Phi.
\end{equation}
The above equation is reminiscent of the Floquet treatment of periodically driven systems, where now the role of the Floquet operator is taken by the unitary matrix ${\cal S}$, and its eigenphases $\varepsilon$ play the role of the quasienergies. 
By solving Eq.~\eqref{eq:Floquet}, we gain access to the spectrum of the network model, enabling us to identify gaps in the spectrum of eigenphases and potential corner modes. 
Furthermore, by considering an infinite, translationally invariant network model, Eq.~\eqref{eq:Floquet} provides access to the "bandstructure" $\varepsilon({\bm k})$ of the system with ${\bm k}=(k_x, k_y)$ the two wavenumbers.

In our network model, the momentum space Ho-Chalker operator in the translationally invariant setting is a $16\times16$ matrix, due to the fact that  there are 16 links per unit cell. 
Using the labeling convention of Fig.~\ref{fig:setup}, it reads
\begin{equation}\label{eq:Ho-Chalker}
\mathcal{S}(\bm{k}) = 
 \begin{pmatrix}
  0 & 0 & 0 & \mathcal{S}_4(\bm{k}) \\ 
  \mathcal{S}_1(\bm{k}) & 0 & 0 & 0 \\
  0 & \mathcal{S}_2(\bm{k}) & 0 & 0 \\
  0 & 0 & \mathcal{S}_3(\bm{k}) & 0 
  \end{pmatrix}, 
\end{equation}
where the blocks are given by
\begin{equation}\label{eq:HCblock1}
\mathcal{S}_1(\bm{k}) = 
 \begin{pmatrix}
 e^{-ik_y}\sin\theta_3 & \cos\theta_3 & 0 & 0 \\
 \cos\theta_3 & -e^{ik_y}\sin\theta_3 & 0 & 0 \\
 0 & 0 & \cos\theta_1 & \sin\theta_1 \\
 0 & 0 & \sin\theta_1 & -\cos\theta_1
 \end{pmatrix},
\end{equation}
\begin{equation}\label{eq:HCblock2}
\mathcal{S}_2(\bm{k}) = 
 \begin{pmatrix}
 e^{ik_x}\sin\theta_4 & 0 & 0 & \cos\theta_4 \\
 0 & \cos\theta_2 & \sin\theta_2 & 0 \\
 0 & \sin\theta_2 & -\cos\theta_2 & 0 \\
 \cos\theta_4 & 0 & 0 & -e^{-ik_x}\sin\theta_4
 \end{pmatrix},
\end{equation}
\begin{equation}\label{eq:HCblock3}
  \mathcal{S}_3(\bm{k}) = \mathcal{S}_1(\bm{k}) \;[ k_y\mapsto-k_y,\theta_3\mapsto-\theta_3,\theta_1\mapsto\pi-\theta_1],
\end{equation}
\begin{equation}\label{eq:HCblock4}
\mathcal{S}_4(\bm{k}) = \mathcal{S}_2(\bm{k}) \;[k_x\mapsto-k_x].
\end{equation}

In the translationally invariant system, particle-hole symmetry is expressed as ${\cal S}({\bm k})={\cal S}^*(-{\bm k})$. 
This is reminiscent of Floquet systems and indicates that for any eigenstate of the network model at eigenphase $\varepsilon$ and momentum ${\bm k}$, there exists another eigenstate at $-\varepsilon$ and $-{\bm k}$. Further, along the line $\theta_1=\theta_2$, $\theta_3=\theta_4$, the Ho-Chalker operator has $C_4$ symmetry $\mathcal{R}$ with ${\cal R} {\cal S}(k_x, k_y) {\cal R}^\dag = {\cal S}(k_y, -k_x)$, where
\begin{equation}\label{eq:C4operator}
{\cal R} = 
 \begin{pmatrix}
  0 & 0 & 0 & {\cal R}_4 \\
  {\cal R}_3 & 0 & 0 & 0 \\
  0 & {\cal R}_2 & 0 & 0 \\
  0 & 0 & {\cal R}_1 & 0 
  \end{pmatrix}, \quad
{\cal R}_i = K_i 
 \begin{pmatrix}
  1 & 0 & 0 & 0 \\
  0 & 0 & 0 & 1 \\
  0 & 0 & 1 & 0 \\
   0 & 1 & 0 & 0
 \end{pmatrix};
\end{equation}
$K_i$ is a $4\times 4$ diagonal matrix whose $i$-th diagonal element is $-1$ while the others are $1$.

Finally, the Ho-Chalker operator ${\cal S}$ also possesses a so-called phase-rotation symmetry (PRS) ${\cal Z} {\cal S}(\bm{k})  {\cal Z}^\dag = e^{i\pi/2} {\cal S} (\bm{k})$, with ${\cal Z} = \mathop{\rm diag} (i,-1,-i,1 ) \otimes \mathbbm{1}_4$ the phase-rotation operator \cite{Delplace2017}. 
The latter symmetry implies that for any eigenstate with eigenphase $\varepsilon$ there also exists an eigenstate at $\varepsilon+\pi/2$. 
As such, the spectrum of the network model repeats four times in $\varepsilon\in [-\pi,\pi)$, allowing to focus on the "fundamental phase domain" with $\varepsilon\in [-\pi/4,\pi/4)$.

The phase-rotation symmetry arises due to the geometry of the network model, as explained in Ref.~\cite{Delplace2017}. Specifically, it is due to the fact that it consists of unidirectional modes which are coupled to each other in a cyclic way, as shown in Fig.~\ref{fig:setup}. Thus, it can be thought of as a structural constraint, similar to the more familiar lattice symmetries such as rotation. Unlike rotations, however, the phase rotation symmetry is unique to unitary operators, since in a Hermitian Hamiltonian the energy shift associated to it would be unphysical.

\begin{figure*}[tb]
\centering
\includegraphics[width=\textwidth]{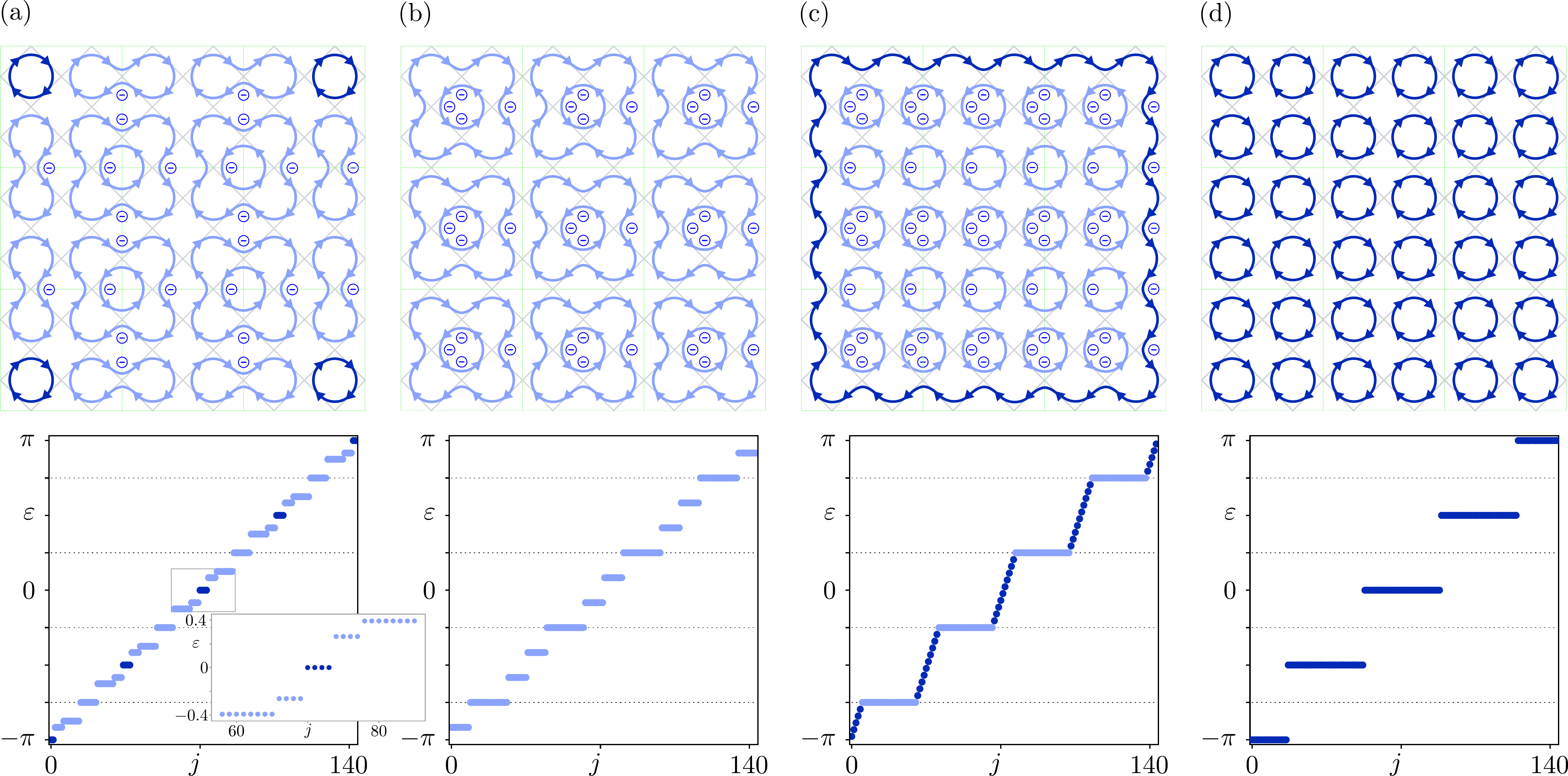}
\caption{The top panels show the network model, consisting of $3\times3$ unit cells, in the four decoupled limits: HOTP ($\theta_1=\theta_3=\pi/2$) in panel (a), trivial ($\theta_1=\theta_3=0$) in panel (b), STP ($\theta_1=0$, $\theta_3=\pi/2$) in panel (c), and Majorana flat band ($\theta_1=\pi/2$, $\theta_3=0$) in panel (d). 
The bottom panels show the spectra of the Ho-Chalker operators in the same limits. 
The inset in panel (a) is a closeup of the midgap corner modes. 
Gapped (gapless) modes and the Majorana loops producing them are shown in light (dark) blue. 
The horizontal dotted lines mark the boundaries of the fundamental phase domain, which repeats four times due to phase-rotation symmetry. 
Notice that all gapped Majorana loops contain an odd number of minus signs (denoted by $\ominus$) in a closed loop. 
See Fig.~\ref{fig:setup} for the minus sign conventions within a unit cell.\label{fig:limit_cases}
}
\end{figure*}

\section{Decoupled limits}
\label{sec:decoupled}

We study the network model by first focusing on the $C_4$-symmetric case where the network properties are controlled by the two angles $\theta_1(=\theta_2)$ and $\theta_3(=\theta_4)$, corresponding to the mixing angle at the inner and outer nodes of the unit cell, see Fig.~\ref{fig:setup}. 
To determine the properties of the edge and corner states, we construct finite-sized network models by imposing open boundary conditions (OBC) in both directions. 
As shown in Fig.~\ref{fig:limit_cases}, the OBC are obtained by demanding that the Majorana wavefunctions impinging on the boundary are reflected with unit probability, which implies a vanishing of the probability current across the boundary.

First, we investigate the four `decoupled limits' of the network model, obtained when $\theta_i$ are either $0$ or $\pi/2$. 
For these values, the incoming Majorana modes do not get mixed by node scattering but instead turn either clockwise or counterclockwise with unit probability, see Fig.~\ref{fig:limit_cases}. 
The system then consists of Majorana modes propagating along closed loops that are fully decoupled from each other.

In this limit, we can determine the spectrum of the network model entirely from the structure of these closed loops. 
For a closed path consisting of $\ell$ links, a Majorana wavefunction comes back to itself after $\ell$ discrete time steps, i.e., after $\ell$ applications of the Ho-Chalker operator of Eq.~\eqref{eq:discrete_time}. 
Note that there are only two options after a full loop. 
Either the link amplitude picks up a total phase of 0 (periodic) or $\pi$ (antiperiodic), since the Majorana wavefunction is real. 
The spectrum of the system can thus be determined from the lengths of these loops.
All periodic loops of length $\ell$ contribute to the spectrum of the Ho-Chalker operator with the eigenphases $\varepsilon = 2\pi k/\ell$, where $k=0,1,\ldots, \ell-1$. 
As such, they admit a gapless solution with an eigenphase $\varepsilon=0$. 
In contrast, the spectrum of the antiperiodic loops is shifted by $\tfrac12$ with $\varepsilon = 2\pi  (k+ \tfrac12)/\ell$ and no gapless solution is possible in this case. 
In the following, we discuss the four different decoupled limits shown in Fig.~\ref{fig:limit_cases}, in order of appearance.

Setting $\theta_1=\theta_3=\pi/2$, the network model realizes a HOTP, see Fig.~\ref{fig:limit_cases}(a). 
We find that the bulk and edges are gapped, consisting only of antiperiodic Majorana loops. 
At the four corners, however, periodic loops of length 4 lead to midgap modes in the spectrum. 
There are a total of four modes localized at each one of the corners, with eigenphases given by the fourth roots of unity, i.e., $\varepsilon=0$, $\pm\pi/2$, and $\pi$. 
The 0 and $\pi$ modes cannot be shifted away from their value without breaking particle-hole symmetry. 
However, the $\varepsilon=\pm\pi/2$ modes cannot be moved away from their values due to the combination of particle-hole and phase-rotation symmetry. 
As such, the only mechanism through which a corner mode may be removed is by coupling it with another corner mode to form a dimerized pair. 
Since the system obeys a $C_4$ symmetry, the hybridization of the  corner mode cannot happen along an edge of the system, but requires all four midgap states to be simultaneously moved to the center (or bulk) of the network. 
Therefore, the $C_4$ symmetry protects the midgap corner states of the bulk HOTP through a mechanism that is fully analogous to that of the BBH model (see Appendix \ref{app:BBH}).

For $\theta_1=\theta_3=0$, the network model is topologically trivial, see Fig.~\ref{fig:limit_cases}(b). 
All Majorana loops are antiperiodic, so that no midgap states exist in the spectrum. 
Notice, however, that the pattern of closed loops is the same as that of the HOTP network in Fig.~\ref{fig:limit_cases}(a). 
The main difference between the two limits is that for the trivial system the loops are fully contained inside each unit cell, whereas in the HOTP they extend across the boundaries of the unit cells. 
This feature is analogous to the BBH model, where the HOTP is obtained when sites are dimerized across the unit cell boundary and the trivial phase contains sites which are dimerized within the unit cell. 
As an additional common feature, this means that the HOTP and trivial phase can be mapped onto each other by a redefinition of the unit cell (see Appendix \ref{app:BBH}).

For $\theta_1=0$ and $\theta_3=\pi/2$, the network model is in a STP, see Fig.~\ref{fig:limit_cases}(c), equivalent to that realized in the Cho-Fisher model~\cite{Cho1996}. 
The bulk is gapped, and a single chiral Majorana mode propagates along the edge, akin to topological $p$-wave superconductors~\cite{Alicea2012}.
The large Majorana loop extending along the perimeter of the network is antiperiodic, such that it does not admit exact zero-eigenphase states. 
Instead, like for the chiral Majorana mode on the boundary of $p$-wave superconductors, there is a finite-sized gap in the spectrum of the edge mode \cite{Nayak2008}.
The minigap $\pi/U$ is inversely proportional to the perimeter $U$ and vanishes in the thermodynamic limit. 
Note that the spectrum of the Ho-Chalker operator is periodic. 
As a result, every bulk band has an edge mode in the gap above and an edge mode in the gap below. 
Due to this fact, which is reminiscent of so-called anomalous Floquet topological phases~\cite{Rudner2013, Titum2016, Maczewsky2017, Mukherjee2017}, the bulk bands all have a vanishing Chern number at this point \footnote{See the code in our Supplemental Material for an explicit calculation of the Chern numbers.}, despite the presence of the chiral edge mode, which winds around the torus formed by the momentum (along the edge) and eigenphase. We note that away from this fine-tuned point, the bulk band degeneracy is lifted: band gaps appear at the boundaries of the fundamental phase domain, such that the STP loses its anomalous nature and each bulk band instead carries a Chern number $\pm 1$.

Finally, setting $\theta_1=\pi/2$ and $\theta_3=0$, the network model realizes a gapless, Majorana flat band, as shown in Fig.~\ref{fig:limit_cases}(d). 
All closed loops are periodic, such that the number of zero-eigenphase states is extensive and scales with the system size. 
In the other three decoupled limits, the presence of a bulk gap protects the resulting phase against small, symmetry preserving parameter changes. 
Here, in contrast, the Majorana flat bands may be gapped out by  small changes of the parameters. 
In fact, as we will show in the following, the decoupled limit of Fig.~\ref{fig:limit_cases}(d) represents a tricritical point in the phase diagram of the network model at which the other three phases meet. 

\begin{figure}[tb]
\centering
\includegraphics[width=\linewidth]{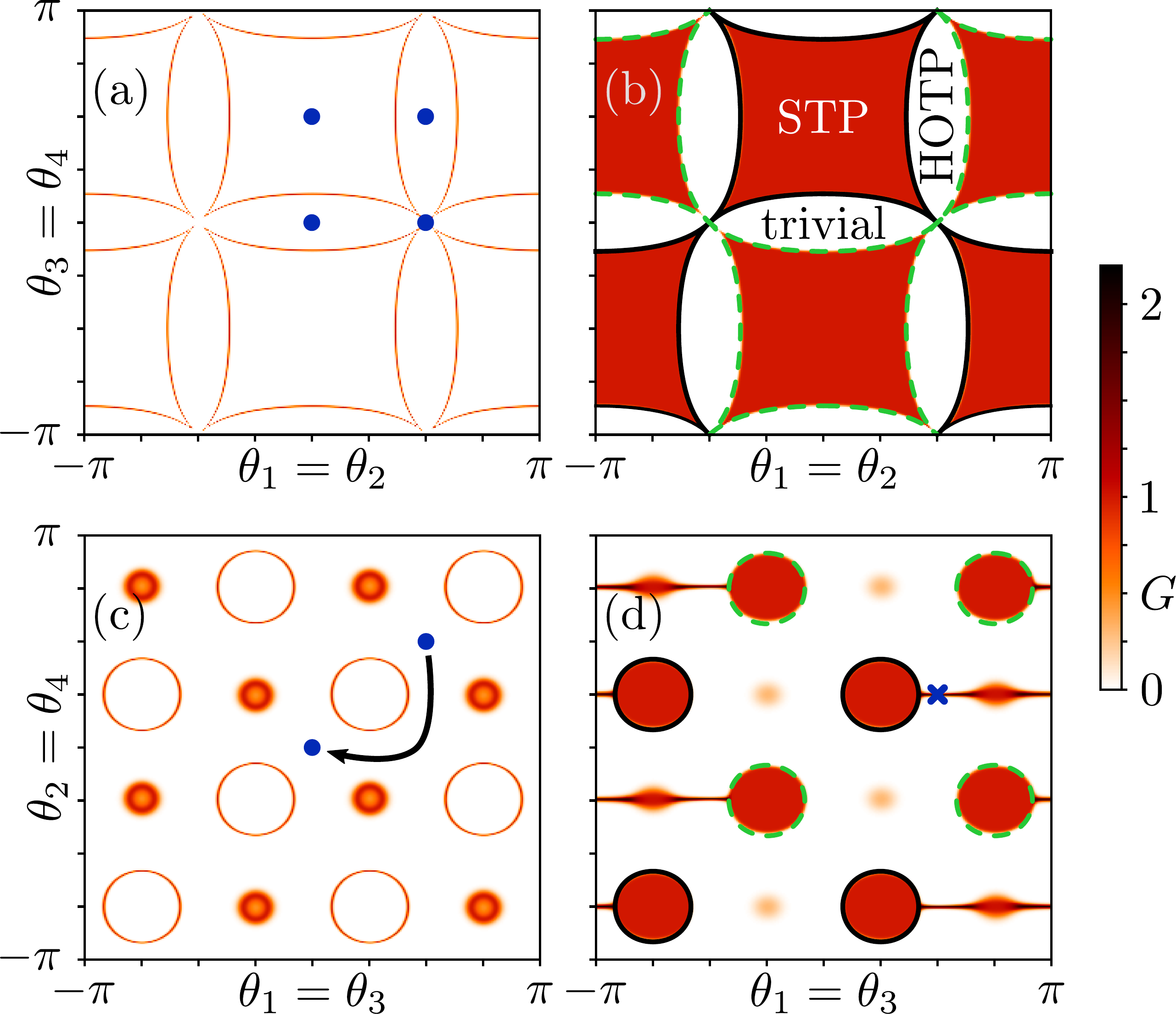}
\caption{The dimensionless conductance $G$, computed for a network of $20\times20$ unit cells, is plotted as a function of  the angles $\theta_1=\theta_2$ and $\theta_3=\theta_4$ ($C_4$ symmetric) in panels (a) and (b). 
Similarly, panels (c) and (d) show the case with $\theta_1=\theta_3$ and $\theta_2=\theta_4$ (broken $C_4$ symmetry). 
Note that in panels (a) and (c) we use periodic boundary conditions, whereas in panels (b) and (d) are computed using open boundary conditions.
The black solid and green dashed lines denote the gap closings at $\bm{k}=(0, 0)$ and $(\pi, \pi)$, respectively. 
The four blue dots in panel (a) mark the decoupled limits of Fig.~\ref{fig:limit_cases}. 
The two blue dots in panel (c) indicate the trivial (all $\theta=0$) and HOTP (all $\theta=\pi/2$) decoupled limits. 
The black arrow connecting them is an example of a $C_4$ symmetry breaking path in parameter space that allows to transition from the HOTP to the trivial phase without closing of the bulk gap. 
The blue cross in panel (d) shows a point at which only the top and bottom edges of the network conduct.
\label{fig:phase_diag}
}
\end{figure}

\section{Phase diagram}
\label{sec:pd}

We determine the phase diagram of the network model by computing its two-terminal dimensionless conductance $G$, given by the total transmission probability. 
To this end, we attach semi-infinite leads composed of Majorana modes to its left and right boundaries and calculate the two-lead scattering matrix associated to the whole network (see Appendix~\ref{app:leads} for details). 
We consider systems with either OBC or with periodic boundary conditions (PBC) in the transverse direction. 
This enables us to distinguish between bulk and edge contributions to the two-terminal transmission probability.

We begin by preserving the $C_4$ symmetry and determining the phase diagram in the $\theta_1$--$\theta_3$ plane, shown in Figs.~\ref{fig:phase_diag}(a) and \ref{fig:phase_diag}(b). 
We observe that the trivial phase and the HOTP have a vanishing two-terminal transmission probability $G$ for both OBC and PBC. 
The reason is that both the bulk and the edges are gapped. 
There are four topologically trivial regions in the $\theta_1$--$\theta_3$ plane, appearing as horizontally elongated regions of $G=0$ in Fig.~\ref{fig:phase_diag}(b), centered around $\theta_3=0,\pi$. 
The four HOTP regions appear in the same panel as vertically elongated $G=0$ regions centered around $\theta_1=\pm\pi/2$.

Additionally, we observe a total of four regions of STP, which show a quantized value $G=1$ with OBC, Fig.~\ref{fig:phase_diag}(b), but $G=0$ when imposing PBC, Fig.~\ref{fig:phase_diag}(a). 
This is consistent with edge transport due to the presence of a single chiral edge mode. 
Notice how the decoupled limits shown in Fig.~\ref{fig:limit_cases} correspond to the centers of the HOTP, trivial phase, and STP regions, with the Majorana flat band of Fig.~\ref{fig:limit_cases}(d) forming a critical point at which the three phases meet. 
The position of these phases relative to each other in parameter space is a consequence of several symmetries of the Ho-Chalker operator, as we explain in Appendix~\ref{app:sym}.

Along the phase transition lines, the bulk gap around $\varepsilon=0$ closes, as do all of the other gaps (at $\varepsilon = \pm \pi/2$, $\pi$) related to it by the phase-rotation symmetry. 
The gap closing condition is given by
\begin{equation}\label{eq:gap_closing}
  \det \Biggl( \prod_{j=1}^4 {\cal S}_j - \mathbbm{1}_{4\times4} \Biggr) = 0,
\end{equation}
with solutions of the form
\begin{equation}
|\cos \theta_1 \mp \sin \theta_3| = \sqrt{2}|\sin \theta_3 \cos \theta_1|, 
\end{equation}
at $\bm{k}=(0,0)$ and $\bm{k}=(\pi,\pi)$, respectively. 
Note that the two solutions, shown as solid black and dashed green lines in Fig.~\ref{fig:phase_diag}(b), nicely match the numerics.

In Figs.~\ref{fig:phase_diag}(a) and \ref{fig:phase_diag}(b), there is no path in the phase diagram which connects the HOTP to the trivial phase without a bulk gap closing. 
As discussed in the previous section, this is a consequence of $C_4$ symmetry, which forbids the dimerization of corner modes along the edges of the system. 
Conversely, if $C_4$ symmetry is broken, it becomes possible to connect the HOTP and trivial phase while preserving the bulk gap. 
We explore this possibility in Figs.~\ref{fig:phase_diag}(c) and \ref{fig:phase_diag}(d), where the transmission probability is plotted as a function of $\theta_1=\theta_3$ and $\theta_2=\theta_4$. 
These two axes correspond to different dimerizations between Majorana loops of the network in the horizontal and vertical directions, analogous to how dimerization can be varied independently for the horizontal and vertical hoppings of the BBH model (see Appendix~\ref{app:BBH}). 

Along the diagonals of Fig.~\ref{fig:phase_diag}(c), $\theta_1=\theta_2$ and $\theta_1=-\theta_2$, a $C_4$ symmetry is preserved (albeit with a different symmetry operator; see Appendix~\ref{app:c4}). 
When all $\theta$s are equal, there is an intermediate STP centered around $\theta=\pi/4$ which separates the trivial phase ($\theta=0$) and HOTP ($\theta=\pi/2$). 
In this case, the analytical expression for the gap closing is given by  $|\sin(\pi/4\pm\theta_1)\sin(\pi/4\pm\theta_2)|=\sqrt{3}/2$ at ${\bm k}=(0,0),(\pi,\pi)$. 
The corresponding contours are shown as black solid and green dashed lines in Fig.~\ref{fig:phase_diag}(d). 
Note that along the antidiagonal with $\theta_1=-\theta_2$, the system remains trivial. 
However, we observe bulk gap closings at the points  $(\theta_1,\theta_2)=(\pi/4, -\pi/4)$ and $(-\pi/4, \pi/4)$ [corresponding to  $\bm{k}=(0,\pi)$ and $(\pi, 0)$]. 
These are visible  as small spots of increased transmission probability in the numerical data of Fig.~\ref{fig:phase_diag}(c). 

Away from the diagonal, there exists a continuous path, shown as a black arrow in Fig.~\ref{fig:phase_diag}(c), which connects the HOTP to the trivial phase without closing the bulk gap. 
Note that along this path the topology is changed by an edge gap closing. Indeed, for any path connecting the HOTP and trivial regions either the bulk or the edge gap must close, since the Majorana corner modes can only be gapped out by pairwise coupling. 
When adjacent corner modes hybridize through the top and bottom edges, they form counterpropagating edge modes which are visible in the transmission map of Fig.~\ref{fig:phase_diag}(d); for instance at $\theta_1=\pi/2$, $\theta_2=\pi/4$ (blue cross). 
In fact, the closing of the edge gap is the reason for the thin horizontal lines of Fig.~\ref{fig:phase_diag}(d) at which we observe $G=2$, consistent with the presence of counterpropagating edge modes on both the top and bottom edges. 
In contrast, corner modes overlapping via the left and right edges of the network (such as at the point $\theta_1=\pi/4$, $\theta_2=\pi/2$) do not contribute to transmission, since the leads are attached to the left and right sides of the network and thus we only probe transport along the top and bottom edges.

Note that at the point $(\theta_1,\theta_2)=(\pi/4, \pi/4)$, the links of the network are maximally coupled, in the sense that each incoming Majorana mode has an equal probability of turning clockwise or counterclockwise at each node of the network. 
In the CC model and, as far as we know, in all of its generalizations 
\citep{mirlin2008review, chalker1999super, chalker2001super, fulga2012thermal, murdy2014TI, chalker2002qshe, murdy2007qshe, chong2014floquet, Son2020}, 
the maximally coupled limit is a gapless critical point separating topologically distinct phases. 
Interestingly, here it corresponds to a gapped, STP. 
For the parametrization shown in Figs.~\ref{fig:phase_diag}(c) and \ref{fig:phase_diag}(d), in which $\theta_1=\theta_3$ and $\theta_2=\theta_4$, the point $(\theta_1,\theta_2)=(\pi/4, \pi/4)$ is in the middle of the strong topological region of the phase diagram, at which the bulk gap is the largest.

\begin{figure}[tb]
\centering
\includegraphics[width=\linewidth]{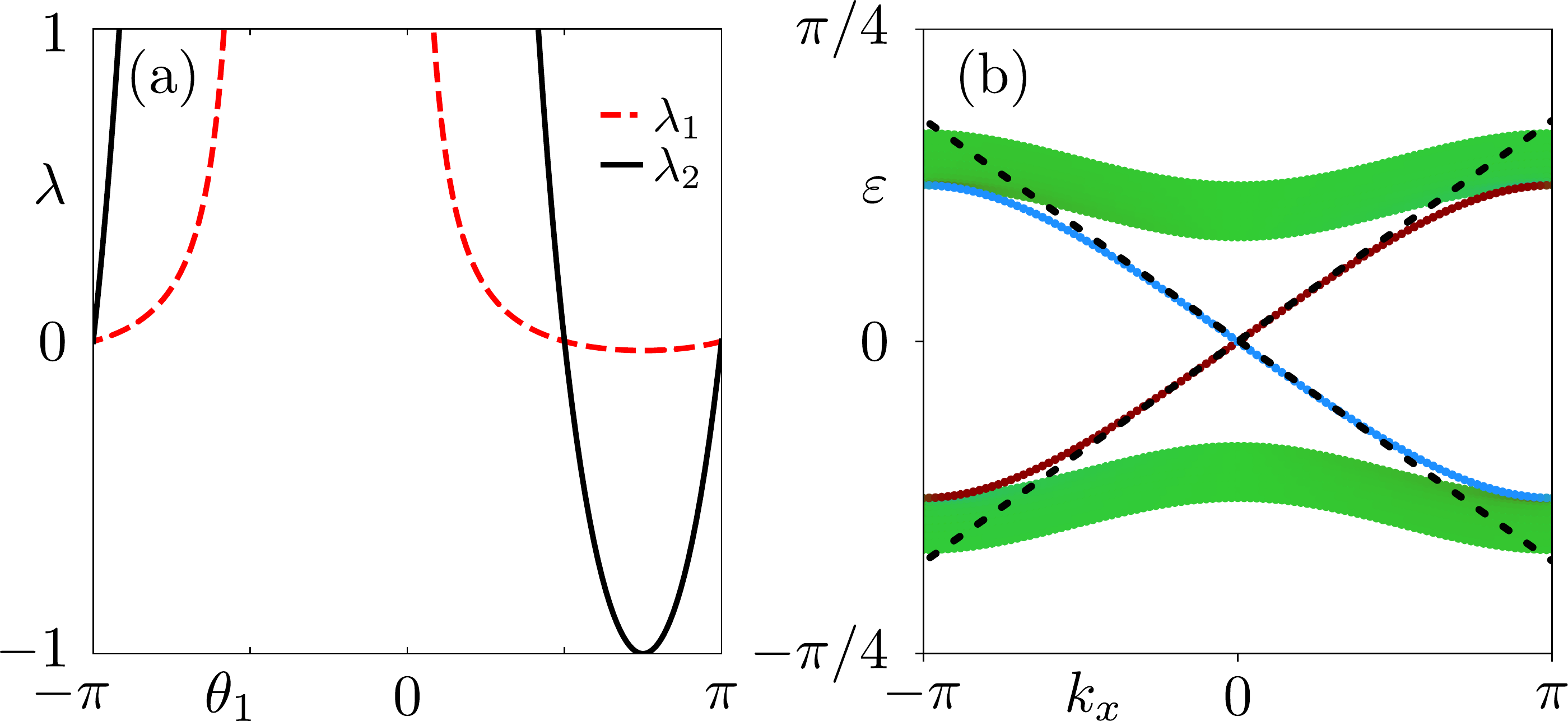}
\caption{(a) Decay parameters as a function of $\theta_1$ at  $\theta_2=\pi/4$. 
Note that the number of solutions with $\lambda < 1$ correspond to the number of edge modes present.
(b) Bandstructure of the network model at the maximally coupled point, $(\theta_1,\theta_2)=(\pi/4, \pi/4)$, computed in a ribbon geometry, infinite in the horizontal direction and consisting of 20 unit cells in the vertical direction. 
The bulk states are shown in green, whereas red and blue colors denote modes localized on the top and bottom boundaries, respectively. 
The dashed black lines show the velocities $v_E=\pm \sqrt{2}/8$ which is the perturbative expression valid for small wavenumbers.
\label{fig:max_coupling}
}
\end{figure}

To gain insight into the behavior of the network in the vicinity of $(\theta_1,\theta_2)=(\pi/4, \pi/4)$, we study a half-plane geometry (on $y<0$). 
We use an ansatz wavefunction $\Phi_{\rm tot}=(\Phi,\lambda\Phi,\lambda^2\Phi,\cdots)^T$ for the edge state, with a decay factor $\lambda$ between successive unit cells along the $y$ direction. 
The eigenvalue equation yielding stationary states then becomes
\begin{align}
  [\mathcal{S}_0(k_x)+\lambda \mathcal{S}^{'}]\Phi &= e^{-i\varepsilon(k_x)}\Phi, \label{eigen_equation1}\\
[\lambda^{-1}\mathcal{S}^{''}+\mathcal{S}(k_x)+\lambda\mathcal{S}^{'}]\Phi &= e^{-i\varepsilon(k_x)}\Phi, \label{eigen_equation2}
\end{align} 
where $\mathcal{S}_0(k_x)$ and $\mathcal{S}(k_x)$ represent scattering terms within each unit cell, located at the boundary of the network ($y=0$) and away from it, respectively.
The two other Ho-Chalker operators $\mathcal{S}^{'}$ and $\mathcal{S}^{''}$ represent coupling to the adjacent unit cells, located respectively below and above.   
As discussed before, the open boundary condition is implemented by having the scattering nodes on the edge of the system completely reflect the incoming waves with unit probability amplitude (this corresponds to setting $\theta_3=0$ for those particular nodes). 
Considering $\varepsilon (k_x) = 0$ and fixing one angle as $\theta_2=\pi/2$, we find two possible values for the decay constant, $\lambda_{\pm}=(\sqrt{2}\pm 1)^2\,\tan(\pi/4-\theta_1/2)/\tan(\theta_1/2)$ which are shown in Fig.~\ref{fig:max_coupling}(a). 
Since normalizable edge modes require $|\lambda|<1$, this criterion leads to one edge state for $\theta_1\in (\frac{1}{12}\pi, \frac{5}{12}\pi) \cup (-\frac{11}{12}\pi,-\frac{7}{12}\pi)$, and two counterpropagating edge states for $\theta_1\in(\frac{5}{12}\pi,\frac{13}{12}\pi)$.
The ranges for $\theta_1$ correspond to the regions of the STP and the horizontal line of $G=2$ emanating from it, respectively, as shown in Fig.~\ref{fig:phase_diag}(d).
Furthermore, by performing perturbation theory in $k_x$, we obtain the linear dispersion $\varepsilon(k_x)= v_E k_x$, with the velocity $|v_E|=\sqrt{2}/8 \approx 0.18$ of the edge state, see Fig.~\ref{fig:max_coupling}(b).

\section{Topological invariants}
\label{sec:inv}

In this section, we classify the different topological phases realized by the network model using topological invariants based on the scattering matrix of the system. 
In the HOTP, this approach is augmented with a symmetry indicator analysis whenever possible.

We first discuss the HOTP with its limiting case shown in Fig.~\ref{fig:limit_cases}(a). 
Note that, as discussed in the previous section, phase transitions separating the trivial phase from the HOTP can occur either via a bulk gap closing or via an edge gap closing, depending on whether rotation symmetry is preserved or not.
This marks a distinction between the topological phases classified as \emph{intrinsic} and those classified as \emph{extrinsic} in Ref.~\cite{Geier2018}. The intrinsic HOTPs host corner states due to bulk topological invariants enabled by lattice symmetries, and thus can only transition to a trivial phase by closing the bulk gap or by lattice symmetry breaking. In contrast, extrinsic HOTPs owe their topological midgap states to the strong nontrivial topology of the system's surface. Thus, in an extrinsic HOTP, corner modes are robust against lattice symmetry breaking (e.g. due to disorder), provided that the surface gap does not close. As we shall see in the following, the network model we introduce has a dual topology: When rotation symmetry is preserved it realizes an intrinsic HOTP protected by a bulk invariant, and when rotation is broken it realizes an extrinsic HOTP which does not rely on lattice symmetries.

A set of topological invariants detecting the presence of corner modes can be found in a four-terminal geometry, in which one lead is attached to each of the corners of the system, as shown in Refs.~\cite{Geier2018, Franca2019}. We provide more details on this transport geometry in Appendix~\ref{app:leads}. 
This setup only allows us to characterize the HOTP as an \emph{extrinsic} topological phase~\cite{Geier2018}, as corner leads cannot distinguish between topology due to a nontrivial bulk and due to nontrivial edges. 

In the HOTP, both bulk and edges are gapped, as exemplified in Fig.~\ref{fig:limit_cases}(a). 
As such, in each of the four corner leads, incoming modes are fully back-reflected and the corresponding corner reflection matrices $\mathfrak{r}_j$ ($j=1,2,3,4$) are unitary. 
Furthermore, the reflection blocks $\mathfrak{r}_j$ are also real due to particle-hole symmetry, leading to the $\mathbb{Z}_2$ invariants \citep{Akhmerov2011, Fulga2012}
\begin{equation} \label{eq:detR_inv}
\nu_j = \mathop{\rm sgn} \det \mathfrak{r}_j\,.
\end{equation}  
The values $\nu_j=\pm 1$ are topologically distinct since the only way to interpolate between them is by going through a point for which $\det{\mathfrak{r}_j}=0$, indicating that there is at least one fully transmitting mode connecting the leads at the different corners, either through the bulk or through the edges.
Since in the HOTP the corner modes simultaneously appear at all four corners, we will only investigate $\mathfrak{r}_1$ in the following.

Conventionally, a value $\nu_1 = 1$ indicates the trivial phase and $\nu_1 = -1$ the topological phase, due to the  $\pi$ phase shift caused by the resonant reflection of waves on a zero-energy state~\citep{Akhmerov2011, Fulga2012, Franca2019}. 
However, the relation is exactly inverted in our model, cf.\ Fig.~\ref{fig:limit_cases}(a). 
In particular, we see that the loop containing the corner mode is periodic and thus does not lead to a $\pi$ phase shift. 
As a result, we find $\nu_1 =1$ in the topological phase.
Analogously, the invariant is $\nu_1 = -1$ in the trivial phase, with decoupled limit shown in Fig.~\ref{fig:limit_cases}(b) due to the fact that the loop at the corner is antiperiodic. Therefore, $\nu_1 = -1$ describes the trivial phase of the network model 
\footnote{Finally, note that the scattering invariant does not distinguish a HOTP from the point $\theta_1 = \theta_2 = \pi/2; \theta_3 = \theta_4 = 0$, corresponding to the Majorana flat band limit Fig.~\ref{fig:limit_cases}(d), which is an isolated point in the phase diagram.}.

\begin{figure}[tb]
\centering
\includegraphics[width=\linewidth]{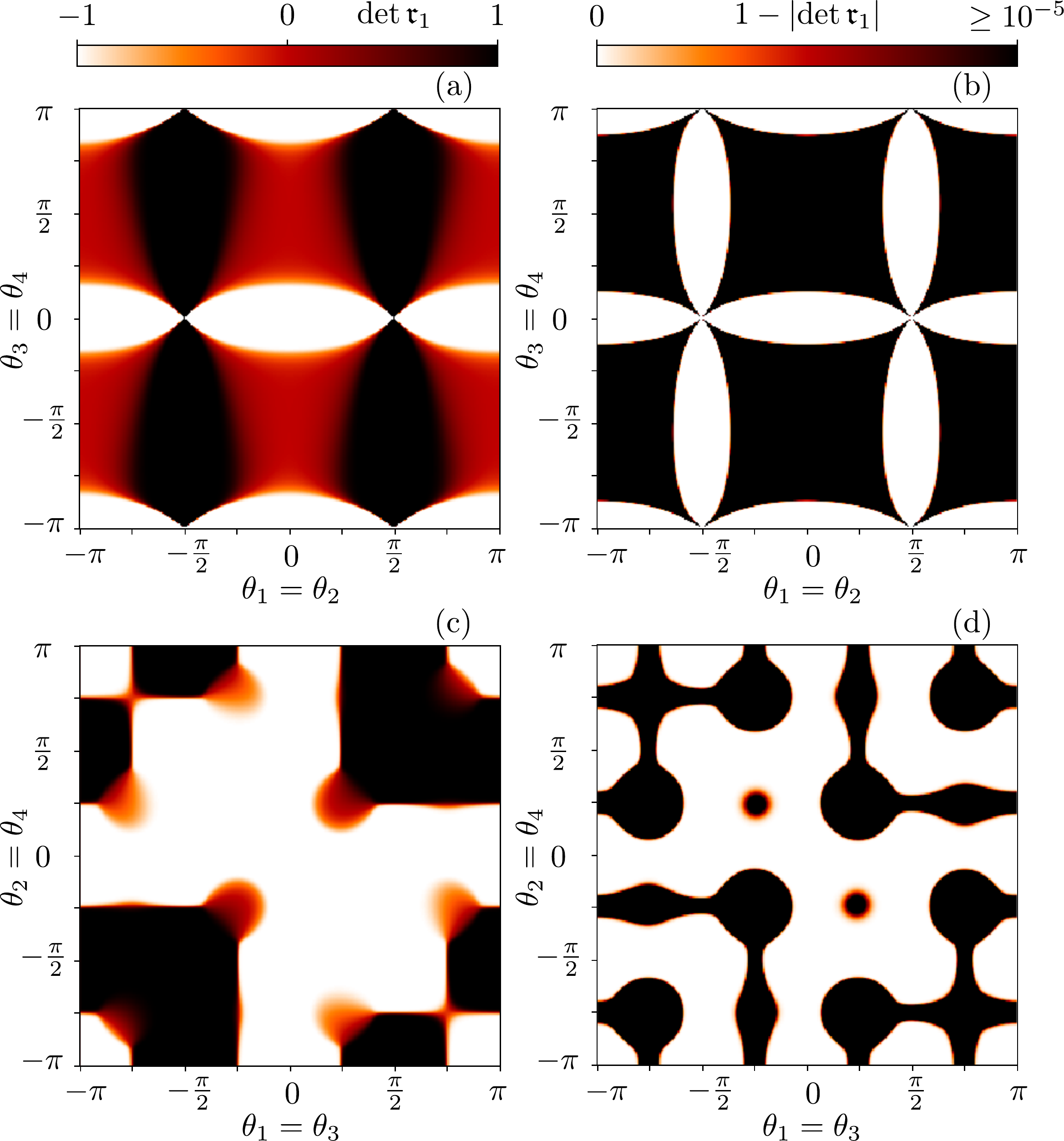}
\caption{In panel (a), $\det \mathfrak{r}_1$ is plotted for $\theta_1=\theta_2$ and $\theta_3=\theta_4$. 
Panel (b) shows the deviation of $|\det \mathfrak{r}_j|$ from unity, such that the phase transitions are clearly visible. 
Panel (c) [(d)]  depicts $\det \mathfrak{r}_1$ [$1-|\det \mathfrak{r}_1|$] for $\theta_1=\theta_3$ and $\theta_2=\theta_4$ such that the $C_4$ symmetry is broken. 
\label{fig:topological_invariant}
}
\end{figure}

In Fig.~\ref{fig:topological_invariant}, we show how $\det \mathfrak{r}_1$ depends on the angles $\theta_1, \theta_2, \theta_3$, and $\theta_4$. 
In Fig.~\ref{fig:topological_invariant}(a), we consider the $C_4$ symmetric network model. 
We observe a good agreement with the phase diagram of Fig.~\ref{fig:limit_cases}(b), which has been obtained from the two-terminal transmission probability. 
The invariant correctly identifies the HOTP and trivial regions, while the STP regions have $\det \mathfrak{r}_1 \approx 0$. 
However, the phase boundaries between the HOTP (or trivial phase) and the STP are not as easily seen as in the two terminal calculations. 
For this reason, we plot $1-|\det \mathfrak{r}_1|$ in Fig.~\ref{fig:topological_invariant}(b), and look at small deviations from $0$. 
The resulting phase diagram is now in excellent agreement with Fig.~\ref{fig:limit_cases}(b).   

Once $\theta_1=\theta_3$ and $\theta_2=\theta_4$, such that $C_4$ is broken, the system produces the phase diagram shown in Fig.~\ref{fig:topological_invariant}(c). 
Comparing to Fig.~\ref{fig:limit_cases}(c), it differs significantly. 
The difference occurs because the two-terminal transmission probability is only sensitive to gap closings along one set of edges, while the four-terminal geometry detects gap closings along all edges. 
For this reason, in Fig.~\ref{fig:limit_cases}(c), we only observe the gap closings on the top and bottom edges, while Fig.~\ref{fig:topological_invariant}(c) contains information on gap closings also on the left and right edges. 
Again the features are better visible in Fig.~\ref{fig:topological_invariant}(d), where we plot $1-|\det \mathfrak{r}_1|$. 
We see that the HOTP is separated from the trivial phase by a gap closing either in the bulk or along the edges, as is characteristic of extrinsic topological phases.

In the presence of a $C_4$ symmetry, however, the HOTP is an intrinsic, bulk topological phase, such that a transition to a different phase requires the closing and reopening of the bulk (and not just the edge) gap~\cite{Benalcazar2017prb}. 
To show this fact mathematically, we define a bulk $\mathbb{Z}_2$ topological invariant $Q=\pm 1$ based on the $C_4$ symmetry eigenvalues of the Ho-Chalker eigenstates for a translationally invariant system at the $C_4$ invariant points of Brillouin zone $\Gamma = (0,0)$ and $M = (\pi, \pi)$, similar to how this is done for static Hamiltonian systems \cite{Benalcazar2017prb, Benalcazar2018, Schindler2019, Geier2020, Roberts2020}. It reads 
\begin{equation} \label{eq:symm_indicator}
  \mathcal{Q} = d_4^+ (M) d_4^+ (\Gamma)^* =  d_4^- (M) d_4^- (\Gamma)^*,
\end{equation} 
where $d_4^\pm$ are the $\mathcal{C}_4$ eigenvalues that square to $\pm i$ in the "occupied bands" of the Ho-Chalker operator. 
Even though the latter has a periodic spectrum, the phase-rotation and particle-hole symmetries enable us to unambiguously define the occupied bands as those in the interval $-\pi/4 < \varepsilon < 0$ (which is half of the fundamental phase domain).

Since the $C_4$ symmetry operator of Eq.~\eqref{eq:C4operator} obeys $\mathcal{R}^4 = -1$, its eigenvalues are $e^{\pm i \pi/4}$ and $e^{\pm i 3 \pi/4}$.
In the HOTP, we find that the two occupied bands in the fundamental domain have eigenvalues $e^{-i \pi/4}$ and $e^{-i 3\pi/4}$ at the $\Gamma$ point. At the $M$ point, these bands change their symmetry eigenvalues to $e^{i 3\pi/4}$ and $e^{i\pi/4}$, respectively. 
We thus obtain $\mathcal{Q} = -1$ which corresponds to the HOTP. In the trivial phase, we find the $C_4$ symmetry indicators of occupied bands do not change between $\Gamma$ and $M$, which leads to a value of $\mathcal{Q} = 1$. 

Finally, the topological invariant of the STP can be expressed as the winding number~\cite{Braunlich2010, Fulga2012, Fulga2016}
\begin{equation}\label{eq:winding}
\mathcal{W}=\frac{1}{2\pi i}\int_{0}^{2\pi}\!d\varphi\,\frac{d}{d\varphi}\text{log}\:\text{det}\:\mathfrak{r}(\varphi),
\end{equation}
where $\mathfrak{r}(\varphi)$ is the reflection block of the two-terminal scattering matrix. 
Here, we impose twisted boundary conditions in the vertical direction, with twist angle $\varphi$, such that $\varphi=0$ corresponds to PBC. 
We have checked that all phases of Fig.~\ref{fig:phase_diag} in which $G=0$ with PBC and $G=1$ with OBC have $\mathcal{W}= -1$, confirming that they are indeed STPs.

\section{Disorder}
\label{sec:disorder}

We study the effect of disorder on the network model by adding a random term $\delta \theta_i$ to each $\theta_i$, drawn uniformly for each angle from the uniform distribution $[-w, w]$, with the disorder strength $w$, $0\leq w \leq \pi$.
The disorder breaks all lattice symmetries but preserves particle-hole symmetry since the disordered scattering matrices remain orthogonal. As the system is still in the class D, we expect that, with sufficiently strong disorder, the phase boundaries separating the HOTP, STP, and trivial phase will evolve into a delocalized phase similar to the `thermal metal' which occurs in static disordered systems \cite{mirlin2008review, chalker1999super, chalker2001super, Cho1996, medvedyeva2010, Fulga2020}.
In contrast, for weak disorder, the system should be insensitive to disorder, with only minor modifications of the phase diagram with respect to the clean case.

\begin{figure}[tb]
\centering
\includegraphics[width=\linewidth]{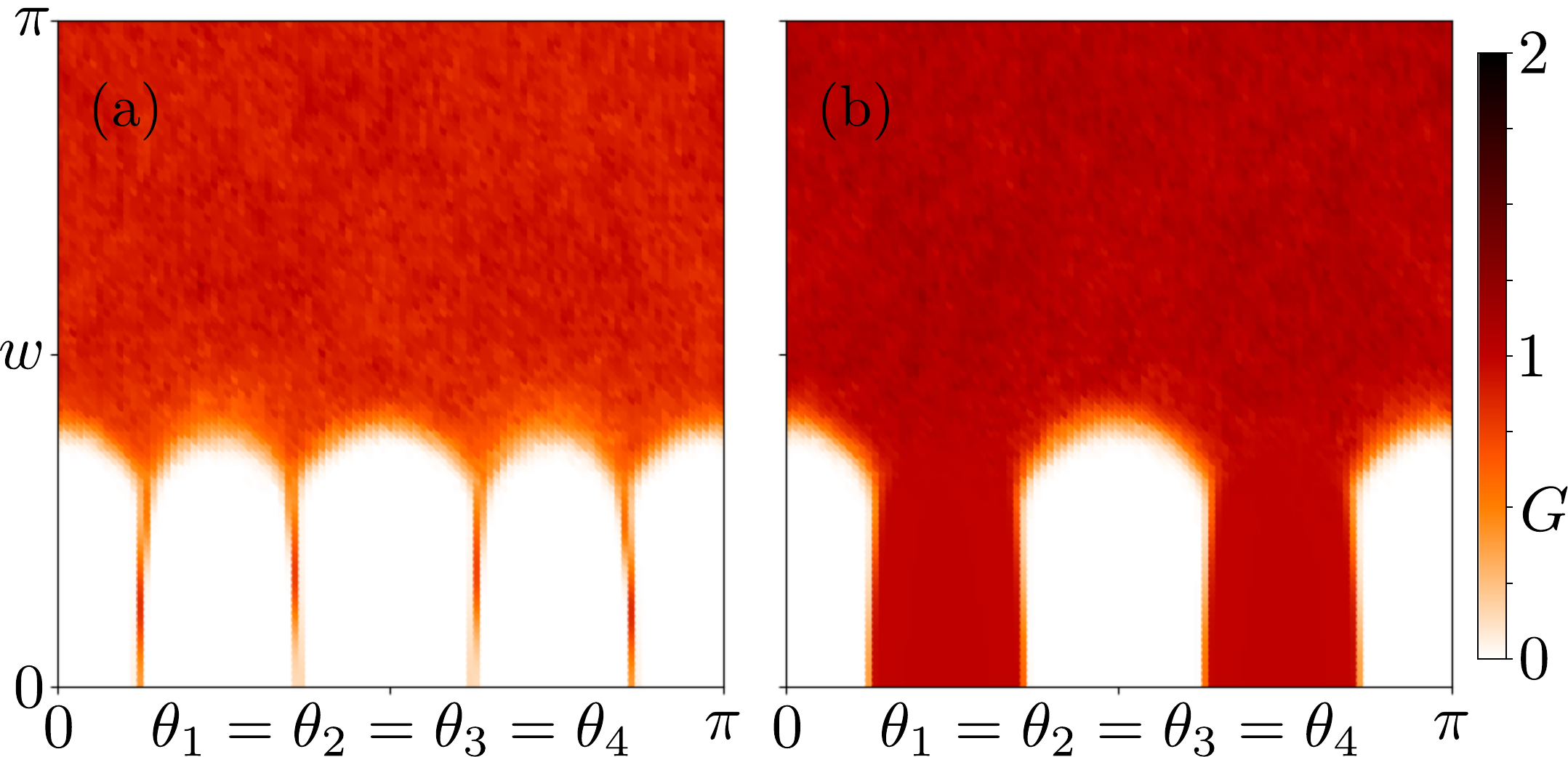}
\caption{Dimensionless conductance $G$ computed for a network model with $20\times 20$ unit cells as a function of $w$ and $\theta_i$ ($i=1,2,3,4$) using PBC (a) and OBC (b). 
Each point is obtained by averaging over $100$ disorder realizations.
\label{fig:disorder_phase_diagram_all_theta}
}
\end{figure}

We verify this behavior by computing the average two-terminal transmission of the disordered network model. 
In Fig.~\ref{fig:disorder_phase_diagram_all_theta}, without disorder ($w=0$), we observe all three phases: trivial, STP, and HOTP. 
All of them have a localized bulk and thus are robust against disorder up to a moderate disorder strength, $w<1$.
For the HOTP, this robustness is a consequence of the extrinsic nature of the topological phase. Even though the rotation symmetry enabling the bulk invariant Eq.~\eqref{eq:symm_indicator} is broken, the particle-hole symmetric corner modes cannot be removed as long as the bulk and surfaces remain localized.
However, as disorder strength is increased beyond $w\simeq 1$, each phase undergoes a bulk localization-delocalization transition, and its topological properties are lost.

\begin{figure}[tb]
\centering
\includegraphics[width=\linewidth]{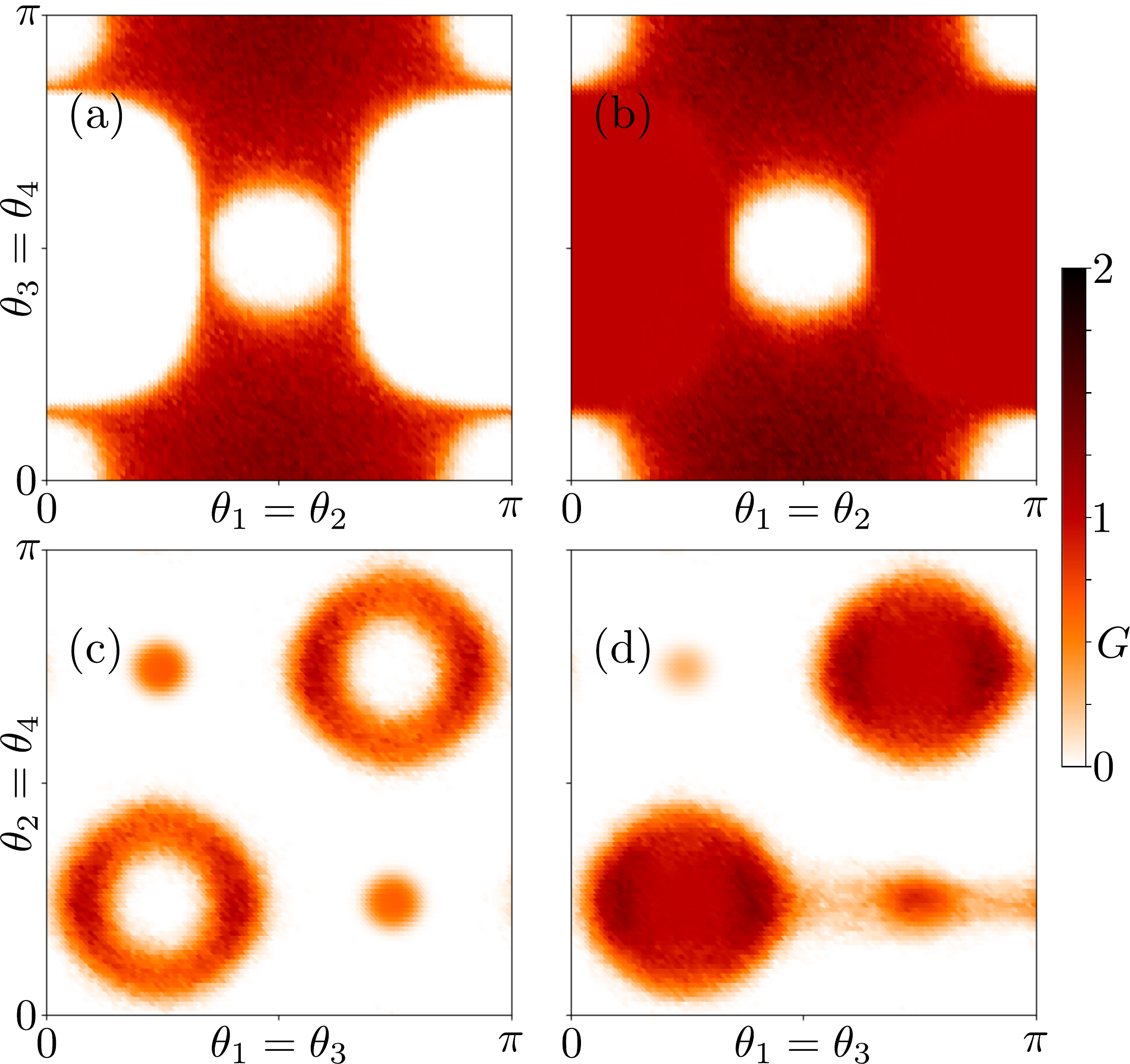}
\caption{Dimensionless conductance $G$ as a function of $\theta_1=\theta_2$ and $\theta_3=\theta_4$ ($C_4$ symmetric) in panels (a) and (b), and as a function of $\theta_1=\theta_3$ and $\theta_2=\theta_4$ (broken $C_4$ symmetry) in panels (c) and (d). 
In panels (a) and (c), we use periodic boundary conditions, whereas panels (b) and (d) we use open boundary conditions. 
All plots are computed for a network model consisting of $50\times 50$ unit cells, with disorder strength $w=1.1$, and each point is obtained by averaging over 50 disorder realizations.
\label{fig:disorder_phase_diagram}
}
\end{figure}

Next, we fix $w=1.1$ and in Fig.~\ref{fig:disorder_phase_diagram} plot the average transmission probability as a function of different $\theta$s, to show how the phase diagram of Fig.~\ref{fig:phase_diag} changes in the presence of moderate disorder.
Among the four decoupled limits of the clean $C_4$ symmetric case, the Majorana flat band is first to be delocalized due to its gapless bandstructure. 
In addition, the network model near the phase boundaries (or any other gap closing points) can also be easily delocalized, as seen in Figs.~\ref{fig:disorder_phase_diagram}(a) and \ref{fig:disorder_phase_diagram}(b). 
As such, we observe that in the disordered phase diagram the phase transition lines separating HOTP, STP, and trivial phases evolve into finite-width delocalized regions, as do the gapless points at $(\theta_1, \theta_2)=(\pi/4,-\pi/4)$ and $(-\pi/4,\pi/4)$. 
Nevertheless, the STP and HOTP retain their nontrivial nature, despite the presence of disorder. 
The transmission probability still changes from $G=0$ with PBC to $G=1$ with OBC in the STP, signaling that the chiral Majorana edge mode is robust against disorder. 
We have confirmed that the scattering matrix invariants characterizing the HOTP and STP retain their nontrivial values.
Finally, we observe that the counterpropagating edge modes separating HOTP and trivial phase in the $C_4$ broken case [Fig.~\ref{fig:disorder_phase_diagram}(d)] remain conducting, albeit with a conductance $G<2$. 
In fact they are protected by an average translation symmetry of the ensemble of disorder realizations \cite{Brouwer2000, Brouwer2003, Fulga2014, Diez2014, Diez2015reflection}.

\section{ Network models in experiment }
\label{sec:xp}

The first network model was introduced by Chalker and Coddington to describe the low-energy, course-grained physics of a static system: the quantum Hall effect \cite{chalker1988}. 
Unidirectional links represented chiral quantum Hall edge modes and the scattering centers were the saddle points of the chemical potential at which these edge modes mix. 
Using a similar language, we have introduced our particle-hole symmetric network model at the beginning of Sec.~\ref{sec:nm} in terms of chiral Majorana modes. 
These are connected to each other by scattering processes, as would occur in a static $p$-wave superconductor, for instance.

Network models, however, are dynamical systems. 
This fact, as pointed out in Refs.~\cite{zirnbauer1999, chong2014floquet}, is related to the unitary nature of the operator used to describe network models, the Ho-Chalker operator. 
The latter encodes all of the scattering processes acting on the network wavefunction, which can be modeled as a discrete-time evolution [see Eq.~\eqref{eq:discrete_time}]. 
Its eigenstates are the steady states of the network, those which preserve their shape even upon multiple scatterings [see Eq.~\eqref{eq:Ho-Chalker}].

The understanding of the network model as a dynamical system has led to its experimental realization in different platforms, such as microwave circuits \cite{Hu2015}, or systems of coupled photonic \cite{Hafezi2011, Afzal2020} and plasmonic \cite{Gao2016, Gao2018} ring resonators. 
Excitations propagate along the ring resonators (or along optical fibers, in the microwave circuit implementation), and have the chance of scattering to neighboring resonators (optical fibers), depending on the distance, or amount of coupling between them. 
The entire array can then be described by a matrix of scattering amplitudes (the Ho-Chalker operator), and the dynamical phases acquired by its steady states after they scatter represent the eigenphases of this matrix. 
This means that the eigenphases of a network model are experimentally tunable parameters, similar to how the energies of states can be tuned by varying the chemical potential of static systems. 
The eigenphases can be changed by altering the ratio between the wavelength of the photons (plasmons) and the length of the optical fiber (resonator). 
In fact, recent experiments managed to probe the network model over the entire range $\varepsilon \in [0, 2\pi)$ \cite{Afzal2020}. 
In practice, these measurements oftentimes involve directly visualizing states as they scatter through the network, similar to how Floquet topological phases are measured in photonic crystals \cite{elHassan2019corner}.

Finally, we note that the large degree of control available in meta-material platforms enables the experimental realization of network models subject to various constraints, including particle-hole symmetry. 
In a traditional condensed matter setup, the latter would require superconductivity, hence our discussion in terms of Majorana modes in Sec.~\ref{sec:nm}. 
On the network model level, however, particle-hole symmetry simply means that the scattering matrices are real. 
This can be ensured in experiment by coupling the optical fibers or ring resonators to each other such that each scattering process involves the same phase difference between incoming and outgoing waves. 
In fact, network models with particle-hole symmetric spectra have already been measured in Refs.~\cite{Gao2016, Afzal2020}.

\section{Conclusion}
\label{sec:conc}

We have introduced a network model 
realization of a higher-order topological phase. 
The system shows many similarities to the HOTP present in static or periodically driven Hamiltonian systems, namely it hosts zero-modes localized at the corners, which are protected by the bulk and edge gaps due to a combination of a particle-hole symmetry and a fourfold rotation symmetry. 
However, the network model additionally features  a phase-rotation symmetry. As a result, there are a total of 16 topologically protected corner states, more than for a fourfold symmetric static or Floquet system. 
The HOTP is robust against disorder, being characterized by nontrivial scattering matrix invariants as well as by a more conventional invariant based on symmetry indicators.

We hope that our work will motivate further theoretical and experimental research into network models. 
The latter (as well as HOTPs) have been successfully realized using optical fibers and coupled ring resonators \cite{Ozawa2019, Hafezi2011, Hafezi2013, Liang2013}. 
In some of these platforms it is also possible to directly measure the scattering matrix invariants \cite{Hu2015, Wang2017}. 
Therefore, the HOTP network proposed here, and its invariants could be experimentally probed in these platforms.
Finally, our work provides an example of a network model for a point group symmetry protected topological phase. 
These are by now the most studied types of topology in condensed-matter, Hamiltonian systems, but have not yet been explored on the level of network models. We believe that different types of topological crystalline phases protected by different lattice symmetries may be realized using arrays of scattering matrices, and we plan to further explore this direction in the future.

\begin{acknowledgments}
We thank Ulrike Nitzsche for technical assistance, as well as Anton Akhmerov, Cristoph Groth, and Michael Wimmer for useful discussions and for sharing their code with us.
This work was supported by the Deutsche Forschungsgemeinschaft (DFG, German Research Foundation) under Germany's Excellence Strategy through the W\"{u}rzburg-Dresden Cluster of Excellence on Complexity and Topology in Quantum Matter -- \emph{ct.qmat} (EXC 2147, project-id 390858490) and under Germany's Excellence Strategy -- Cluster of Excellence Matter and Light for Quantum Computing (ML4Q) EXC 2004/1 -- 390534769.
\end{acknowledgments}

\appendix

\section{BBH Model} \label{app:BBH}

In this section, we briefly review the main features of the BBH model and explain its similarities to our network model. 

\begin{figure}[tb]
\centering
\includegraphics[width=\linewidth]{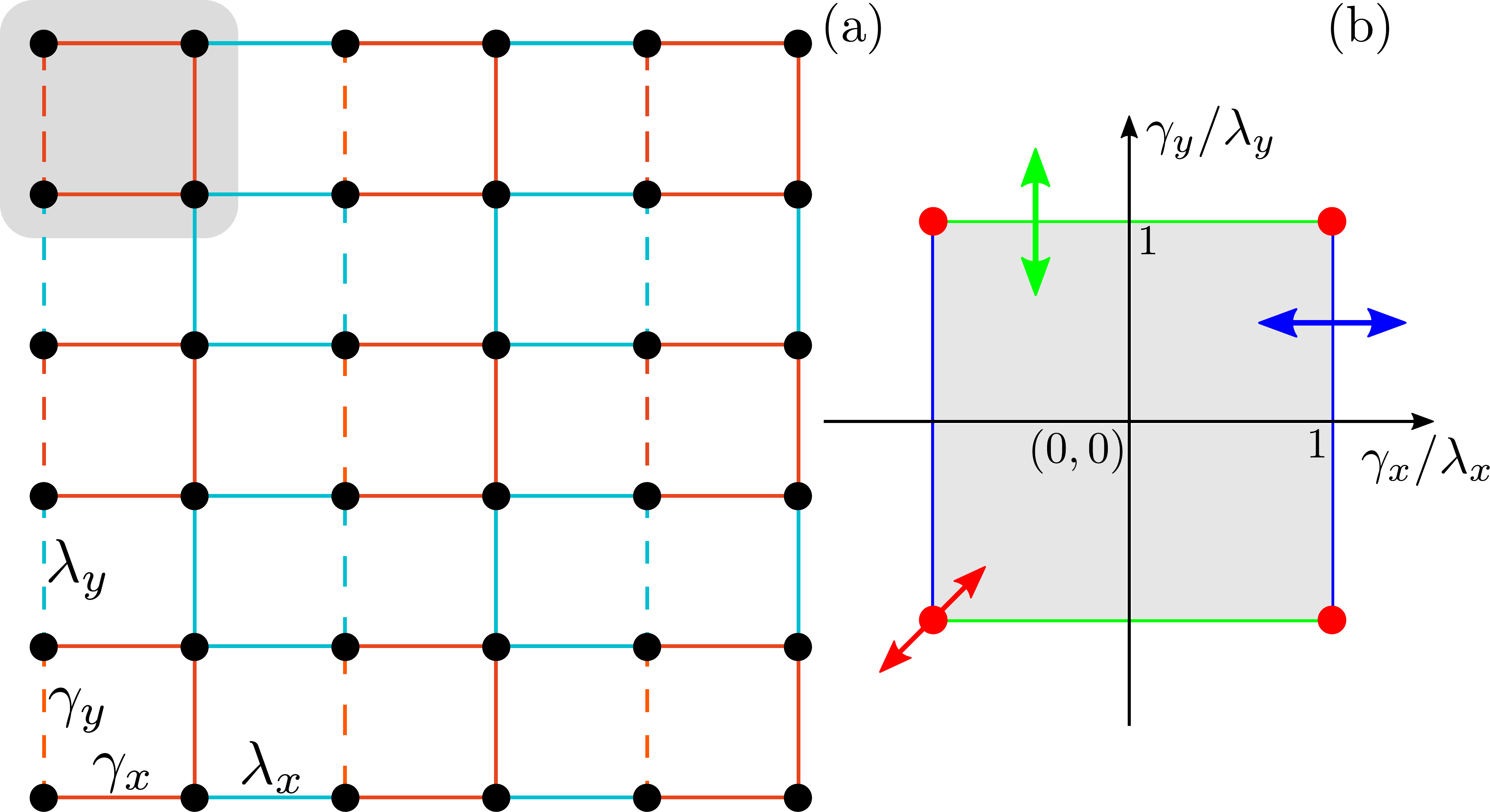}
\caption{(a) Sketch of the BBH model. 
The black dots represent sites. The intracell/intercell hoppings are shown in red/blue, respectively, and dashed lines denote negative hopping strengths (used to implement the $\pi$ fluxes). 
The gray rectangle depicts a unit cell. 
(b) Phase diagram of the BBH model Eq.~\eqref{eq:SSH}. 
The gray area indicates the HOTP, while the red circles denote the points of the bulk gap closings. 
Moreover, the gap closings along the  $x$- and $y$-edges are shown in blue and green, respectively. 
The arrows indicate possible paths in parameter space that lead to bulk gap (red), $x$-edge (blue) and $y$-edge (green) gap closings.
\label{fig:BBHmodel}
}
\end{figure}

The BBH model consists of noninteracting spinless electrons hopping on a square lattice. 
Only the nearest-neighbor hoppings are nonzero, and they are staggered in both directions, consisting of alternating weak and strong hopping strengths, as shown in Fig.~\ref{fig:BBHmodel}(a). 
A magnetic $\pi$-flux pierces each plaquette, ensuring that the bulk and edges are gapped.

The momentum-space Hamiltonian reads
\begin{align}\label{eq:SSH}
\begin{split}
h(\bm{k})  =& (\gamma_x  + \lambda_x \cos{k_x}) \tau_x \sigma_0 - \lambda_x \sin{k_x} \tau_y \sigma_z \\
& -(\gamma_y  + \lambda_y \cos{k_y}) \tau_y \sigma_y - \lambda_y \sin{k_y} \tau_y \sigma_x.
\end{split}
\end{align}
Here, $\bm{k}=(k_x,k_y)$ are the two momenta and Pauli matrices $\tau$ and $\sigma$ represent the sublattice degree of freedom and the site degree of freedom, respectively. 
Intracell hoppings are $\gamma_x$ and $\gamma_y$, while $\lambda_x$ and $\lambda_y$ denote intercell couplings. 

The system has a particle-hole symmetry $\mathcal{P} =\tau_z{\cal K}$, where ${\cal K}$ denotes complex-conjugation. 
Once $\gamma_x = \gamma_y \equiv \gamma$ and $\lambda_x = \lambda_y \equiv \lambda$, the system is $C_4$ symmetric. 
It obeys $\mathcal{R} h(k_x,k_y) \mathcal{R}^{-1} = h(k_y, -k_x)$, where the $C_4$ symmetry operator reads~\cite{Benalcazar2017} 
\begin{equation}
\mathcal{R}=\begin{pmatrix}
  0 & \sigma_0 \\
  -i \sigma_y & 0
  \end{pmatrix}.
\end{equation}

We now discuss the similarities between the BBH model and network model, and focus first on the $C_4$ symmetric limit. 
If the intracell coupling strength is much weaker than intercell coupling strength, then both systems with open boundaries would support four topologically protected zero-energy corner states. 
Then, in the opposite limit, both systems would also yield a trivially gapped bulk.
Furthermore, for the BBH model, the bulk gap closing point is at $\bm{k} = (0,0)$ [$\bm{k} = (\pi,\pi)$] and occurs when $\gamma = - \lambda$ [$\gamma = \lambda$], as shown on Fig.~\ref{fig:BBHmodel}(b). 
Similarly, the network model also possesses two types of gap closing at $\bm{k} = (0,0)$ and $\bm{k} = (\pi,\pi)$, shown in Fig.~\ref{fig:phase_diag}.

Once the $C_4$ symmetry is broken, the BBH model Eq.~\eqref{eq:SSH} is in a HOTP as long as $\gamma_x < \lambda_x$ and $\gamma_y < \lambda_y$. 
As denoted in Fig.~\ref{fig:BBHmodel}(b), the phase without corner states can
be reached by the $x$-edge gap closing ($\gamma_x = \lambda_x$) or the $y$-edge gap closing ($\gamma_y = \lambda_y$). 
The consequence of the edge gap closing is the appearance of two counterpropagating modes per edge. 
Then in the network model, once the $C_4$ symmetry is broken, the system is also in a HOTP as long as this phase does not reach edge gap closing lines, which correspond to a counterpropagating modes with $G=2$, see Figs.~\ref{fig:phase_diag}(d), \ref{fig:topological_invariant}(c), and \ref{fig:topological_invariant}(d).

\section{Transport geometries}
\label{app:leads}

In this section, we detail two transport geometries used for 
obtaining the phase diagrams in Figs. \ref{fig:phase_diag} and \ref{fig:topological_invariant}, respectively.
We attach the code used to perform these calculations as an ancillary file.

\begin{figure}[tb]
\centering
\includegraphics[width=\linewidth]{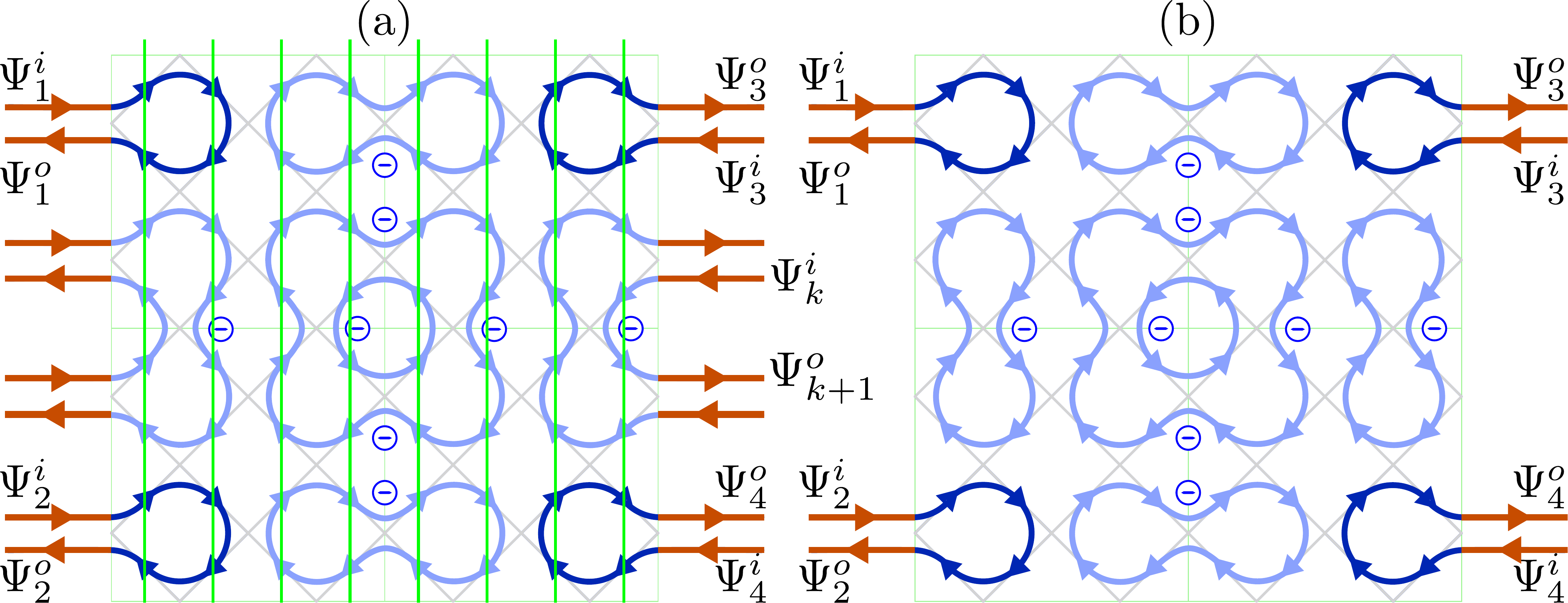}
\caption{In panel (a) we show the two-terminal transport geometry. 
For simplicity, we consider a $N_x \times N_y = 2 \times 2$ system in a HOTP. 
The lead modes are depicted in red. 
The modes at corners are indexed with $k = 1,2,3,4$, while all other modes are represented with $k$ $(k> 4)$. 
The vertical green lines mark the different 1D slices which are combined to yield the full scattering matrix (see text). 
In panel (b), we show the four-terminal transport geometry with the leads  only connected to the corners of the system.
}
\label{fig:scattering_geometry}
\end{figure}

For the conductance calculations in Fig.~\ref{fig:limit_cases}, we use two semi-infinite leads attached to opposite vertical edges of the 2D system, as shown in Fig.~\ref{fig:scattering_geometry}(a). 
Each lead is composed of an array of Majorana modes, thus preserving particle-hole symmetry. 
As such, in the Majorana basis the scattering matrix is real, and it has a size of $4N_y \times 4N_y$ for an $N_x \times N_y$ unit cell system. 
To compute this scattering matrix, we consider that the system consists of a sequence of 1D slices, as shown for a $2 \times 2$ system in Fig.~\ref{fig:scattering_geometry}(a). 
Each slice contains an array of scattering nodes arranged vertically. 
Every slice has $4N_y$ incoming/outgoing modes, related by the scattering matrix
\begin{equation}\label{eq:Sslice}
S_{\rm slice} =\begin{pmatrix}
\mathfrak{r}_{\rm slice} & \mathfrak{t}_{\rm slice}'\\ 
\mathfrak{t}_{\rm slice} & \mathfrak{r}_{\rm slice}'
\end{pmatrix},
\end{equation}
where $\mathfrak{r}_{\rm slice}$ and $\mathfrak{r}_{\rm slice}'$ are reflection matrices of the slice from left to left and from right to right, respectively. 
The transmission matrix from left to right, and vice versa are denoted by $\mathfrak{t}_{\rm slice}$ and $\mathfrak{t}_{\rm slice}'$.  

The final scattering matrix of the full network is obtained by combining scattering matrices of all $2N_x$ slices. 
If we label the scattering matrices for two adjacent slices as $S_L$ (left) and $S_R$ (right), respectively, then the  scattering matrix $S_C$ of the new slice after combination has matrix blocks~\footnote{P. W. Brouwer, Ph.D. thesis, Leiden University, 1997.} 
\begin{align}\label{eq:combiningS}
\begin{split}
& \mathfrak{r}_C = \mathfrak{r}_L + \mathfrak{t}_L' \mathfrak{r}_R (1 - \mathfrak{r}_L' \mathfrak{r}_R)^{-1} \mathfrak{t}_L, \\
& \mathfrak{t}_C = \mathfrak{t}_R (1 - \mathfrak{r}_L' \mathfrak{r}_R)^{-1} \mathfrak{t}_L, \\
& \mathfrak{t}_C' = \mathfrak{t}_L' (1 - \mathfrak{r}_R \mathfrak{r}_L')^{-1} \mathfrak{t}_R', \\
& \mathfrak{r}_C' = \mathfrak{r}_R' + \mathfrak{t}_R \mathfrak{r}_L' (1 - \mathfrak{r}_R \mathfrak{r}_L')^{-1} \mathfrak{t}_R'.
\end{split}
\end{align}
Therefore, after multiple combination processes, one can gradually include all slices and obtain the combined scattering matrix for the whole system. 
With this, we can calculate the conductance as $G =  \text{tr}(\mathfrak{t} \mathfrak{t}^{\dagger})$, where $\mathfrak{t}$ is the transmission block of the full, two-terminal scattering matrix: $S_{\rm 2-term}$. 

For the calculation of the topological invariant 
$\nu_1$ to capture the presence of corner modes, we use a four-terminal geometry depicted in Fig.~\ref{fig:scattering_geometry}(b). 
The associated scattering matrix, $S_{\rm 4-term}$, can be calculated from $S_{\rm 2-term}$ in the following manner. 
We reorder the incoming and outgoing modes of $S_{\rm 2-term}$ such that it has the structure
\begin{equation}\label{eq:S2x2_intermediate}
S_{\rm 2-term} =\begin{pmatrix}
S_A & S_B \\
S_C & S_D
\end{pmatrix}.
\end{equation}   
Here, $S_A$ is a $4\times4$ matrix that relates four corner incoming and outgoing modes, denoted in Fig.~\ref{fig:scattering_geometry}(b). 
Therefore $(\Psi_{1}^o, \Psi_2^o, \Psi_3^o , \Psi_4^o)^T = S_A (\Psi_1^i, \Psi_2^i, \Psi_3^i, \Psi_4^i)^T$. 
In general, $S_A$ is not necessarily a unitary matrix as the corner incoming modes can contribute to the outgoing modes not pinned at corners. 
Matrix $S_B$ has $4 \times (4N_y-4)$ entries, and relates four outgoing corner modes with all remaining incoming amplitudes, while $S_C$ does the opposite. 
Finally, the matrix $S_D$ has $(4N_y-4) \times (4N_y-4)$ entries, and relates all noncorner incoming to noncorner outgoing modes. 

To remove leads from the $x$-edges of the network model, except at the points of corner terminals, we assume $\Psi^i_k = \Psi_k^o$, for $4<k< 4N_y$. 
Then, solving the set of these $4N_y-4$ number of equations, we can eliminate all but four incoming/outgoing modes. 
The resulting scattering matrix reads 
\begin{equation}
S_{\rm 4-term} = S_A + S_B (1 - S_D)^{-1} S_C.
\end{equation}

\section{Symmetries of the network model}
\label{app:sym}

We discuss here how different symmetries of the Ho-Chalker operator can be used to fully characterize the phase diagram discussed in the main text.

In the Ho-Chalker operator, each nonzero element is proportional to either $\sin\theta_i$ or $\cos\theta_i$ ($i=1, 2, 3, 4$). 
This leads to
\begin{equation}
\mathcal{S}(\theta_1,\theta_2,\theta_3,\theta_4)=-\mathcal{S}(\theta_1+\pi,\theta_2+\pi,\theta_3+\pi,\theta_4+\pi).
\label{eq:sym1}
\end{equation}
The minus sign corresponds to a $\pi$ shift of all eigenphases, which, together with the $\pi/2$ shift caused by phase-rotation symmetry, means that both $\mathcal{S}(\theta_1,\theta_2,\theta_3,\theta_4)$ and $\mathcal{S}(\theta_1+\pi,\theta_2+\pi,\theta_3+\pi,\theta_4+\pi)$ have identical spectra and therefore share the same topological properties. 

Another symmetry constraining the phase diagram is
\begin{equation}
U_1\mathcal{S}(\theta_1,\theta_2,\theta_3,\theta_4)U_1^\dagger=\mathcal{S}(-\theta_1,-\theta_2,\theta_3,\theta_4)
\label{eq:sym2}
\end{equation}
with $U_1=\mathbbm{1}_{4\times 4}\otimes {\rm diag}(1, 1, -1, 1)$. 
Because $\mathcal{S}(\theta_1,\theta_2,\theta_3,\theta_4)$ and $\mathcal{S}(-\theta_1,-\theta_2,\theta_3,\theta_4)$ have identical spectra, they are in the same topological phase.

Next, one can find that
\begin{multline}
\mathcal{S}(\theta_1,\theta_2,\theta_3,\theta_4,k_x,k_y)\\
=\mathcal{S}(\theta_1,\theta_2,-\theta_3,-\theta_4, k_x +\pi,k_y+\pi).\label{eq:sym3}
\end{multline}
Therefore, these two systems are also in the same topological class: STP, HOTP, or trivial.

Finally, we introduce a fourth symmetry,
\begin{equation}
U_2\mathcal{S}(\theta_1,\theta_2,\theta_3,\theta_4)U_2^\dagger=\mathcal{S}(\theta_1+\pi,\theta_2+\pi,-\theta_3,-\theta_4),
\end{equation}
where
\begin{equation}
U_2=\begin{pmatrix}
1&0&0&0\\0&-1&0&0\\0&0&1&0\\0&0&0&-1
\end{pmatrix}
\otimes
\begin{pmatrix}
1&0&0&0\\0&-1&0&0\\0&0&-1&0\\0&0&0&-1
\end{pmatrix}.
\end{equation}

Taken together, the symmetries listed above help to explain how the different topological phases of Fig.~\ref{fig:phase_diag} are related to each other.

\section{$C_4$ symmetry when $\theta_1=\theta_3=-\theta_2=-\theta_4$}
\label{app:c4}
Along the diagonal line of $\theta_1=-\theta_2$ in Figs.~\ref{fig:phase_diag}(c) and \ref{fig:phase_diag}(d), the system still preserves $C_4$ symmetry, but the $C_4$ symmetry condition becomes,
\begin{equation}
\mathcal{R}'\mathcal{S}(k_x,k_y)\mathcal{R}'^\dagger=\mathcal{S}(-k_y+\pi,k_x+\pi),
\label{eq:def_newC4}
\end{equation}
with
\begin{equation}\label{eq:newC4operator}
  {\cal R}' =
 \begin{pmatrix}
  0 & 0 & 0 & \!{\cal R}'_4 \\
  \!{\cal R}'_3\! & 0 & 0 & 0 \\
  0 & \!{\cal R}'_2\! & 0 & 0 \\
  0 & 0 & \!{\cal R}'_1\! & 0 
  \end{pmatrix}\!,\;
{\cal R}'_i = K_i 
 \begin{pmatrix}
  1 & 0 & 0 & 0 \\
  0 & 0 & 0 & 1 \\
  0 & 0 & -1 & 0 \\
   0 & 1 & 0 & 0
 \end{pmatrix}\!,
\end{equation}
with the $K_i$ matrices defined as in Eq.~\eqref{eq:C4operator}.
If momentum space is discretized (for instance in a torus geometry), then this symmetry constraint needs an even number of discretized momenta $k_x$ and $k_y$ to be preserved. 
Correspondingly, in real space the rotation axis is positioned at the corner of the unit cell. 

At the two high symmetry points $X=(\pi,0)$ and $Y=(0,\pi)$ we have $[\mathcal{R}', \mathcal{S}] = 0$ in momentum space. 
Like Section \ref{sec:inv}, we can again determine the $C_4$ eigenvalues of the occupied bands. 
Since $(\mathcal{R}')^4=-1$, these eigenvalues are $e^{\pm i\pi/4}$ and $e^{\pm i3\pi/4}$.
When changing $\theta_1=-\theta_2$ so as to move on the diagonal of Fig. \ref{fig:phase_diag}(c), we observe that there is a bulk gap closing and reopening which occurs at the point $\pi/4$. 
This gap closing is parabolic as a function of the $\theta$s, and so does not involve a band inversion in which states with different $C_4$ eigenvalues cross $\varepsilon=0$ and are interchanged. 
As such, no topological phase transition takes place at $\theta_1=-\theta_2=\pi/4$.

\bibliography{Refs}

\begin{thebibliography}{98}%
\makeatletter
\providecommand \@ifxundefined [1]{%
 \@ifx{#1\undefined}
}%
\providecommand \@ifnum [1]{%
 \ifnum #1\expandafter \@firstoftwo
 \else \expandafter \@secondoftwo
 \fi
}%
\providecommand \@ifx [1]{%
 \ifx #1\expandafter \@firstoftwo
 \else \expandafter \@secondoftwo
 \fi
}%
\providecommand \natexlab [1]{#1}%
\providecommand \enquote  [1]{``#1''}%
\providecommand \bibnamefont  [1]{#1}%
\providecommand \bibfnamefont [1]{#1}%
\providecommand \citenamefont [1]{#1}%
\providecommand \href@noop [0]{\@secondoftwo}%
\providecommand \href [0]{\begingroup \@sanitize@url \@href}%
\providecommand \@href[1]{\@@startlink{#1}\@@href}%
\providecommand \@@href[1]{\endgroup#1\@@endlink}%
\providecommand \@sanitize@url [0]{\catcode `\\12\catcode `\$12\catcode
  `\&12\catcode `\#12\catcode `\^12\catcode `\_12\catcode `\%12\relax}%
\providecommand \@@startlink[1]{}%
\providecommand \@@endlink[0]{}%
\providecommand \url  [0]{\begingroup\@sanitize@url \@url }%
\providecommand \@url [1]{\endgroup\@href {#1}{\urlprefix }}%
\providecommand \urlprefix  [0]{URL }%
\providecommand \Eprint [0]{\href }%
\providecommand \doibase [0]{http://dx.doi.org/}%
\providecommand \selectlanguage [0]{\@gobble}%
\providecommand \bibinfo  [0]{\@secondoftwo}%
\providecommand \bibfield  [0]{\@secondoftwo}%
\providecommand \translation [1]{[#1]}%
\providecommand \BibitemOpen [0]{}%
\providecommand \bibitemStop [0]{}%
\providecommand \bibitemNoStop [0]{.\EOS\space}%
\providecommand \EOS [0]{\spacefactor3000\relax}%
\providecommand \BibitemShut  [1]{\csname bibitem#1\endcsname}%
\let\auto@bib@innerbib\@empty
\bibitem [{\citenamefont {Schnyder}\ \emph {et~al.}(2008)\citenamefont
  {Schnyder}, \citenamefont {Ryu}, \citenamefont {Furusaki},\ and\
  \citenamefont {Ludwig}}]{Schnyder2008}%
  \BibitemOpen
  \bibfield  {author} {\bibinfo {author} {\bibfnamefont {Andreas~P.}\
  \bibnamefont {Schnyder}}, \bibinfo {author} {\bibfnamefont {Shinsei}\
  \bibnamefont {Ryu}}, \bibinfo {author} {\bibfnamefont {Akira}\ \bibnamefont
  {Furusaki}}, \ and\ \bibinfo {author} {\bibfnamefont {Andreas W.~W.}\
  \bibnamefont {Ludwig}},\ }\bibfield  {title} {\enquote {\bibinfo {title}
  {Classification of topological insulators and superconductors in three
  spatial dimensions},}\ }\href {\doibase 10.1103/PhysRevB.78.195125}
  {\bibfield  {journal} {\bibinfo  {journal} {Phys. Rev. B}\ }\textbf {\bibinfo
  {volume} {78}},\ \bibinfo {pages} {195125} (\bibinfo {year}
  {2008})}\BibitemShut {NoStop}%
\bibitem [{\citenamefont {Ryu}\ \emph {et~al.}(2010)\citenamefont {Ryu},
  \citenamefont {Schnyder}, \citenamefont {Furusaki},\ and\ \citenamefont
  {Ludwig}}]{Ryu2010}%
  \BibitemOpen
  \bibfield  {author} {\bibinfo {author} {\bibfnamefont {Shinsei}\ \bibnamefont
  {Ryu}}, \bibinfo {author} {\bibfnamefont {Andreas~P}\ \bibnamefont
  {Schnyder}}, \bibinfo {author} {\bibfnamefont {Akira}\ \bibnamefont
  {Furusaki}}, \ and\ \bibinfo {author} {\bibfnamefont {Andreas W.~W.}\
  \bibnamefont {Ludwig}},\ }\bibfield  {title} {\enquote {\bibinfo {title}
  {Topological insulators and superconductors: tenfold way and dimensional
  hierarchy},}\ }\href {\doibase https://doi.org/10.1088/1367-2630/12/6/065010}
  {\bibfield  {journal} {\bibinfo  {journal} {New J. Phys.}\ }\textbf {\bibinfo
  {volume} {12}},\ \bibinfo {pages} {065010} (\bibinfo {year}
  {2010})}\BibitemShut {NoStop}%
\bibitem [{\citenamefont {Chiu}\ \emph {et~al.}(2016)\citenamefont {Chiu},
  \citenamefont {Teo}, \citenamefont {Schnyder},\ and\ \citenamefont
  {Ryu}}]{Chiu2016}%
  \BibitemOpen
  \bibfield  {author} {\bibinfo {author} {\bibfnamefont {Ching-Kai}\
  \bibnamefont {Chiu}}, \bibinfo {author} {\bibfnamefont {Jeffrey C.~Y.}\
  \bibnamefont {Teo}}, \bibinfo {author} {\bibfnamefont {Andreas~P.}\
  \bibnamefont {Schnyder}}, \ and\ \bibinfo {author} {\bibfnamefont {Shinsei}\
  \bibnamefont {Ryu}},\ }\bibfield  {title} {\enquote {\bibinfo {title}
  {Classification of topological quantum matter with symmetries},}\ }\href
  {\doibase 10.1103/RevModPhys.88.035005} {\bibfield  {journal} {\bibinfo
  {journal} {Rev. Mod. Phys.}\ }\textbf {\bibinfo {volume} {88}},\ \bibinfo
  {pages} {035005} (\bibinfo {year} {2016})}\BibitemShut {NoStop}%
\bibitem [{\citenamefont {Mong}\ \emph {et~al.}(2010)\citenamefont {Mong},
  \citenamefont {Essin},\ and\ \citenamefont {Moore}}]{Mong2010}%
  \BibitemOpen
  \bibfield  {author} {\bibinfo {author} {\bibfnamefont {Roger S.~K.}\
  \bibnamefont {Mong}}, \bibinfo {author} {\bibfnamefont {Andrew~M.}\
  \bibnamefont {Essin}}, \ and\ \bibinfo {author} {\bibfnamefont {Joel~E.}\
  \bibnamefont {Moore}},\ }\bibfield  {title} {\enquote {\bibinfo {title}
  {Antiferromagnetic topological insulators},}\ }\href {\doibase
  10.1103/PhysRevB.81.245209} {\bibfield  {journal} {\bibinfo  {journal} {Phys.
  Rev. B}\ }\textbf {\bibinfo {volume} {81}},\ \bibinfo {pages} {245209}
  (\bibinfo {year} {2010})}\BibitemShut {NoStop}%
\bibitem [{\citenamefont {Fu}(2011)}]{fu2011prl}%
  \BibitemOpen
  \bibfield  {author} {\bibinfo {author} {\bibfnamefont {Liang}\ \bibnamefont
  {Fu}},\ }\bibfield  {title} {\enquote {\bibinfo {title} {Topological
  crystalline insulators},}\ }\href {\doibase 10.1103/PhysRevLett.106.106802}
  {\bibfield  {journal} {\bibinfo  {journal} {Phys. Rev. Lett.}\ }\textbf
  {\bibinfo {volume} {106}},\ \bibinfo {pages} {106802} (\bibinfo {year}
  {2011})}\BibitemShut {NoStop}%
\bibitem [{\citenamefont {Slager}\ \emph {et~al.}(2013)\citenamefont {Slager},
  \citenamefont {Mesaros}, \citenamefont {Juri{\v{c}}i{\'c}},\ and\
  \citenamefont {Zaanen}}]{slager2013natphys}%
  \BibitemOpen
  \bibfield  {author} {\bibinfo {author} {\bibfnamefont {Robert-Jan}\
  \bibnamefont {Slager}}, \bibinfo {author} {\bibfnamefont {Andrej}\
  \bibnamefont {Mesaros}}, \bibinfo {author} {\bibfnamefont {Vladimir}\
  \bibnamefont {Juri{\v{c}}i{\'c}}}, \ and\ \bibinfo {author} {\bibfnamefont
  {Jan}\ \bibnamefont {Zaanen}},\ }\bibfield  {title} {\enquote {\bibinfo
  {title} {The space group classification of topological band-insulators},}\
  }\href {\doibase 10.1038/nphys2513} {\bibfield  {journal} {\bibinfo
  {journal} {Nat. Phys.}\ }\textbf {\bibinfo {volume} {9}},\ \bibinfo {pages}
  {98} (\bibinfo {year} {2013})}\BibitemShut {NoStop}%
\bibitem [{\citenamefont {Ando}\ and\ \citenamefont {Fu}(2015)}]{fu2015review}%
  \BibitemOpen
  \bibfield  {author} {\bibinfo {author} {\bibfnamefont {Yoichi}\ \bibnamefont
  {Ando}}\ and\ \bibinfo {author} {\bibfnamefont {Liang}\ \bibnamefont {Fu}},\
  }\bibfield  {title} {\enquote {\bibinfo {title} {Topological crystalline
  insulators and topological superconductors: From concepts to materials},}\
  }\href {\doibase 10.1146/annurev-conmatphys-031214-014501} {\bibfield
  {journal} {\bibinfo  {journal} {Annu. Rev. Condens. Matter Phys.}\ }\textbf
  {\bibinfo {volume} {6}},\ \bibinfo {pages} {361} (\bibinfo {year}
  {2015})}\BibitemShut {NoStop}%
\bibitem [{\citenamefont {Chiu}\ \emph {et~al.}(2013)\citenamefont {Chiu},
  \citenamefont {Yao},\ and\ \citenamefont {Ryu}}]{Chiu2013}%
  \BibitemOpen
  \bibfield  {author} {\bibinfo {author} {\bibfnamefont {Ching-Kai}\
  \bibnamefont {Chiu}}, \bibinfo {author} {\bibfnamefont {Hong}\ \bibnamefont
  {Yao}}, \ and\ \bibinfo {author} {\bibfnamefont {Shinsei}\ \bibnamefont
  {Ryu}},\ }\bibfield  {title} {\enquote {\bibinfo {title} {Classification of
  topological insulators and superconductors in the presence of reflection
  symmetry},}\ }\href {\doibase 10.1103/PhysRevB.88.075142} {\bibfield
  {journal} {\bibinfo  {journal} {Phys. Rev. B}\ }\textbf {\bibinfo {volume}
  {88}},\ \bibinfo {pages} {075142} (\bibinfo {year} {2013})}\BibitemShut
  {NoStop}%
\bibitem [{\citenamefont {Benalcazar}\ \emph
  {et~al.}(2017{\natexlab{a}})\citenamefont {Benalcazar}, \citenamefont
  {Bernevig},\ and\ \citenamefont {Hughes}}]{Benalcazar2017}%
  \BibitemOpen
  \bibfield  {author} {\bibinfo {author} {\bibfnamefont {Wladimir~A.}\
  \bibnamefont {Benalcazar}}, \bibinfo {author} {\bibfnamefont {B.~Andrei}\
  \bibnamefont {Bernevig}}, \ and\ \bibinfo {author} {\bibfnamefont
  {Taylor~L.}\ \bibnamefont {Hughes}},\ }\bibfield  {title} {\enquote {\bibinfo
  {title} {Quantized electric multipole insulators},}\ }\href {\doibase
  10.1126/science.aah6442} {\bibfield  {journal} {\bibinfo  {journal}
  {Science}\ }\textbf {\bibinfo {volume} {357}},\ \bibinfo {pages} {61}
  (\bibinfo {year} {2017}{\natexlab{a}})}\BibitemShut {NoStop}%
\bibitem [{\citenamefont {Benalcazar}\ \emph
  {et~al.}(2017{\natexlab{b}})\citenamefont {Benalcazar}, \citenamefont
  {Bernevig},\ and\ \citenamefont {Hughes}}]{Benalcazar2017prb}%
  \BibitemOpen
  \bibfield  {author} {\bibinfo {author} {\bibfnamefont {Wladimir~A.}\
  \bibnamefont {Benalcazar}}, \bibinfo {author} {\bibfnamefont {B.~Andrei}\
  \bibnamefont {Bernevig}}, \ and\ \bibinfo {author} {\bibfnamefont
  {Taylor~L.}\ \bibnamefont {Hughes}},\ }\bibfield  {title} {\enquote {\bibinfo
  {title} {Electric multipole moments, topological multipole moment pumping,
  and chiral hinge states in crystalline insulators},}\ }\href {\doibase
  10.1103/PhysRevB.96.245115} {\bibfield  {journal} {\bibinfo  {journal} {Phys.
  Rev. B}\ }\textbf {\bibinfo {volume} {96}},\ \bibinfo {pages} {245115}
  (\bibinfo {year} {2017}{\natexlab{b}})}\BibitemShut {NoStop}%
\bibitem [{\citenamefont {Langbehn}\ \emph {et~al.}(2017)\citenamefont
  {Langbehn}, \citenamefont {Peng}, \citenamefont {Trifunovic}, \citenamefont
  {von Oppen},\ and\ \citenamefont {Brouwer}}]{Langbehn2017}%
  \BibitemOpen
  \bibfield  {author} {\bibinfo {author} {\bibfnamefont {Josias}\ \bibnamefont
  {Langbehn}}, \bibinfo {author} {\bibfnamefont {Yang}\ \bibnamefont {Peng}},
  \bibinfo {author} {\bibfnamefont {Luka}\ \bibnamefont {Trifunovic}}, \bibinfo
  {author} {\bibfnamefont {Felix}\ \bibnamefont {von Oppen}}, \ and\ \bibinfo
  {author} {\bibfnamefont {Piet~W.}\ \bibnamefont {Brouwer}},\ }\bibfield
  {title} {\enquote {\bibinfo {title} {Reflection-symmetric second-order
  topological insulators and superconductors},}\ }\href {\doibase
  10.1103/PhysRevLett.119.246401} {\bibfield  {journal} {\bibinfo  {journal}
  {Phys. Rev. Lett.}\ }\textbf {\bibinfo {volume} {119}},\ \bibinfo {pages}
  {246401} (\bibinfo {year} {2017})}\BibitemShut {NoStop}%
\bibitem [{\citenamefont {Song}\ \emph {et~al.}(2017)\citenamefont {Song},
  \citenamefont {Fang},\ and\ \citenamefont {Fang}}]{Song2017}%
  \BibitemOpen
  \bibfield  {author} {\bibinfo {author} {\bibfnamefont {Zhida}\ \bibnamefont
  {Song}}, \bibinfo {author} {\bibfnamefont {Zhong}\ \bibnamefont {Fang}}, \
  and\ \bibinfo {author} {\bibfnamefont {Chen}\ \bibnamefont {Fang}},\
  }\bibfield  {title} {\enquote {\bibinfo {title}
  {$(d\ensuremath{-}2)$-dimensional edge states of rotation symmetry protected
  topological states},}\ }\href {\doibase 10.1103/PhysRevLett.119.246402}
  {\bibfield  {journal} {\bibinfo  {journal} {Phys. Rev. Lett.}\ }\textbf
  {\bibinfo {volume} {119}},\ \bibinfo {pages} {246402} (\bibinfo {year}
  {2017})}\BibitemShut {NoStop}%
\bibitem [{\citenamefont {Schindler}\ \emph
  {et~al.}(2018{\natexlab{a}})\citenamefont {Schindler}, \citenamefont {Wang},
  \citenamefont {Vergniory}, \citenamefont {Cook}, \citenamefont {Murani},
  \citenamefont {Sengupta}, \citenamefont {Kasumov}, \citenamefont {Deblock},
  \citenamefont {Jeon}, \citenamefont {Drozdov} \emph
  {et~al.}}]{Schindler2018natphys}%
  \BibitemOpen
  \bibfield  {author} {\bibinfo {author} {\bibfnamefont {Frank}\ \bibnamefont
  {Schindler}}, \bibinfo {author} {\bibfnamefont {Zhijun}\ \bibnamefont
  {Wang}}, \bibinfo {author} {\bibfnamefont {Maia~G}\ \bibnamefont
  {Vergniory}}, \bibinfo {author} {\bibfnamefont {Ashley~M}\ \bibnamefont
  {Cook}}, \bibinfo {author} {\bibfnamefont {Anil}\ \bibnamefont {Murani}},
  \bibinfo {author} {\bibfnamefont {Shamashis}\ \bibnamefont {Sengupta}},
  \bibinfo {author} {\bibfnamefont {Alik~Yu}\ \bibnamefont {Kasumov}}, \bibinfo
  {author} {\bibfnamefont {Richard}\ \bibnamefont {Deblock}}, \bibinfo {author}
  {\bibfnamefont {Sangjun}\ \bibnamefont {Jeon}}, \bibinfo {author}
  {\bibfnamefont {Ilya}\ \bibnamefont {Drozdov}},  \emph {et~al.},\ }\bibfield
  {title} {\enquote {\bibinfo {title} {Higher-order topology in bismuth},}\
  }\href {\doibase 10.1038/s41567-018-0224-7} {\bibfield  {journal} {\bibinfo
  {journal} {Nat. Phys.}\ }\textbf {\bibinfo {volume} {14}},\ \bibinfo {pages}
  {918} (\bibinfo {year} {2018}{\natexlab{a}})}\BibitemShut {NoStop}%
\bibitem [{\citenamefont {Schindler}\ \emph
  {et~al.}(2018{\natexlab{b}})\citenamefont {Schindler}, \citenamefont {Cook},
  \citenamefont {Vergniory}, \citenamefont {Wang}, \citenamefont {Parkin},
  \citenamefont {Bernevig},\ and\ \citenamefont
  {Neupert}}]{Schindler2018sciadv}%
  \BibitemOpen
  \bibfield  {author} {\bibinfo {author} {\bibfnamefont {Frank}\ \bibnamefont
  {Schindler}}, \bibinfo {author} {\bibfnamefont {Ashley~M.}\ \bibnamefont
  {Cook}}, \bibinfo {author} {\bibfnamefont {Maia~G.}\ \bibnamefont
  {Vergniory}}, \bibinfo {author} {\bibfnamefont {Zhijun}\ \bibnamefont
  {Wang}}, \bibinfo {author} {\bibfnamefont {Stuart S.~P.}\ \bibnamefont
  {Parkin}}, \bibinfo {author} {\bibfnamefont {B.~Andrei}\ \bibnamefont
  {Bernevig}}, \ and\ \bibinfo {author} {\bibfnamefont {Titus}\ \bibnamefont
  {Neupert}},\ }\bibfield  {title} {\enquote {\bibinfo {title} {Higher-order
  topological insulators},}\ }\href {\doibase 10.1126/sciadv.aat0346}
  {\bibfield  {journal} {\bibinfo  {journal} {Sci. Adv.}\ }\textbf {\bibinfo
  {volume} {4}},\ \bibinfo {pages} {0346} (\bibinfo {year}
  {2018}{\natexlab{b}})}\BibitemShut {NoStop}%
\bibitem [{\citenamefont {Geier}\ \emph {et~al.}(2018)\citenamefont {Geier},
  \citenamefont {Trifunovic}, \citenamefont {Hoskam},\ and\ \citenamefont
  {Brouwer}}]{Geier2018}%
  \BibitemOpen
  \bibfield  {author} {\bibinfo {author} {\bibfnamefont {Max}\ \bibnamefont
  {Geier}}, \bibinfo {author} {\bibfnamefont {Luka}\ \bibnamefont
  {Trifunovic}}, \bibinfo {author} {\bibfnamefont {Max}\ \bibnamefont
  {Hoskam}}, \ and\ \bibinfo {author} {\bibfnamefont {Piet~W.}\ \bibnamefont
  {Brouwer}},\ }\bibfield  {title} {\enquote {\bibinfo {title} {Second-order
  topological insulators and superconductors with an order-two crystalline
  symmetry},}\ }\href {\doibase 10.1103/physrevb.97.205135} {\bibfield
  {journal} {\bibinfo  {journal} {Phys. Rev. B}\ }\textbf {\bibinfo {volume}
  {97}},\ \bibinfo {pages} {205135} (\bibinfo {year} {2018})}\BibitemShut
  {NoStop}%
\bibitem [{\citenamefont {Khalaf}(2018)}]{Khalaf2018}%
  \BibitemOpen
  \bibfield  {author} {\bibinfo {author} {\bibfnamefont {Eslam}\ \bibnamefont
  {Khalaf}},\ }\bibfield  {title} {\enquote {\bibinfo {title} {Higher-order
  topological insulators and superconductors protected by inversion
  symmetry},}\ }\href {\doibase 10.1103/physrevb.97.205136} {\bibfield
  {journal} {\bibinfo  {journal} {Phys. Rev. B}\ }\textbf {\bibinfo {volume}
  {97}},\ \bibinfo {pages} {205136} (\bibinfo {year} {2018})}\BibitemShut
  {NoStop}%
\bibitem [{\citenamefont {Trifunovic}\ and\ \citenamefont
  {Brouwer}(2019)}]{Trifunovic2019}%
  \BibitemOpen
  \bibfield  {author} {\bibinfo {author} {\bibfnamefont {Luka}\ \bibnamefont
  {Trifunovic}}\ and\ \bibinfo {author} {\bibfnamefont {Piet~W.}\ \bibnamefont
  {Brouwer}},\ }\bibfield  {title} {\enquote {\bibinfo {title} {Higher-order
  bulk-boundary correspondence for topological crystalline phases},}\ }\href
  {\doibase 10.1103/PhysRevX.9.011012} {\bibfield  {journal} {\bibinfo
  {journal} {Phys. Rev. X}\ }\textbf {\bibinfo {volume} {9}},\ \bibinfo {pages}
  {011012} (\bibinfo {year} {2019})}\BibitemShut {NoStop}%
\bibitem [{\citenamefont {Wang}\ \emph
  {et~al.}(2018{\natexlab{a}})\citenamefont {Wang}, \citenamefont {Lin},\ and\
  \citenamefont {Hughes}}]{Hughes2018}%
  \BibitemOpen
  \bibfield  {author} {\bibinfo {author} {\bibfnamefont {Yuxuan}\ \bibnamefont
  {Wang}}, \bibinfo {author} {\bibfnamefont {Mao}\ \bibnamefont {Lin}}, \ and\
  \bibinfo {author} {\bibfnamefont {Taylor~L.}\ \bibnamefont {Hughes}},\
  }\bibfield  {title} {\enquote {\bibinfo {title} {Weak-pairing higher order
  topological superconductors},}\ }\href {\doibase 10.1103/PhysRevB.98.165144}
  {\bibfield  {journal} {\bibinfo  {journal} {Phys. Rev. B}\ }\textbf {\bibinfo
  {volume} {98}},\ \bibinfo {pages} {165144} (\bibinfo {year}
  {2018}{\natexlab{a}})}\BibitemShut {NoStop}%
\bibitem [{\citenamefont {van Miert}\ and\ \citenamefont
  {Ortix}(2018)}]{Ortix2018}%
  \BibitemOpen
  \bibfield  {author} {\bibinfo {author} {\bibfnamefont {Guido}\ \bibnamefont
  {van Miert}}\ and\ \bibinfo {author} {\bibfnamefont {Carmine}\ \bibnamefont
  {Ortix}},\ }\bibfield  {title} {\enquote {\bibinfo {title} {Higher-order
  topological insulators protected by inversion and rotoinversion
  symmetries},}\ }\href {\doibase 10.1103/PhysRevB.98.081110} {\bibfield
  {journal} {\bibinfo  {journal} {Phys. Rev. B}\ }\textbf {\bibinfo {volume}
  {98}},\ \bibinfo {pages} {081110} (\bibinfo {year} {2018})}\BibitemShut
  {NoStop}%
\bibitem [{\citenamefont {You}\ \emph {et~al.}(2018)\citenamefont {You},
  \citenamefont {Devakul}, \citenamefont {Burnell},\ and\ \citenamefont
  {Neupert}}]{You2018}%
  \BibitemOpen
  \bibfield  {author} {\bibinfo {author} {\bibfnamefont {Yizhi}\ \bibnamefont
  {You}}, \bibinfo {author} {\bibfnamefont {Trithep}\ \bibnamefont {Devakul}},
  \bibinfo {author} {\bibfnamefont {F.~J.}\ \bibnamefont {Burnell}}, \ and\
  \bibinfo {author} {\bibfnamefont {Titus}\ \bibnamefont {Neupert}},\
  }\bibfield  {title} {\enquote {\bibinfo {title} {Higher-order
  symmetry-protected topological states for interacting bosons and fermions},}\
  }\href {\doibase 10.1103/PhysRevB.98.235102} {\bibfield  {journal} {\bibinfo
  {journal} {Phys. Rev. B}\ }\textbf {\bibinfo {volume} {98}},\ \bibinfo
  {pages} {235102} (\bibinfo {year} {2018})}\BibitemShut {NoStop}%
\bibitem [{\citenamefont {C\ifmmode \u{a}\else \u{a}\fi{}lug\ifmmode~\u{a}\else
  \u{a}\fi{}ru}\ \emph {et~al.}(2019)\citenamefont {C\ifmmode \u{a}\else
  \u{a}\fi{}lug\ifmmode~\u{a}\else \u{a}\fi{}ru}, \citenamefont {Juri\ifmmode
  \check{c}\else \v{c}\fi{}i\ifmmode~\acute{c}\else \'{c}\fi{}},\ and\
  \citenamefont {Roy}}]{Roy2019}%
  \BibitemOpen
  \bibfield  {author} {\bibinfo {author} {\bibfnamefont {Dumitru}\ \bibnamefont
  {C\ifmmode \u{a}\else \u{a}\fi{}lug\ifmmode~\u{a}\else \u{a}\fi{}ru}},
  \bibinfo {author} {\bibfnamefont {Vladimir}\ \bibnamefont {Juri\ifmmode
  \check{c}\else \v{c}\fi{}i\ifmmode~\acute{c}\else \'{c}\fi{}}}, \ and\
  \bibinfo {author} {\bibfnamefont {Bitan}\ \bibnamefont {Roy}},\ }\bibfield
  {title} {\enquote {\bibinfo {title} {Higher-order topological phases: A
  general principle of construction},}\ }\href {\doibase
  10.1103/PhysRevB.99.041301} {\bibfield  {journal} {\bibinfo  {journal} {Phys.
  Rev. B}\ }\textbf {\bibinfo {volume} {99}},\ \bibinfo {pages} {041301}
  (\bibinfo {year} {2019})}\BibitemShut {NoStop}%
\bibitem [{\citenamefont {Araki}\ \emph {et~al.}(2019)\citenamefont {Araki},
  \citenamefont {Mizoguchi},\ and\ \citenamefont {Hatsugai}}]{Araki2019}%
  \BibitemOpen
  \bibfield  {author} {\bibinfo {author} {\bibfnamefont {Hiromu}\ \bibnamefont
  {Araki}}, \bibinfo {author} {\bibfnamefont {Tomonari}\ \bibnamefont
  {Mizoguchi}}, \ and\ \bibinfo {author} {\bibfnamefont {Yasuhiro}\
  \bibnamefont {Hatsugai}},\ }\bibfield  {title} {\enquote {\bibinfo {title}
  {Phase diagram of a disordered higher-order topological insulator: A machine
  learning study},}\ }\href {\doibase 10.1103/PhysRevB.99.085406} {\bibfield
  {journal} {\bibinfo  {journal} {Phys. Rev. B}\ }\textbf {\bibinfo {volume}
  {99}},\ \bibinfo {pages} {085406} (\bibinfo {year} {2019})}\BibitemShut
  {NoStop}%
\bibitem [{\citenamefont {Kudo}\ \emph {et~al.}(2019)\citenamefont {Kudo},
  \citenamefont {Yoshida},\ and\ \citenamefont {Hatsugai}}]{Hatsugai2019}%
  \BibitemOpen
  \bibfield  {author} {\bibinfo {author} {\bibfnamefont {Koji}\ \bibnamefont
  {Kudo}}, \bibinfo {author} {\bibfnamefont {Tsuneya}\ \bibnamefont {Yoshida}},
  \ and\ \bibinfo {author} {\bibfnamefont {Yasuhiro}\ \bibnamefont
  {Hatsugai}},\ }\bibfield  {title} {\enquote {\bibinfo {title} {Higher-order
  topological {M}ott insulators},}\ }\href {\doibase
  10.1103/PhysRevLett.123.196402} {\bibfield  {journal} {\bibinfo  {journal}
  {Phys. Rev. Lett.}\ }\textbf {\bibinfo {volume} {123}},\ \bibinfo {pages}
  {196402} (\bibinfo {year} {2019})}\BibitemShut {NoStop}%
\bibitem [{\citenamefont {Tiwari}\ \emph {et~al.}(2020)\citenamefont {Tiwari},
  \citenamefont {Li}, \citenamefont {Bernevig}, \citenamefont {Neupert},\ and\
  \citenamefont {Parameswaran}}]{Tiwari2020}%
  \BibitemOpen
  \bibfield  {author} {\bibinfo {author} {\bibfnamefont {Apoorv}\ \bibnamefont
  {Tiwari}}, \bibinfo {author} {\bibfnamefont {Ming-Hao}\ \bibnamefont {Li}},
  \bibinfo {author} {\bibfnamefont {B.~A.}\ \bibnamefont {Bernevig}}, \bibinfo
  {author} {\bibfnamefont {Titus}\ \bibnamefont {Neupert}}, \ and\ \bibinfo
  {author} {\bibfnamefont {S.~A.}\ \bibnamefont {Parameswaran}},\ }\bibfield
  {title} {\enquote {\bibinfo {title} {Unhinging the surfaces of higher-order
  topological insulators and superconductors},}\ }\href {\doibase
  10.1103/PhysRevLett.124.046801} {\bibfield  {journal} {\bibinfo  {journal}
  {Phys. Rev. Lett.}\ }\textbf {\bibinfo {volume} {124}},\ \bibinfo {pages}
  {046801} (\bibinfo {year} {2020})}\BibitemShut {NoStop}%
\bibitem [{\citenamefont {Ezawa}(2018{\natexlab{a}})}]{Ezawa2018kagome}%
  \BibitemOpen
  \bibfield  {author} {\bibinfo {author} {\bibfnamefont {Motohiko}\
  \bibnamefont {Ezawa}},\ }\bibfield  {title} {\enquote {\bibinfo {title}
  {Higher-order topological insulators and semimetals on the breathing {K}agome
  and pyrochlore lattices},}\ }\href {\doibase 10.1103/physrevlett.120.026801}
  {\bibfield  {journal} {\bibinfo  {journal} {Phys. Rev. Lett.}\ }\textbf
  {\bibinfo {volume} {120}},\ \bibinfo {pages} {026801} (\bibinfo {year}
  {2018}{\natexlab{a}})}\BibitemShut {NoStop}%
\bibitem [{\citenamefont {Ezawa}(2018{\natexlab{b}})}]{Ezawa2018phosphorene}%
  \BibitemOpen
  \bibfield  {author} {\bibinfo {author} {\bibfnamefont {Motohiko}\
  \bibnamefont {Ezawa}},\ }\bibfield  {title} {\enquote {\bibinfo {title}
  {Minimal models for {W}annier-type higher-order topological insulators and
  phosphorene},}\ }\href {\doibase 10.1103/physrevb.98.045125} {\bibfield
  {journal} {\bibinfo  {journal} {Phys. Rev. B}\ }\textbf {\bibinfo {volume}
  {98}},\ \bibinfo {pages} {045125} (\bibinfo {year}
  {2018}{\natexlab{b}})}\BibitemShut {NoStop}%
\bibitem [{\citenamefont {Hsu}\ \emph {et~al.}(2018)\citenamefont {Hsu},
  \citenamefont {Stano}, \citenamefont {Klinovaja},\ and\ \citenamefont
  {Loss}}]{loss2018}%
  \BibitemOpen
  \bibfield  {author} {\bibinfo {author} {\bibfnamefont {Chen-Hsuan}\
  \bibnamefont {Hsu}}, \bibinfo {author} {\bibfnamefont {Peter}\ \bibnamefont
  {Stano}}, \bibinfo {author} {\bibfnamefont {Jelena}\ \bibnamefont
  {Klinovaja}}, \ and\ \bibinfo {author} {\bibfnamefont {Daniel}\ \bibnamefont
  {Loss}},\ }\bibfield  {title} {\enquote {\bibinfo {title} {{M}ajorana
  {K}ramers pairs in higher-order topological insulators},}\ }\href {\doibase
  10.1103/PhysRevLett.121.196801} {\bibfield  {journal} {\bibinfo  {journal}
  {Phys. Rev. Lett.}\ }\textbf {\bibinfo {volume} {121}},\ \bibinfo {pages}
  {196801} (\bibinfo {year} {2018})}\BibitemShut {NoStop}%
\bibitem [{\citenamefont {Wang}\ \emph
  {et~al.}(2018{\natexlab{b}})\citenamefont {Wang}, \citenamefont {Liu},
  \citenamefont {Lu},\ and\ \citenamefont {Zhang}}]{Wang2018prl}%
  \BibitemOpen
  \bibfield  {author} {\bibinfo {author} {\bibfnamefont {Qiyue}\ \bibnamefont
  {Wang}}, \bibinfo {author} {\bibfnamefont {Cheng-Cheng}\ \bibnamefont {Liu}},
  \bibinfo {author} {\bibfnamefont {Yuan-Ming}\ \bibnamefont {Lu}}, \ and\
  \bibinfo {author} {\bibfnamefont {Fan}\ \bibnamefont {Zhang}},\ }\bibfield
  {title} {\enquote {\bibinfo {title} {High-temperature {M}ajorana corner
  states},}\ }\href {\doibase 10.1103/PhysRevLett.121.186801} {\bibfield
  {journal} {\bibinfo  {journal} {Phys. Rev. Lett.}\ }\textbf {\bibinfo
  {volume} {121}},\ \bibinfo {pages} {186801} (\bibinfo {year}
  {2018}{\natexlab{b}})}\BibitemShut {NoStop}%
\bibitem [{\citenamefont {Yan}\ \emph {et~al.}(2018)\citenamefont {Yan},
  \citenamefont {Song},\ and\ \citenamefont {Wang}}]{Yan2018prl}%
  \BibitemOpen
  \bibfield  {author} {\bibinfo {author} {\bibfnamefont {Zhongbo}\ \bibnamefont
  {Yan}}, \bibinfo {author} {\bibfnamefont {Fei}\ \bibnamefont {Song}}, \ and\
  \bibinfo {author} {\bibfnamefont {Zhong}\ \bibnamefont {Wang}},\ }\bibfield
  {title} {\enquote {\bibinfo {title} {{M}ajorana corner modes in a
  high-temperature platform},}\ }\href {\doibase
  10.1103/PhysRevLett.121.096803} {\bibfield  {journal} {\bibinfo  {journal}
  {Phys. Rev. Lett.}\ }\textbf {\bibinfo {volume} {121}},\ \bibinfo {pages}
  {096803} (\bibinfo {year} {2018})}\BibitemShut {NoStop}%
\bibitem [{\citenamefont {Ghorashi}\ \emph
  {et~al.}(2020{\natexlab{a}})\citenamefont {Ghorashi}, \citenamefont {Li},\
  and\ \citenamefont {Hughes}}]{sayed2020}%
  \BibitemOpen
  \bibfield  {author} {\bibinfo {author} {\bibfnamefont {Sayed Ali~Akbar}\
  \bibnamefont {Ghorashi}}, \bibinfo {author} {\bibfnamefont {Tianhe}\
  \bibnamefont {Li}}, \ and\ \bibinfo {author} {\bibfnamefont {Taylor~L.}\
  \bibnamefont {Hughes}},\ }\bibfield  {title} {\enquote {\bibinfo {title}
  {Higher-order {W}eyl semimetals},}\ }\href {\doibase
  10.1103/PhysRevLett.125.266804} {\bibfield  {journal} {\bibinfo  {journal}
  {Phys. Rev. Lett.}\ }\textbf {\bibinfo {volume} {125}},\ \bibinfo {pages}
  {266804} (\bibinfo {year} {2020}{\natexlab{a}})}\BibitemShut {NoStop}%
\bibitem [{\citenamefont {Ghorashi}\ \emph
  {et~al.}(2020{\natexlab{b}})\citenamefont {Ghorashi}, \citenamefont
  {Hughes},\ and\ \citenamefont {Rossi}}]{sayed2020_1}%
  \BibitemOpen
  \bibfield  {author} {\bibinfo {author} {\bibfnamefont {Sayed Ali~Akbar}\
  \bibnamefont {Ghorashi}}, \bibinfo {author} {\bibfnamefont {Taylor~L.}\
  \bibnamefont {Hughes}}, \ and\ \bibinfo {author} {\bibfnamefont {Enrico}\
  \bibnamefont {Rossi}},\ }\bibfield  {title} {\enquote {\bibinfo {title}
  {Vortex and surface phase transitions in superconducting higher-order
  topological insulators},}\ }\href {\doibase 10.1103/PhysRevLett.125.037001}
  {\bibfield  {journal} {\bibinfo  {journal} {Phys. Rev. Lett.}\ }\textbf
  {\bibinfo {volume} {125}},\ \bibinfo {pages} {037001} (\bibinfo {year}
  {2020}{\natexlab{b}})}\BibitemShut {NoStop}%
\bibitem [{\citenamefont {Ghorashi}\ \emph {et~al.}(2019)\citenamefont
  {Ghorashi}, \citenamefont {Hu}, \citenamefont {Hughes},\ and\ \citenamefont
  {Rossi}}]{sayed2019}%
  \BibitemOpen
  \bibfield  {author} {\bibinfo {author} {\bibfnamefont {Sayed Ali~Akbar}\
  \bibnamefont {Ghorashi}}, \bibinfo {author} {\bibfnamefont {Xiang}\
  \bibnamefont {Hu}}, \bibinfo {author} {\bibfnamefont {Taylor~L.}\
  \bibnamefont {Hughes}}, \ and\ \bibinfo {author} {\bibfnamefont {Enrico}\
  \bibnamefont {Rossi}},\ }\bibfield  {title} {\enquote {\bibinfo {title}
  {Second-order {D}irac superconductors and magnetic field induced {M}ajorana
  hinge modes},}\ }\href {\doibase 10.1103/PhysRevB.100.020509} {\bibfield
  {journal} {\bibinfo  {journal} {Phys. Rev. B}\ }\textbf {\bibinfo {volume}
  {100}},\ \bibinfo {pages} {020509} (\bibinfo {year} {2019})}\BibitemShut
  {NoStop}%
\bibitem [{Note1()}]{Note1}%
  \BibitemOpen
  \bibinfo {note} {Even though the BBH model is probably well known by now, we
  briefly summarize its main features and compare it with our network model in
  Appendix \ref {app:BBH}.}\BibitemShut {Stop}%
\bibitem [{\citenamefont {Serra-Garcia}\ \emph {et~al.}(2018)\citenamefont
  {Serra-Garcia}, \citenamefont {Peri}, \citenamefont {S{\"u}sstrunk},
  \citenamefont {Bilal}, \citenamefont {Larsen}, \citenamefont {Villanueva},\
  and\ \citenamefont {Huber}}]{Serra-Garcia2018}%
  \BibitemOpen
  \bibfield  {author} {\bibinfo {author} {\bibfnamefont {Marc}\ \bibnamefont
  {Serra-Garcia}}, \bibinfo {author} {\bibfnamefont {Valerio}\ \bibnamefont
  {Peri}}, \bibinfo {author} {\bibfnamefont {Roman}\ \bibnamefont
  {S{\"u}sstrunk}}, \bibinfo {author} {\bibfnamefont {Osama~R}\ \bibnamefont
  {Bilal}}, \bibinfo {author} {\bibfnamefont {Tom}\ \bibnamefont {Larsen}},
  \bibinfo {author} {\bibfnamefont {Luis~Guillermo}\ \bibnamefont
  {Villanueva}}, \ and\ \bibinfo {author} {\bibfnamefont {Sebastian~D}\
  \bibnamefont {Huber}},\ }\bibfield  {title} {\enquote {\bibinfo {title}
  {Observation of a phononic quadrupole topological insulator},}\ }\href
  {\doibase 10.1038/nature25156} {\bibfield  {journal} {\bibinfo  {journal}
  {Nature}\ }\textbf {\bibinfo {volume} {555}},\ \bibinfo {pages} {342}
  (\bibinfo {year} {2018})}\BibitemShut {NoStop}%
\bibitem [{\citenamefont {Peterson}\ \emph {et~al.}(2018)\citenamefont
  {Peterson}, \citenamefont {Benalcazar}, \citenamefont {Hughes},\ and\
  \citenamefont {Bahl}}]{Peterson2018}%
  \BibitemOpen
  \bibfield  {author} {\bibinfo {author} {\bibfnamefont {Christopher~W.}\
  \bibnamefont {Peterson}}, \bibinfo {author} {\bibfnamefont {Wladimir~A.}\
  \bibnamefont {Benalcazar}}, \bibinfo {author} {\bibfnamefont {Taylor~L.}\
  \bibnamefont {Hughes}}, \ and\ \bibinfo {author} {\bibfnamefont {Gaurav}\
  \bibnamefont {Bahl}},\ }\bibfield  {title} {\enquote {\bibinfo {title} {A
  quantized microwave quadrupole insulator with topologically protected corner
  states},}\ }\href {\doibase 10.1038/nature25777} {\bibfield  {journal}
  {\bibinfo  {journal} {Nature}\ }\textbf {\bibinfo {volume} {555}},\ \bibinfo
  {pages} {346} (\bibinfo {year} {2018})}\BibitemShut {NoStop}%
\bibitem [{\citenamefont {Imhof}\ \emph {et~al.}(2018)\citenamefont {Imhof},
  \citenamefont {Berger}, \citenamefont {Bayer}, \citenamefont {Brehm},
  \citenamefont {Molenkamp}, \citenamefont {Kiessling}, \citenamefont
  {Schindler}, \citenamefont {Lee}, \citenamefont {Greiter}, \citenamefont
  {Neupert} \emph {et~al.}}]{Imhof2018}%
  \BibitemOpen
  \bibfield  {author} {\bibinfo {author} {\bibfnamefont {Stefan}\ \bibnamefont
  {Imhof}}, \bibinfo {author} {\bibfnamefont {Christian}\ \bibnamefont
  {Berger}}, \bibinfo {author} {\bibfnamefont {Florian}\ \bibnamefont {Bayer}},
  \bibinfo {author} {\bibfnamefont {Johannes}\ \bibnamefont {Brehm}}, \bibinfo
  {author} {\bibfnamefont {Laurens~W}\ \bibnamefont {Molenkamp}}, \bibinfo
  {author} {\bibfnamefont {Tobias}\ \bibnamefont {Kiessling}}, \bibinfo
  {author} {\bibfnamefont {Frank}\ \bibnamefont {Schindler}}, \bibinfo {author}
  {\bibfnamefont {Ching~Hua}\ \bibnamefont {Lee}}, \bibinfo {author}
  {\bibfnamefont {Martin}\ \bibnamefont {Greiter}}, \bibinfo {author}
  {\bibfnamefont {Titus}\ \bibnamefont {Neupert}},  \emph {et~al.},\ }\bibfield
   {title} {\enquote {\bibinfo {title} {Topolectrical-circuit realization of
  topological corner modes},}\ }\href {\doibase 10.1038/s41567-018-0246-1}
  {\bibfield  {journal} {\bibinfo  {journal} {Nat. Phys.}\ }\textbf {\bibinfo
  {volume} {14}},\ \bibinfo {pages} {925} (\bibinfo {year} {2018})}\BibitemShut
  {NoStop}%
\bibitem [{\citenamefont {Serra-Garcia}\ \emph {et~al.}(2019)\citenamefont
  {Serra-Garcia}, \citenamefont {S\"usstrunk},\ and\ \citenamefont
  {Huber}}]{Serra-Garcia2019prb}%
  \BibitemOpen
  \bibfield  {author} {\bibinfo {author} {\bibfnamefont {Marc}\ \bibnamefont
  {Serra-Garcia}}, \bibinfo {author} {\bibfnamefont {Roman}\ \bibnamefont
  {S\"usstrunk}}, \ and\ \bibinfo {author} {\bibfnamefont {Sebastian~D.}\
  \bibnamefont {Huber}},\ }\bibfield  {title} {\enquote {\bibinfo {title}
  {Observation of quadrupole transitions and edge mode topology in an{ LC}
  circuit network},}\ }\href {\doibase 10.1103/PhysRevB.99.020304} {\bibfield
  {journal} {\bibinfo  {journal} {Phys. Rev. B}\ }\textbf {\bibinfo {volume}
  {99}},\ \bibinfo {pages} {020304} (\bibinfo {year} {2019})}\BibitemShut
  {NoStop}%
\bibitem [{\citenamefont {Zhang}\ \emph {et~al.}(2019)\citenamefont {Zhang},
  \citenamefont {Wang}, \citenamefont {Lin}, \citenamefont {Tian},
  \citenamefont {Xie}, \citenamefont {Lu}, \citenamefont {Chen},\ and\
  \citenamefont {Jiang}}]{zhang2019natphys}%
  \BibitemOpen
  \bibfield  {author} {\bibinfo {author} {\bibfnamefont {Xiujuan}\ \bibnamefont
  {Zhang}}, \bibinfo {author} {\bibfnamefont {Hai-Xiao}\ \bibnamefont {Wang}},
  \bibinfo {author} {\bibfnamefont {Zhi-Kang}\ \bibnamefont {Lin}}, \bibinfo
  {author} {\bibfnamefont {Yuan}\ \bibnamefont {Tian}}, \bibinfo {author}
  {\bibfnamefont {Biye}\ \bibnamefont {Xie}}, \bibinfo {author} {\bibfnamefont
  {Ming-Hui}\ \bibnamefont {Lu}}, \bibinfo {author} {\bibfnamefont {Yan-Feng}\
  \bibnamefont {Chen}}, \ and\ \bibinfo {author} {\bibfnamefont {Jian-Hua}\
  \bibnamefont {Jiang}},\ }\bibfield  {title} {\enquote {\bibinfo {title}
  {Second-order topology and multidimensional topological transitions in sonic
  crystals},}\ }\href {\doibase 10.1038/s41567-019-0472-1} {\bibfield
  {journal} {\bibinfo  {journal} {Nat. Phys.}\ }\textbf {\bibinfo {volume}
  {15}},\ \bibinfo {pages} {582} (\bibinfo {year} {2019})}\BibitemShut
  {NoStop}%
\bibitem [{\citenamefont {Kempkes}\ \emph {et~al.}(2019)\citenamefont
  {Kempkes}, \citenamefont {Slot}, \citenamefont {van Den~Broeke},
  \citenamefont {Capiod}, \citenamefont {Benalcazar}, \citenamefont
  {Vanmaekelbergh}, \citenamefont {Bercioux}, \citenamefont {Swart},\ and\
  \citenamefont {Smith}}]{Kempkes2019}%
  \BibitemOpen
  \bibfield  {author} {\bibinfo {author} {\bibfnamefont {S.~N.}\ \bibnamefont
  {Kempkes}}, \bibinfo {author} {\bibfnamefont {M.~R.}\ \bibnamefont {Slot}},
  \bibinfo {author} {\bibfnamefont {J.~J.}\ \bibnamefont {van Den~Broeke}},
  \bibinfo {author} {\bibfnamefont {P.}~\bibnamefont {Capiod}}, \bibinfo
  {author} {\bibfnamefont {W.~A.}\ \bibnamefont {Benalcazar}}, \bibinfo
  {author} {\bibfnamefont {D.}~\bibnamefont {Vanmaekelbergh}}, \bibinfo
  {author} {\bibfnamefont {D.}~\bibnamefont {Bercioux}}, \bibinfo {author}
  {\bibfnamefont {I.}~\bibnamefont {Swart}}, \ and\ \bibinfo {author}
  {\bibfnamefont {C.~Morais}\ \bibnamefont {Smith}},\ }\bibfield  {title}
  {\enquote {\bibinfo {title} {Robust zero-energy modes in an electronic
  higher-order topological insulator},}\ }\href {\doibase
  10.1038/s41563-019-0483-4} {\bibfield  {journal} {\bibinfo  {journal} {Nat.
  Mater.}\ }\textbf {\bibinfo {volume} {18}},\ \bibinfo {pages} {1292}
  (\bibinfo {year} {2019})}\BibitemShut {NoStop}%
\bibitem [{\citenamefont {Yue}\ \emph {et~al.}(2019)\citenamefont {Yue},
  \citenamefont {Xu}, \citenamefont {Song}, \citenamefont {Weng}, \citenamefont
  {Lu}, \citenamefont {Fang},\ and\ \citenamefont {Dai}}]{Yue2019natphys}%
  \BibitemOpen
  \bibfield  {author} {\bibinfo {author} {\bibfnamefont {Changming}\
  \bibnamefont {Yue}}, \bibinfo {author} {\bibfnamefont {Yuanfeng}\
  \bibnamefont {Xu}}, \bibinfo {author} {\bibfnamefont {Zhida}\ \bibnamefont
  {Song}}, \bibinfo {author} {\bibfnamefont {Hongming}\ \bibnamefont {Weng}},
  \bibinfo {author} {\bibfnamefont {Yuan-Ming}\ \bibnamefont {Lu}}, \bibinfo
  {author} {\bibfnamefont {Chen}\ \bibnamefont {Fang}}, \ and\ \bibinfo
  {author} {\bibfnamefont {Xi}~\bibnamefont {Dai}},\ }\bibfield  {title}
  {\enquote {\bibinfo {title} {Symmetry-enforced chiral hinge states and
  surface quantum anomalous hall effect in the magnetic axion insulator
  bi2-xsmxse3},}\ }\href {https://doi.org/10.1038/s41567-019-0457-0} {\bibfield
   {journal} {\bibinfo  {journal} {Nature Physics}\ }\textbf {\bibinfo {volume}
  {15}},\ \bibinfo {pages} {577--581} (\bibinfo {year} {2019})}\BibitemShut
  {NoStop}%
\bibitem [{\citenamefont {Mittal}\ \emph {et~al.}(2019)\citenamefont {Mittal},
  \citenamefont {Orre}, \citenamefont {Zhu}, \citenamefont {Gorlach},
  \citenamefont {Poddubny},\ and\ \citenamefont {Hafezi}}]{mittal2019photonic}%
  \BibitemOpen
  \bibfield  {author} {\bibinfo {author} {\bibfnamefont {Sunil}\ \bibnamefont
  {Mittal}}, \bibinfo {author} {\bibfnamefont {Venkata~Vikram}\ \bibnamefont
  {Orre}}, \bibinfo {author} {\bibfnamefont {Guanyu}\ \bibnamefont {Zhu}},
  \bibinfo {author} {\bibfnamefont {Maxim~A}\ \bibnamefont {Gorlach}}, \bibinfo
  {author} {\bibfnamefont {Alexander}\ \bibnamefont {Poddubny}}, \ and\
  \bibinfo {author} {\bibfnamefont {Mohammad}\ \bibnamefont {Hafezi}},\
  }\bibfield  {title} {\enquote {\bibinfo {title} {Photonic quadrupole
  topological phases},}\ }\href {\doibase 10.1038/s41566-019-0452-0} {\bibfield
   {journal} {\bibinfo  {journal} {Nat. Photonics}\ }\textbf {\bibinfo {volume}
  {13}},\ \bibinfo {pages} {692} (\bibinfo {year} {2019})}\BibitemShut
  {NoStop}%
\bibitem [{\citenamefont {Xue}\ \emph {et~al.}(2019)\citenamefont {Xue},
  \citenamefont {Yang}, \citenamefont {Gao}, \citenamefont {Chong},\ and\
  \citenamefont {Zhang}}]{Xue2018}%
  \BibitemOpen
  \bibfield  {author} {\bibinfo {author} {\bibfnamefont {Haoran}\ \bibnamefont
  {Xue}}, \bibinfo {author} {\bibfnamefont {Yahui}\ \bibnamefont {Yang}},
  \bibinfo {author} {\bibfnamefont {Fei}\ \bibnamefont {Gao}}, \bibinfo
  {author} {\bibfnamefont {Yidong}\ \bibnamefont {Chong}}, \ and\ \bibinfo
  {author} {\bibfnamefont {Baile}\ \bibnamefont {Zhang}},\ }\bibfield  {title}
  {\enquote {\bibinfo {title} {Acoustic higher-order topological insulator on a
  {K}agome lattice},}\ }\href {\doibase 10.1038/s41563-018-0251-x} {\bibfield
  {journal} {\bibinfo  {journal} {Nat. Mater.}\ }\textbf {\bibinfo {volume}
  {18}},\ \bibinfo {pages} {108} (\bibinfo {year} {2019})}\BibitemShut
  {NoStop}%
\bibitem [{\citenamefont {Ni}\ \emph {et~al.}(2019)\citenamefont {Ni},
  \citenamefont {Weiner}, \citenamefont {Al{\`u}},\ and\ \citenamefont
  {Khanikaev}}]{ni2019observation}%
  \BibitemOpen
  \bibfield  {author} {\bibinfo {author} {\bibfnamefont {Xiang}\ \bibnamefont
  {Ni}}, \bibinfo {author} {\bibfnamefont {Matthew}\ \bibnamefont {Weiner}},
  \bibinfo {author} {\bibfnamefont {Andrea}\ \bibnamefont {Al{\`u}}}, \ and\
  \bibinfo {author} {\bibfnamefont {Alexander~B}\ \bibnamefont {Khanikaev}},\
  }\bibfield  {title} {\enquote {\bibinfo {title} {Observation of higher-order
  topological acoustic states protected by generalized chiral symmetry},}\
  }\href {\doibase 10.1038/s41563-018-0252-9} {\bibfield  {journal} {\bibinfo
  {journal} {Nat. Mater.}\ }\textbf {\bibinfo {volume} {18}},\ \bibinfo {pages}
  {113} (\bibinfo {year} {2019})}\BibitemShut {NoStop}%
\bibitem [{\citenamefont {El~Hassan}\ \emph {et~al.}(2019)\citenamefont
  {El~Hassan}, \citenamefont {Kunst}, \citenamefont {Moritz}, \citenamefont
  {Andler}, \citenamefont {Bergholtz},\ and\ \citenamefont
  {Bourennane}}]{elHassan2019corner}%
  \BibitemOpen
  \bibfield  {author} {\bibinfo {author} {\bibfnamefont {Ashraf}\ \bibnamefont
  {El~Hassan}}, \bibinfo {author} {\bibfnamefont {Flore~K}\ \bibnamefont
  {Kunst}}, \bibinfo {author} {\bibfnamefont {Alexander}\ \bibnamefont
  {Moritz}}, \bibinfo {author} {\bibfnamefont {Guillermo}\ \bibnamefont
  {Andler}}, \bibinfo {author} {\bibfnamefont {Emil~J}\ \bibnamefont
  {Bergholtz}}, \ and\ \bibinfo {author} {\bibfnamefont {Mohamed}\ \bibnamefont
  {Bourennane}},\ }\bibfield  {title} {\enquote {\bibinfo {title} {Corner
  states of light in photonic waveguides},}\ }\href {\doibase
  10.1038/s41566-019-0519-y} {\bibfield  {journal} {\bibinfo  {journal} {Nat.
  Photonics}\ }\textbf {\bibinfo {volume} {13}},\ \bibinfo {pages} {697}
  (\bibinfo {year} {2019})}\BibitemShut {NoStop}%
\bibitem [{\citenamefont {Evers}\ and\ \citenamefont
  {Mirlin}(2008)}]{mirlin2008review}%
  \BibitemOpen
  \bibfield  {author} {\bibinfo {author} {\bibfnamefont {Ferdinand}\
  \bibnamefont {Evers}}\ and\ \bibinfo {author} {\bibfnamefont {Alexander~D.}\
  \bibnamefont {Mirlin}},\ }\bibfield  {title} {\enquote {\bibinfo {title}
  {{A}nderson transitions},}\ }\href {\doibase 10.1103/RevModPhys.80.1355}
  {\bibfield  {journal} {\bibinfo  {journal} {Rev. Mod. Phys.}\ }\textbf
  {\bibinfo {volume} {80}},\ \bibinfo {pages} {1355} (\bibinfo {year}
  {2008})}\BibitemShut {NoStop}%
\bibitem [{\citenamefont {Chalker}\ and\ \citenamefont
  {Coddington}(1988)}]{chalker1988}%
  \BibitemOpen
  \bibfield  {author} {\bibinfo {author} {\bibfnamefont {J.~T.}\ \bibnamefont
  {Chalker}}\ and\ \bibinfo {author} {\bibfnamefont {P.~D.}\ \bibnamefont
  {Coddington}},\ }\bibfield  {title} {\enquote {\bibinfo {title} {Percolation,
  quantum tunnelling and the integer {H}all effect},}\ }\href {\doibase
  10.1088/0022-3719/21/14/008} {\bibfield  {journal} {\bibinfo  {journal} {J.
  Phys. C: Solid State Physics}\ }\textbf {\bibinfo {volume} {21}},\ \bibinfo
  {pages} {2665} (\bibinfo {year} {1988})}\BibitemShut {NoStop}%
\bibitem [{\citenamefont {Kramer}\ \emph {et~al.}(2005)\citenamefont {Kramer},
  \citenamefont {Ohtsuki},\ and\ \citenamefont {Kettemann}}]{kramer2005review}%
  \BibitemOpen
  \bibfield  {author} {\bibinfo {author} {\bibfnamefont {B.}~\bibnamefont
  {Kramer}}, \bibinfo {author} {\bibfnamefont {T.}~\bibnamefont {Ohtsuki}}, \
  and\ \bibinfo {author} {\bibfnamefont {S.}~\bibnamefont {Kettemann}},\
  }\bibfield  {title} {\enquote {\bibinfo {title} {Random network models and
  quantum phase transitions in two dimensions},}\ }\href {\doibase
  https://doi.org/10.1016/j.physrep.2005.07.001} {\bibfield  {journal}
  {\bibinfo  {journal} {Phys. Rep.}\ }\textbf {\bibinfo {volume} {417}},\
  \bibinfo {pages} {211} (\bibinfo {year} {2005})}\BibitemShut {NoStop}%
\bibitem [{\citenamefont {Kagalovsky}\ \emph {et~al.}(1999)\citenamefont
  {Kagalovsky}, \citenamefont {Horovitz}, \citenamefont {Avishai},\ and\
  \citenamefont {Chalker}}]{chalker1999super}%
  \BibitemOpen
  \bibfield  {author} {\bibinfo {author} {\bibfnamefont {V.}~\bibnamefont
  {Kagalovsky}}, \bibinfo {author} {\bibfnamefont {B.}~\bibnamefont
  {Horovitz}}, \bibinfo {author} {\bibfnamefont {Y.}~\bibnamefont {Avishai}}, \
  and\ \bibinfo {author} {\bibfnamefont {J.~T.}\ \bibnamefont {Chalker}},\
  }\bibfield  {title} {\enquote {\bibinfo {title} {Quantum {H}all plateau
  transitions in disordered superconductors},}\ }\href {\doibase
  10.1103/PhysRevLett.82.3516} {\bibfield  {journal} {\bibinfo  {journal}
  {Phys. Rev. Lett.}\ }\textbf {\bibinfo {volume} {82}},\ \bibinfo {pages}
  {3516} (\bibinfo {year} {1999})}\BibitemShut {NoStop}%
\bibitem [{\citenamefont {Chalker}\ \emph {et~al.}(2001)\citenamefont
  {Chalker}, \citenamefont {Read}, \citenamefont {Kagalovsky}, \citenamefont
  {Horovitz}, \citenamefont {Avishai},\ and\ \citenamefont
  {Ludwig}}]{chalker2001super}%
  \BibitemOpen
  \bibfield  {author} {\bibinfo {author} {\bibfnamefont {J.~T.}\ \bibnamefont
  {Chalker}}, \bibinfo {author} {\bibfnamefont {N.}~\bibnamefont {Read}},
  \bibinfo {author} {\bibfnamefont {V.}~\bibnamefont {Kagalovsky}}, \bibinfo
  {author} {\bibfnamefont {B.}~\bibnamefont {Horovitz}}, \bibinfo {author}
  {\bibfnamefont {Y.}~\bibnamefont {Avishai}}, \ and\ \bibinfo {author}
  {\bibfnamefont {A.~W.~W.}\ \bibnamefont {Ludwig}},\ }\bibfield  {title}
  {\enquote {\bibinfo {title} {Thermal metal in network models of a disordered
  two-dimensional superconductor},}\ }\href {\doibase
  10.1103/PhysRevB.65.012506} {\bibfield  {journal} {\bibinfo  {journal} {Phys.
  Rev. B}\ }\textbf {\bibinfo {volume} {65}},\ \bibinfo {pages} {012506}
  (\bibinfo {year} {2001})}\BibitemShut {NoStop}%
\bibitem [{\citenamefont {Beamond}\ \emph {et~al.}(2002)\citenamefont
  {Beamond}, \citenamefont {Cardy},\ and\ \citenamefont
  {Chalker}}]{chalker2002qshe}%
  \BibitemOpen
  \bibfield  {author} {\bibinfo {author} {\bibfnamefont {E.~J.}\ \bibnamefont
  {Beamond}}, \bibinfo {author} {\bibfnamefont {John}\ \bibnamefont {Cardy}}, \
  and\ \bibinfo {author} {\bibfnamefont {J.~T.}\ \bibnamefont {Chalker}},\
  }\bibfield  {title} {\enquote {\bibinfo {title} {Quantum and classical
  localization, the spin quantum {H}all effect, and generalizations},}\ }\href
  {\doibase 10.1103/PhysRevB.65.214301} {\bibfield  {journal} {\bibinfo
  {journal} {Phys. Rev. B}\ }\textbf {\bibinfo {volume} {65}},\ \bibinfo
  {pages} {214301} (\bibinfo {year} {2002})}\BibitemShut {NoStop}%
\bibitem [{\citenamefont {Obuse}\ \emph {et~al.}(2007)\citenamefont {Obuse},
  \citenamefont {Furusaki}, \citenamefont {Ryu},\ and\ \citenamefont
  {Mudry}}]{murdy2007qshe}%
  \BibitemOpen
  \bibfield  {author} {\bibinfo {author} {\bibfnamefont {Hideaki}\ \bibnamefont
  {Obuse}}, \bibinfo {author} {\bibfnamefont {Akira}\ \bibnamefont {Furusaki}},
  \bibinfo {author} {\bibfnamefont {Shinsei}\ \bibnamefont {Ryu}}, \ and\
  \bibinfo {author} {\bibfnamefont {Christopher}\ \bibnamefont {Mudry}},\
  }\bibfield  {title} {\enquote {\bibinfo {title} {Two-dimensional
  spin-filtered chiral network model for the $\mathbbm{Z}_{2}$ quantum
  spin-{H}all effect},}\ }\href {\doibase 10.1103/PhysRevB.76.075301}
  {\bibfield  {journal} {\bibinfo  {journal} {Phys. Rev. B}\ }\textbf {\bibinfo
  {volume} {76}},\ \bibinfo {pages} {075301} (\bibinfo {year}
  {2007})}\BibitemShut {NoStop}%
\bibitem [{\citenamefont {Obuse}\ \emph {et~al.}(2014)\citenamefont {Obuse},
  \citenamefont {Ryu}, \citenamefont {Furusaki},\ and\ \citenamefont
  {Mudry}}]{murdy2014TI}%
  \BibitemOpen
  \bibfield  {author} {\bibinfo {author} {\bibfnamefont {Hideaki}\ \bibnamefont
  {Obuse}}, \bibinfo {author} {\bibfnamefont {Shinsei}\ \bibnamefont {Ryu}},
  \bibinfo {author} {\bibfnamefont {Akira}\ \bibnamefont {Furusaki}}, \ and\
  \bibinfo {author} {\bibfnamefont {Christopher}\ \bibnamefont {Mudry}},\
  }\bibfield  {title} {\enquote {\bibinfo {title} {Spin-directed network model
  for the surface states of weak three-dimensional $\mathbbm{Z}_{2}$
  topological insulators},}\ }\href {\doibase 10.1103/PhysRevB.89.155315}
  {\bibfield  {journal} {\bibinfo  {journal} {Phys. Rev. B}\ }\textbf {\bibinfo
  {volume} {89}},\ \bibinfo {pages} {155315} (\bibinfo {year}
  {2014})}\BibitemShut {NoStop}%
\bibitem [{\citenamefont {Fulga}\ \emph
  {et~al.}(2012{\natexlab{a}})\citenamefont {Fulga}, \citenamefont {Akhmerov},
  \citenamefont {Tworzyd\l{}o}, \citenamefont {B\'eri},\ and\ \citenamefont
  {Beenakker}}]{fulga2012thermal}%
  \BibitemOpen
  \bibfield  {author} {\bibinfo {author} {\bibfnamefont {I.~C.}\ \bibnamefont
  {Fulga}}, \bibinfo {author} {\bibfnamefont {A.~R.}\ \bibnamefont {Akhmerov}},
  \bibinfo {author} {\bibfnamefont {J.}~\bibnamefont {Tworzyd\l{}o}}, \bibinfo
  {author} {\bibfnamefont {B.}~\bibnamefont {B\'eri}}, \ and\ \bibinfo {author}
  {\bibfnamefont {C.~W.~J.}\ \bibnamefont {Beenakker}},\ }\bibfield  {title}
  {\enquote {\bibinfo {title} {Thermal metal-insulator transition in a helical
  topological superconductor},}\ }\href {\doibase 10.1103/PhysRevB.86.054505}
  {\bibfield  {journal} {\bibinfo  {journal} {Phys. Rev. B}\ }\textbf {\bibinfo
  {volume} {86}},\ \bibinfo {pages} {054505} (\bibinfo {year}
  {2012}{\natexlab{a}})}\BibitemShut {NoStop}%
\bibitem [{\citenamefont {Pasek}\ and\ \citenamefont
  {Chong}(2014)}]{chong2014floquet}%
  \BibitemOpen
  \bibfield  {author} {\bibinfo {author} {\bibfnamefont {Michael}\ \bibnamefont
  {Pasek}}\ and\ \bibinfo {author} {\bibfnamefont {Y.~D.}\ \bibnamefont
  {Chong}},\ }\bibfield  {title} {\enquote {\bibinfo {title} {Network models of
  photonic {F}loquet topological insulators},}\ }\href {\doibase
  10.1103/PhysRevB.89.075113} {\bibfield  {journal} {\bibinfo  {journal} {Phys.
  Rev. B}\ }\textbf {\bibinfo {volume} {89}},\ \bibinfo {pages} {075113}
  (\bibinfo {year} {2014})}\BibitemShut {NoStop}%
\bibitem [{\citenamefont {De~Beule}\ \emph {et~al.}(2020)\citenamefont
  {De~Beule}, \citenamefont {Dominguez},\ and\ \citenamefont
  {Recher}}]{debeule2020}%
  \BibitemOpen
  \bibfield  {author} {\bibinfo {author} {\bibfnamefont {C.}~\bibnamefont
  {De~Beule}}, \bibinfo {author} {\bibfnamefont {F.}~\bibnamefont {Dominguez}},
  \ and\ \bibinfo {author} {\bibfnamefont {P.}~\bibnamefont {Recher}},\
  }\bibfield  {title} {\enquote {\bibinfo {title} {Aharonov-{B}ohm oscillations
  in minimally twisted bilayer graphene},}\ }\href {\doibase
  10.1103/PhysRevLett.125.096402} {\bibfield  {journal} {\bibinfo  {journal}
  {Phys. Rev. Lett.}\ }\textbf {\bibinfo {volume} {125}},\ \bibinfo {pages}
  {096402} (\bibinfo {year} {2020})}\BibitemShut {NoStop}%
\bibitem [{\citenamefont {Rodriguez-Vega}\ \emph {et~al.}(2019)\citenamefont
  {Rodriguez-Vega}, \citenamefont {Kumar},\ and\ \citenamefont
  {Seradjeh}}]{Rodriguez-Vega2019}%
  \BibitemOpen
  \bibfield  {author} {\bibinfo {author} {\bibfnamefont {Martin}\ \bibnamefont
  {Rodriguez-Vega}}, \bibinfo {author} {\bibfnamefont {Abhishek}\ \bibnamefont
  {Kumar}}, \ and\ \bibinfo {author} {\bibfnamefont {Babak}\ \bibnamefont
  {Seradjeh}},\ }\bibfield  {title} {\enquote {\bibinfo {title} {Higher-order
  {F}loquet topological phases with corner and bulk bound states},}\ }\href
  {\doibase 10.1103/PhysRevB.100.085138} {\bibfield  {journal} {\bibinfo
  {journal} {Phys. Rev. B}\ }\textbf {\bibinfo {volume} {100}},\ \bibinfo
  {pages} {085138} (\bibinfo {year} {2019})}\BibitemShut {NoStop}%
\bibitem [{\citenamefont {Bomantara}\ \emph {et~al.}(2019)\citenamefont
  {Bomantara}, \citenamefont {Zhou}, \citenamefont {Pan},\ and\ \citenamefont
  {Gong}}]{Bomentara2019}%
  \BibitemOpen
  \bibfield  {author} {\bibinfo {author} {\bibfnamefont {Raditya~Weda}\
  \bibnamefont {Bomantara}}, \bibinfo {author} {\bibfnamefont {Longwen}\
  \bibnamefont {Zhou}}, \bibinfo {author} {\bibfnamefont {Jiaxin}\ \bibnamefont
  {Pan}}, \ and\ \bibinfo {author} {\bibfnamefont {Jiangbin}\ \bibnamefont
  {Gong}},\ }\bibfield  {title} {\enquote {\bibinfo {title} {Coupled-wire
  construction of static and {F}loquet second-order topological insulators},}\
  }\href {\doibase 10.1103/PhysRevB.99.045441} {\bibfield  {journal} {\bibinfo
  {journal} {Phys. Rev. B}\ }\textbf {\bibinfo {volume} {99}},\ \bibinfo
  {pages} {045441} (\bibinfo {year} {2019})}\BibitemShut {NoStop}%
\bibitem [{\citenamefont {Huang}\ and\ \citenamefont {Liu}(2020)}]{Huang2020}%
  \BibitemOpen
  \bibfield  {author} {\bibinfo {author} {\bibfnamefont {Biao}\ \bibnamefont
  {Huang}}\ and\ \bibinfo {author} {\bibfnamefont {W.~Vincent}\ \bibnamefont
  {Liu}},\ }\bibfield  {title} {\enquote {\bibinfo {title} {{F}loquet
  higher-order topological insulators with anomalous dynamical polarization},}\
  }\href {\doibase 10.1103/physrevlett.124.216601} {\bibfield  {journal}
  {\bibinfo  {journal} {Phys. Rev. Lett.}\ }\textbf {\bibinfo {volume} {124}},\
  \bibinfo {pages} {216601} (\bibinfo {year} {2020})}\BibitemShut {NoStop}%
\bibitem [{\citenamefont {Delplace}\ \emph {et~al.}(2017)\citenamefont
  {Delplace}, \citenamefont {Fruchart},\ and\ \citenamefont
  {Tauber}}]{Delplace2017}%
  \BibitemOpen
  \bibfield  {author} {\bibinfo {author} {\bibfnamefont {Pierre}\ \bibnamefont
  {Delplace}}, \bibinfo {author} {\bibfnamefont {Michel}\ \bibnamefont
  {Fruchart}}, \ and\ \bibinfo {author} {\bibfnamefont {Cl\'ement}\
  \bibnamefont {Tauber}},\ }\bibfield  {title} {\enquote {\bibinfo {title}
  {Phase rotation symmetry and the topology of oriented scattering networks},}\
  }\href {\doibase 10.1103/PhysRevB.95.205413} {\bibfield  {journal} {\bibinfo
  {journal} {Phys. Rev. B}\ }\textbf {\bibinfo {volume} {95}},\ \bibinfo
  {pages} {205413} (\bibinfo {year} {2017})}\BibitemShut {NoStop}%
\bibitem [{\citenamefont {Altland}\ and\ \citenamefont
  {Zirnbauer}(1997)}]{Altland1997}%
  \BibitemOpen
  \bibfield  {author} {\bibinfo {author} {\bibfnamefont {Alexander}\
  \bibnamefont {Altland}}\ and\ \bibinfo {author} {\bibfnamefont {Martin~R.}\
  \bibnamefont {Zirnbauer}},\ }\bibfield  {title} {\enquote {\bibinfo {title}
  {Nonstandard symmetry classes in mesoscopic normal-superconducting hybrid
  structures},}\ }\href {\doibase 10.1103/PhysRevB.55.1142} {\bibfield
  {journal} {\bibinfo  {journal} {Phys. Rev. B}\ }\textbf {\bibinfo {volume}
  {55}},\ \bibinfo {pages} {1142} (\bibinfo {year} {1997})}\BibitemShut
  {NoStop}%
\bibitem [{\citenamefont {{Son}}\ and\ \citenamefont {{Raghu}}()}]{Son2020}%
  \BibitemOpen
  \bibfield  {author} {\bibinfo {author} {\bibfnamefont {Jun~Ho}\ \bibnamefont
  {{Son}}}\ and\ \bibinfo {author} {\bibfnamefont {S.}~\bibnamefont
  {{Raghu}}},\ }\bibfield  {title} {\enquote {\bibinfo {title} {{3D} network
  model for strong topological insulator transitions},}\ }\href
  {https://arxiv.org/abs/2008.03315} {\bibinfo  {journal} {arXiv:2008.03315}\
  }\BibitemShut {NoStop}%
\bibitem [{\citenamefont {Ho}\ and\ \citenamefont {Chalker}(1996)}]{ho1996}%
  \BibitemOpen
\bibfield  {journal} {  }\bibfield  {author} {\bibinfo {author} {\bibfnamefont
  {C.-M.}\ \bibnamefont {Ho}}\ and\ \bibinfo {author} {\bibfnamefont {J.~T.}\
  \bibnamefont {Chalker}},\ }\bibfield  {title} {\enquote {\bibinfo {title}
  {Models for the integer quantum {H}all effect: The network model, the {D}irac
  equation, and a tight-binding {H}amiltonian},}\ }\href {\doibase
  10.1103/PhysRevB.54.8708} {\bibfield  {journal} {\bibinfo  {journal} {Phys.
  Rev. B}\ }\textbf {\bibinfo {volume} {54}},\ \bibinfo {pages} {8708}
  (\bibinfo {year} {1996})}\BibitemShut {NoStop}%
\bibitem [{\citenamefont {Janssen}\ \emph {et~al.}(1999)\citenamefont
  {Janssen}, \citenamefont {Metzler},\ and\ \citenamefont
  {Zirnbauer}}]{zirnbauer1999}%
  \BibitemOpen
  \bibfield  {author} {\bibinfo {author} {\bibfnamefont {Martin}\ \bibnamefont
  {Janssen}}, \bibinfo {author} {\bibfnamefont {Marcus}\ \bibnamefont
  {Metzler}}, \ and\ \bibinfo {author} {\bibfnamefont {Martin~R.}\ \bibnamefont
  {Zirnbauer}},\ }\bibfield  {title} {\enquote {\bibinfo {title} {Point-contact
  conductances at the quantum {H}all transition},}\ }\href {\doibase
  10.1103/PhysRevB.59.15836} {\bibfield  {journal} {\bibinfo  {journal} {Phys.
  Rev. B}\ }\textbf {\bibinfo {volume} {59}},\ \bibinfo {pages} {15836}
  (\bibinfo {year} {1999})}\BibitemShut {NoStop}%
\bibitem [{\citenamefont {Cho}\ and\ \citenamefont {Fisher}(1997)}]{Cho1996}%
  \BibitemOpen
  \bibfield  {author} {\bibinfo {author} {\bibfnamefont {Sora}\ \bibnamefont
  {Cho}}\ and\ \bibinfo {author} {\bibfnamefont {Matthew P.~A.}\ \bibnamefont
  {Fisher}},\ }\bibfield  {title} {\enquote {\bibinfo {title} {Criticality in
  the two-dimensional random-bond {I}sing model},}\ }\href {\doibase
  10.1103/PhysRevB.55.1025} {\bibfield  {journal} {\bibinfo  {journal} {Phys.
  Rev. B}\ }\textbf {\bibinfo {volume} {55}},\ \bibinfo {pages} {1025}
  (\bibinfo {year} {1997})}\BibitemShut {NoStop}%
\bibitem [{\citenamefont {Alicea}(2012)}]{Alicea2012}%
  \BibitemOpen
  \bibfield  {author} {\bibinfo {author} {\bibfnamefont {Jason}\ \bibnamefont
  {Alicea}},\ }\bibfield  {title} {\enquote {\bibinfo {title} {New directions
  in the pursuit of majorana fermions in solid state systems},}\ }\href
  {\doibase 10.1088/0034-4885/75/7/076501} {\bibfield  {journal} {\bibinfo
  {journal} {Reports on Progress in Physics}\ }\textbf {\bibinfo {volume}
  {75}},\ \bibinfo {pages} {076501} (\bibinfo {year} {2012})}\BibitemShut
  {NoStop}%
\bibitem [{\citenamefont {Nayak}\ \emph {et~al.}(2008)\citenamefont {Nayak},
  \citenamefont {Simon}, \citenamefont {Stern}, \citenamefont {Freedman},\ and\
  \citenamefont {Das~Sarma}}]{Nayak2008}%
  \BibitemOpen
  \bibfield  {author} {\bibinfo {author} {\bibfnamefont {Chetan}\ \bibnamefont
  {Nayak}}, \bibinfo {author} {\bibfnamefont {Steven~H.}\ \bibnamefont
  {Simon}}, \bibinfo {author} {\bibfnamefont {Ady}\ \bibnamefont {Stern}},
  \bibinfo {author} {\bibfnamefont {Michael}\ \bibnamefont {Freedman}}, \ and\
  \bibinfo {author} {\bibfnamefont {Sankar}\ \bibnamefont {Das~Sarma}},\
  }\bibfield  {title} {\enquote {\bibinfo {title} {Non-{A}belian anyons and
  topological quantum computation},}\ }\href {\doibase
  10.1103/RevModPhys.80.1083} {\bibfield  {journal} {\bibinfo  {journal} {Rev.
  Mod. Phys.}\ }\textbf {\bibinfo {volume} {80}},\ \bibinfo {pages} {1083}
  (\bibinfo {year} {2008})}\BibitemShut {NoStop}%
\bibitem [{\citenamefont {Rudner}\ \emph {et~al.}(2013)\citenamefont {Rudner},
  \citenamefont {Lindner}, \citenamefont {Berg},\ and\ \citenamefont
  {Levin}}]{Rudner2013}%
  \BibitemOpen
  \bibfield  {author} {\bibinfo {author} {\bibfnamefont {Mark~S.}\ \bibnamefont
  {Rudner}}, \bibinfo {author} {\bibfnamefont {Netanel~H.}\ \bibnamefont
  {Lindner}}, \bibinfo {author} {\bibfnamefont {Erez}\ \bibnamefont {Berg}}, \
  and\ \bibinfo {author} {\bibfnamefont {Michael}\ \bibnamefont {Levin}},\
  }\bibfield  {title} {\enquote {\bibinfo {title} {Anomalous edge states and
  the bulk-edge correspondence for periodically driven two-dimensional
  systems},}\ }\href {\doibase 10.1103/physrevx.3.031005} {\bibfield  {journal}
  {\bibinfo  {journal} {Phys. Rev. X}\ }\textbf {\bibinfo {volume} {3}},\
  \bibinfo {pages} {031005} (\bibinfo {year} {2013})}\BibitemShut {NoStop}%
\bibitem [{\citenamefont {Titum}\ \emph {et~al.}(2016)\citenamefont {Titum},
  \citenamefont {Berg}, \citenamefont {Rudner}, \citenamefont {Refael},\ and\
  \citenamefont {Lindner}}]{Titum2016}%
  \BibitemOpen
  \bibfield  {author} {\bibinfo {author} {\bibfnamefont {Paraj}\ \bibnamefont
  {Titum}}, \bibinfo {author} {\bibfnamefont {Erez}\ \bibnamefont {Berg}},
  \bibinfo {author} {\bibfnamefont {Mark~S.}\ \bibnamefont {Rudner}}, \bibinfo
  {author} {\bibfnamefont {Gil}\ \bibnamefont {Refael}}, \ and\ \bibinfo
  {author} {\bibfnamefont {Netanel~H.}\ \bibnamefont {Lindner}},\ }\bibfield
  {title} {\enquote {\bibinfo {title} {Anomalous {F}loquet-{A}nderson insulator
  as a nonadiabatic quantized charge pump},}\ }\href {\doibase
  10.1103/physrevx.6.021013} {\bibfield  {journal} {\bibinfo  {journal} {Phys.
  Rev. X}\ }\textbf {\bibinfo {volume} {6}},\ \bibinfo {pages} {021013}
  (\bibinfo {year} {2016})}\BibitemShut {NoStop}%
\bibitem [{\citenamefont {Maczewsky}\ \emph {et~al.}(2017)\citenamefont
  {Maczewsky}, \citenamefont {Zeuner}, \citenamefont {Nolte},\ and\
  \citenamefont {Szameit}}]{Maczewsky2017}%
  \BibitemOpen
  \bibfield  {author} {\bibinfo {author} {\bibfnamefont {Lukas~J.}\
  \bibnamefont {Maczewsky}}, \bibinfo {author} {\bibfnamefont {Julia~M.}\
  \bibnamefont {Zeuner}}, \bibinfo {author} {\bibfnamefont {Stefan}\
  \bibnamefont {Nolte}}, \ and\ \bibinfo {author} {\bibfnamefont {Alexander}\
  \bibnamefont {Szameit}},\ }\bibfield  {title} {\enquote {\bibinfo {title}
  {Observation of photonic anomalous {F}loquet topological insulators},}\
  }\href {\doibase 10.1038/ncomms13756} {\bibfield  {journal} {\bibinfo
  {journal} {Nat. Commun.}\ }\textbf {\bibinfo {volume} {8}},\ \bibinfo {pages}
  {13756} (\bibinfo {year} {2017})}\BibitemShut {NoStop}%
\bibitem [{\citenamefont {Mukherjee}\ \emph {et~al.}(2017)\citenamefont
  {Mukherjee}, \citenamefont {Spracklen}, \citenamefont {Valiente},
  \citenamefont {Andersson}, \citenamefont {{\"O}hberg}, \citenamefont
  {Goldman},\ and\ \citenamefont {Thomson}}]{Mukherjee2017}%
  \BibitemOpen
  \bibfield  {author} {\bibinfo {author} {\bibfnamefont {Sebabrata}\
  \bibnamefont {Mukherjee}}, \bibinfo {author} {\bibfnamefont {Alexander}\
  \bibnamefont {Spracklen}}, \bibinfo {author} {\bibfnamefont {Manuel}\
  \bibnamefont {Valiente}}, \bibinfo {author} {\bibfnamefont {Erika}\
  \bibnamefont {Andersson}}, \bibinfo {author} {\bibfnamefont {Patrik}\
  \bibnamefont {{\"O}hberg}}, \bibinfo {author} {\bibfnamefont {Nathan}\
  \bibnamefont {Goldman}}, \ and\ \bibinfo {author} {\bibfnamefont {Robert~R.}\
  \bibnamefont {Thomson}},\ }\bibfield  {title} {\enquote {\bibinfo {title}
  {Experimental observation of anomalous topological edge modes in a slowly
  driven photonic lattice},}\ }\href {\doibase 10.1038/ncomms13918} {\bibfield
  {journal} {\bibinfo  {journal} {Nat. Commun.}\ }\textbf {\bibinfo {volume}
  {8}},\ \bibinfo {pages} {13918} (\bibinfo {year} {2017})}\BibitemShut
  {NoStop}%
\bibitem [{Note2()}]{Note2}%
  \BibitemOpen
  \bibinfo {note} {See the code in our Supplemental Material for an explicit
  calculation of the Chern numbers.}\BibitemShut {Stop}%
\bibitem [{\citenamefont {Franca}\ \emph {et~al.}(2019)\citenamefont {Franca},
  \citenamefont {Efremov},\ and\ \citenamefont {Fulga}}]{Franca2019}%
  \BibitemOpen
  \bibfield  {author} {\bibinfo {author} {\bibfnamefont {S.}~\bibnamefont
  {Franca}}, \bibinfo {author} {\bibfnamefont {D.~V.}\ \bibnamefont {Efremov}},
  \ and\ \bibinfo {author} {\bibfnamefont {I.~C.}\ \bibnamefont {Fulga}},\
  }\bibfield  {title} {\enquote {\bibinfo {title} {Phase-tunable second-order
  topological superconductor},}\ }\href {\doibase 10.1103/PhysRevB.100.075415}
  {\bibfield  {journal} {\bibinfo  {journal} {Phys. Rev. B}\ }\textbf {\bibinfo
  {volume} {100}},\ \bibinfo {pages} {075415} (\bibinfo {year}
  {2019})}\BibitemShut {NoStop}%
\bibitem [{\citenamefont {Akhmerov}\ \emph {et~al.}(2011)\citenamefont
  {Akhmerov}, \citenamefont {Dahlhaus}, \citenamefont {Hassler}, \citenamefont
  {Wimmer},\ and\ \citenamefont {Beenakker}}]{Akhmerov2011}%
  \BibitemOpen
  \bibfield  {author} {\bibinfo {author} {\bibfnamefont {A.~R.}\ \bibnamefont
  {Akhmerov}}, \bibinfo {author} {\bibfnamefont {J.~P.}\ \bibnamefont
  {Dahlhaus}}, \bibinfo {author} {\bibfnamefont {F.}~\bibnamefont {Hassler}},
  \bibinfo {author} {\bibfnamefont {M.}~\bibnamefont {Wimmer}}, \ and\ \bibinfo
  {author} {\bibfnamefont {C.~W.~J.}\ \bibnamefont {Beenakker}},\ }\bibfield
  {title} {\enquote {\bibinfo {title} {Quantized conductance at the {M}ajorana
  phase transition in a disordered superconducting wire},}\ }\href {\doibase
  10.1103/PhysRevLett.106.057001} {\bibfield  {journal} {\bibinfo  {journal}
  {Phys. Rev. Lett.}\ }\textbf {\bibinfo {volume} {106}},\ \bibinfo {pages}
  {057001} (\bibinfo {year} {2011})}\BibitemShut {NoStop}%
\bibitem [{\citenamefont {Fulga}\ \emph
  {et~al.}(2012{\natexlab{b}})\citenamefont {Fulga}, \citenamefont {Hassler},\
  and\ \citenamefont {Akhmerov}}]{Fulga2012}%
  \BibitemOpen
  \bibfield  {author} {\bibinfo {author} {\bibfnamefont {I.~C.}\ \bibnamefont
  {Fulga}}, \bibinfo {author} {\bibfnamefont {F.}~\bibnamefont {Hassler}}, \
  and\ \bibinfo {author} {\bibfnamefont {A.~R.}\ \bibnamefont {Akhmerov}},\
  }\bibfield  {title} {\enquote {\bibinfo {title} {Scattering theory of
  topological insulators and superconductors},}\ }\href {\doibase
  10.1103/PhysRevB.85.165409} {\bibfield  {journal} {\bibinfo  {journal} {Phys.
  Rev. B}\ }\textbf {\bibinfo {volume} {85}},\ \bibinfo {pages} {165409}
  (\bibinfo {year} {2012}{\natexlab{b}})}\BibitemShut {NoStop}%
\bibitem [{Note3()}]{Note3}%
  \BibitemOpen
  \bibinfo {note} {Finally, note that the scattering invariant does not
  distinguish a HOTP from the point $\theta _1 = \theta _2 = \pi /2; \theta _3
  = \theta _4 = 0$, corresponding to the Majorana flat band limit Fig.~\ref
  {fig:limit_cases}(d), which is an isolated point in the phase
  diagram.}\BibitemShut {Stop}%
\bibitem [{\citenamefont {Benalcazar}\ \emph {et~al.}(2019)\citenamefont
  {Benalcazar}, \citenamefont {Li},\ and\ \citenamefont
  {Hughes}}]{Benalcazar2018}%
  \BibitemOpen
  \bibfield  {author} {\bibinfo {author} {\bibfnamefont {Wladimir~A.}\
  \bibnamefont {Benalcazar}}, \bibinfo {author} {\bibfnamefont {Tianhe}\
  \bibnamefont {Li}}, \ and\ \bibinfo {author} {\bibfnamefont {Taylor~L.}\
  \bibnamefont {Hughes}},\ }\bibfield  {title} {\enquote {\bibinfo {title}
  {Quantization of fractional corner charge in ${C}_{n}$-symmetric higher-order
  topological crystalline insulators},}\ }\href {\doibase
  10.1103/PhysRevB.99.245151} {\bibfield  {journal} {\bibinfo  {journal} {Phys.
  Rev. B}\ }\textbf {\bibinfo {volume} {99}},\ \bibinfo {pages} {245151}
  (\bibinfo {year} {2019})}\BibitemShut {NoStop}%
\bibitem [{\citenamefont {Schindler}\ \emph {et~al.}(2019)\citenamefont
  {Schindler}, \citenamefont {Brzezi\ifmmode~\acute{n}\else \'{n}\fi{}ska},
  \citenamefont {Benalcazar}, \citenamefont {Iraola}, \citenamefont {Bouhon},
  \citenamefont {Tsirkin}, \citenamefont {Vergniory},\ and\ \citenamefont
  {Neupert}}]{Schindler2019}%
  \BibitemOpen
  \bibfield  {author} {\bibinfo {author} {\bibfnamefont {Frank}\ \bibnamefont
  {Schindler}}, \bibinfo {author} {\bibfnamefont {Marta}\ \bibnamefont
  {Brzezi\ifmmode~\acute{n}\else \'{n}\fi{}ska}}, \bibinfo {author}
  {\bibfnamefont {Wladimir~A.}\ \bibnamefont {Benalcazar}}, \bibinfo {author}
  {\bibfnamefont {Mikel}\ \bibnamefont {Iraola}}, \bibinfo {author}
  {\bibfnamefont {Adrien}\ \bibnamefont {Bouhon}}, \bibinfo {author}
  {\bibfnamefont {Stepan~S.}\ \bibnamefont {Tsirkin}}, \bibinfo {author}
  {\bibfnamefont {Maia~G.}\ \bibnamefont {Vergniory}}, \ and\ \bibinfo {author}
  {\bibfnamefont {Titus}\ \bibnamefont {Neupert}},\ }\bibfield  {title}
  {\enquote {\bibinfo {title} {Fractional corner charges in spin-orbit coupled
  crystals},}\ }\href {\doibase 10.1103/PhysRevResearch.1.033074} {\bibfield
  {journal} {\bibinfo  {journal} {Phys. Rev. Research}\ }\textbf {\bibinfo
  {volume} {1}},\ \bibinfo {pages} {033074} (\bibinfo {year}
  {2019})}\BibitemShut {NoStop}%
\bibitem [{\citenamefont {Geier}\ \emph {et~al.}(2020)\citenamefont {Geier},
  \citenamefont {Brouwer},\ and\ \citenamefont {Trifunovic}}]{Geier2020}%
  \BibitemOpen
  \bibfield  {author} {\bibinfo {author} {\bibfnamefont {Max}\ \bibnamefont
  {Geier}}, \bibinfo {author} {\bibfnamefont {Piet~W.}\ \bibnamefont
  {Brouwer}}, \ and\ \bibinfo {author} {\bibfnamefont {Luka}\ \bibnamefont
  {Trifunovic}},\ }\bibfield  {title} {\enquote {\bibinfo {title}
  {Symmetry-based indicators for topological {B}ogoliubov--de {G}ennes
  {H}amiltonians},}\ }\href {\doibase 10.1103/PhysRevB.101.245128} {\bibfield
  {journal} {\bibinfo  {journal} {Phys. Rev. B}\ }\textbf {\bibinfo {volume}
  {101}},\ \bibinfo {pages} {245128} (\bibinfo {year} {2020})}\BibitemShut
  {NoStop}%
\bibitem [{\citenamefont {Roberts}\ \emph {et~al.}(2020)\citenamefont
  {Roberts}, \citenamefont {Behrends},\ and\ \citenamefont
  {B\'eri}}]{Roberts2020}%
  \BibitemOpen
  \bibfield  {author} {\bibinfo {author} {\bibfnamefont {Elis}\ \bibnamefont
  {Roberts}}, \bibinfo {author} {\bibfnamefont {Jan}\ \bibnamefont {Behrends}},
  \ and\ \bibinfo {author} {\bibfnamefont {Benjamin}\ \bibnamefont {B\'eri}},\
  }\bibfield  {title} {\enquote {\bibinfo {title} {Second-order bulk-boundary
  correspondence in rotationally symmetric topological superconductors from
  stacked {D}irac {H}amiltonians},}\ }\href {\doibase
  10.1103/PhysRevB.101.155133} {\bibfield  {journal} {\bibinfo  {journal}
  {Phys. Rev. B}\ }\textbf {\bibinfo {volume} {101}},\ \bibinfo {pages}
  {155133} (\bibinfo {year} {2020})}\BibitemShut {NoStop}%
\bibitem [{\citenamefont {Bräunlich}\ \emph {et~al.}(2010)\citenamefont
  {Bräunlich}, \citenamefont {Graf},\ and\ \citenamefont
  {Ortelli}}]{Braunlich2010}%
  \BibitemOpen
  \bibfield  {author} {\bibinfo {author} {\bibfnamefont {G.}~\bibnamefont
  {Bräunlich}}, \bibinfo {author} {\bibfnamefont {G.M.}\ \bibnamefont {Graf}},
  \ and\ \bibinfo {author} {\bibfnamefont {G}~\bibnamefont {Ortelli}},\
  }\bibfield  {title} {\enquote {\bibinfo {title} {Equivalence of topological
  and scattering approaches to quantum pumping},}\ }\href {\doibase
  https://doi.org/10.1007/s00220-009-0983-1} {\bibfield  {journal} {\bibinfo
  {journal} {Commun. Math. Phys.}\ }\textbf {\bibinfo {volume} {295}},\
  \bibinfo {pages} {243--259} (\bibinfo {year} {2010})}\BibitemShut {NoStop}%
\bibitem [{\citenamefont {Fulga}\ and\ \citenamefont
  {Maksymenko}(2016)}]{Fulga2016}%
  \BibitemOpen
  \bibfield  {author} {\bibinfo {author} {\bibfnamefont {I.~C.}\ \bibnamefont
  {Fulga}}\ and\ \bibinfo {author} {\bibfnamefont {M.}~\bibnamefont
  {Maksymenko}},\ }\bibfield  {title} {\enquote {\bibinfo {title} {Scattering
  matrix invariants of {F}loquet topological insulators},}\ }\href {\doibase
  10.1103/PhysRevB.93.075405} {\bibfield  {journal} {\bibinfo  {journal} {Phys.
  Rev. B}\ }\textbf {\bibinfo {volume} {93}},\ \bibinfo {pages} {075405}
  (\bibinfo {year} {2016})}\BibitemShut {NoStop}%
\bibitem [{\citenamefont {Medvedyeva}\ \emph {et~al.}(2010)\citenamefont
  {Medvedyeva}, \citenamefont {Tworzyd\l{}o},\ and\ \citenamefont
  {Beenakker}}]{medvedyeva2010}%
  \BibitemOpen
  \bibfield  {author} {\bibinfo {author} {\bibfnamefont {M.~V.}\ \bibnamefont
  {Medvedyeva}}, \bibinfo {author} {\bibfnamefont {J.}~\bibnamefont
  {Tworzyd\l{}o}}, \ and\ \bibinfo {author} {\bibfnamefont {C.~W.~J.}\
  \bibnamefont {Beenakker}},\ }\bibfield  {title} {\enquote {\bibinfo {title}
  {Effective mass and tricritical point for lattice fermions localized by a
  random mass},}\ }\href {\doibase 10.1103/PhysRevB.81.214203} {\bibfield
  {journal} {\bibinfo  {journal} {Phys. Rev. B}\ }\textbf {\bibinfo {volume}
  {81}},\ \bibinfo {pages} {214203} (\bibinfo {year} {2010})}\BibitemShut
  {NoStop}%
\bibitem [{\citenamefont {Fulga}\ \emph {et~al.}(2020)\citenamefont {Fulga},
  \citenamefont {Oreg}, \citenamefont {Mirlin}, \citenamefont {Stern},\ and\
  \citenamefont {Mross}}]{Fulga2020}%
  \BibitemOpen
  \bibfield  {author} {\bibinfo {author} {\bibfnamefont
  {I.{\hspace{0.167em}}C.}\ \bibnamefont {Fulga}}, \bibinfo {author}
  {\bibfnamefont {Yuval}\ \bibnamefont {Oreg}}, \bibinfo {author}
  {\bibfnamefont {Alexander~D.}\ \bibnamefont {Mirlin}}, \bibinfo {author}
  {\bibfnamefont {Ady}\ \bibnamefont {Stern}}, \ and\ \bibinfo {author}
  {\bibfnamefont {David~F.}\ \bibnamefont {Mross}},\ }\bibfield  {title}
  {\enquote {\bibinfo {title} {Temperature enhancement of thermal {H}all
  conductance quantization},}\ }\href {\doibase 10.1103/physrevlett.125.236802}
  {\bibfield  {journal} {\bibinfo  {journal} {Phys. Rev. Lett.}\ }\textbf
  {\bibinfo {volume} {125}},\ \bibinfo {pages} {236802} (\bibinfo {year}
  {2020})}\BibitemShut {NoStop}%
\bibitem [{\citenamefont {Brouwer}\ \emph {et~al.}(2000)\citenamefont
  {Brouwer}, \citenamefont {Furusaki}, \citenamefont {Gruzberg},\ and\
  \citenamefont {Mudry}}]{Brouwer2000}%
  \BibitemOpen
  \bibfield  {author} {\bibinfo {author} {\bibfnamefont {P.~W.}\ \bibnamefont
  {Brouwer}}, \bibinfo {author} {\bibfnamefont {A.}~\bibnamefont {Furusaki}},
  \bibinfo {author} {\bibfnamefont {I.~A.}\ \bibnamefont {Gruzberg}}, \ and\
  \bibinfo {author} {\bibfnamefont {C.}~\bibnamefont {Mudry}},\ }\bibfield
  {title} {\enquote {\bibinfo {title} {Localization and delocalization in dirty
  superconducting wires},}\ }\href {\doibase 10.1103/PhysRevLett.85.1064}
  {\bibfield  {journal} {\bibinfo  {journal} {Phys. Rev. Lett.}\ }\textbf
  {\bibinfo {volume} {85}},\ \bibinfo {pages} {1064} (\bibinfo {year}
  {2000})}\BibitemShut {NoStop}%
\bibitem [{\citenamefont {Brouwer}\ \emph {et~al.}(2003)\citenamefont
  {Brouwer}, \citenamefont {Furusaki},\ and\ \citenamefont
  {Mudry}}]{Brouwer2003}%
  \BibitemOpen
  \bibfield  {author} {\bibinfo {author} {\bibfnamefont {P.~W.}\ \bibnamefont
  {Brouwer}}, \bibinfo {author} {\bibfnamefont {A.}~\bibnamefont {Furusaki}}, \
  and\ \bibinfo {author} {\bibfnamefont {C.}~\bibnamefont {Mudry}},\ }\bibfield
   {title} {\enquote {\bibinfo {title} {Universality of delocalization in
  unconventional dirty superconducting wires with broken spin-rotation
  symmetry},}\ }\href {\doibase 10.1103/PhysRevB.67.014530} {\bibfield
  {journal} {\bibinfo  {journal} {Phys. Rev. B}\ }\textbf {\bibinfo {volume}
  {67}},\ \bibinfo {pages} {014530} (\bibinfo {year} {2003})}\BibitemShut
  {NoStop}%
\bibitem [{\citenamefont {Fulga}\ \emph {et~al.}(2014)\citenamefont {Fulga},
  \citenamefont {van Heck}, \citenamefont {Edge},\ and\ \citenamefont
  {Akhmerov}}]{Fulga2014}%
  \BibitemOpen
  \bibfield  {author} {\bibinfo {author} {\bibfnamefont {I.~C.}\ \bibnamefont
  {Fulga}}, \bibinfo {author} {\bibfnamefont {B.}~\bibnamefont {van Heck}},
  \bibinfo {author} {\bibfnamefont {J.~M.}\ \bibnamefont {Edge}}, \ and\
  \bibinfo {author} {\bibfnamefont {A.~R.}\ \bibnamefont {Akhmerov}},\
  }\bibfield  {title} {\enquote {\bibinfo {title} {Statistical topological
  insulators},}\ }\href {\doibase 10.1103/PhysRevB.89.155424} {\bibfield
  {journal} {\bibinfo  {journal} {Phys. Rev. B}\ }\textbf {\bibinfo {volume}
  {89}},\ \bibinfo {pages} {155424} (\bibinfo {year} {2014})}\BibitemShut
  {NoStop}%
\bibitem [{\citenamefont {Diez}\ \emph {et~al.}(2014)\citenamefont {Diez},
  \citenamefont {Fulga}, \citenamefont {Pikulin}, \citenamefont
  {Tworzyd{\l}o},\ and\ \citenamefont {Beenakker}}]{Diez2014}%
  \BibitemOpen
  \bibfield  {author} {\bibinfo {author} {\bibfnamefont {M}~\bibnamefont
  {Diez}}, \bibinfo {author} {\bibfnamefont {I~C}\ \bibnamefont {Fulga}},
  \bibinfo {author} {\bibfnamefont {D~I}\ \bibnamefont {Pikulin}}, \bibinfo
  {author} {\bibfnamefont {J}~\bibnamefont {Tworzyd{\l}o}}, \ and\ \bibinfo
  {author} {\bibfnamefont {C~W~J}\ \bibnamefont {Beenakker}},\ }\bibfield
  {title} {\enquote {\bibinfo {title} {Bimodal conductance distribution of
  {K}itaev edge modes in topological superconductors},}\ }\href {\doibase
  10.1088/1367-2630/16/6/063049} {\bibfield  {journal} {\bibinfo  {journal}
  {New J. Phys.}\ }\textbf {\bibinfo {volume} {16}},\ \bibinfo {pages} {063049}
  (\bibinfo {year} {2014})}\BibitemShut {NoStop}%
\bibitem [{\citenamefont {Diez}\ \emph {et~al.}(2015)\citenamefont {Diez},
  \citenamefont {Pikulin}, \citenamefont {Fulga},\ and\ \citenamefont
  {Tworzyd{\l}o}}]{Diez2015reflection}%
  \BibitemOpen
  \bibfield  {author} {\bibinfo {author} {\bibfnamefont {M}~\bibnamefont
  {Diez}}, \bibinfo {author} {\bibfnamefont {D~I}\ \bibnamefont {Pikulin}},
  \bibinfo {author} {\bibfnamefont {I~C}\ \bibnamefont {Fulga}}, \ and\
  \bibinfo {author} {\bibfnamefont {J}~\bibnamefont {Tworzyd{\l}o}},\
  }\bibfield  {title} {\enquote {\bibinfo {title} {Extended topological group
  structure due to average reflection symmetry},}\ }\href {\doibase
  10.1088/1367-2630/17/4/043014} {\bibfield  {journal} {\bibinfo  {journal}
  {New J. Phys.}\ }\textbf {\bibinfo {volume} {17}},\ \bibinfo {pages} {043014}
  (\bibinfo {year} {2015})}\BibitemShut {NoStop}%
\bibitem [{\citenamefont {Hu}\ \emph {et~al.}(2015)\citenamefont {Hu},
  \citenamefont {Pillay}, \citenamefont {Wu}, \citenamefont {Pasek},
  \citenamefont {Shum},\ and\ \citenamefont {Chong}}]{Hu2015}%
  \BibitemOpen
  \bibfield  {author} {\bibinfo {author} {\bibfnamefont {Wenchao}\ \bibnamefont
  {Hu}}, \bibinfo {author} {\bibfnamefont {Jason~C.}\ \bibnamefont {Pillay}},
  \bibinfo {author} {\bibfnamefont {Kan}\ \bibnamefont {Wu}}, \bibinfo {author}
  {\bibfnamefont {Michael}\ \bibnamefont {Pasek}}, \bibinfo {author}
  {\bibfnamefont {Perry~Ping}\ \bibnamefont {Shum}}, \ and\ \bibinfo {author}
  {\bibfnamefont {Y.~D.}\ \bibnamefont {Chong}},\ }\bibfield  {title} {\enquote
  {\bibinfo {title} {Measurement of a topological edge invariant in a microwave
  network},}\ }\href {\doibase 10.1103/PhysRevX.5.011012} {\bibfield  {journal}
  {\bibinfo  {journal} {Phys. Rev. X}\ }\textbf {\bibinfo {volume} {5}},\
  \bibinfo {pages} {011012} (\bibinfo {year} {2015})}\BibitemShut {NoStop}%
\bibitem [{\citenamefont {Hafezi}\ \emph {et~al.}(2011)\citenamefont {Hafezi},
  \citenamefont {Demler}, \citenamefont {Lukin},\ and\ \citenamefont
  {Taylor}}]{Hafezi2011}%
  \BibitemOpen
  \bibfield  {author} {\bibinfo {author} {\bibfnamefont {Mohammad}\
  \bibnamefont {Hafezi}}, \bibinfo {author} {\bibfnamefont {Eugene~A.}\
  \bibnamefont {Demler}}, \bibinfo {author} {\bibfnamefont {Mikhail~D.}\
  \bibnamefont {Lukin}}, \ and\ \bibinfo {author} {\bibfnamefont {Jacob~M.}\
  \bibnamefont {Taylor}},\ }\bibfield  {title} {\enquote {\bibinfo {title}
  {Robust optical delay lines with topological protection},}\ }\href
  {https://doi.org/10.1038/nphys2063} {\bibfield  {journal} {\bibinfo
  {journal} {Nature Physics}\ }\textbf {\bibinfo {volume} {7}},\ \bibinfo
  {pages} {907} (\bibinfo {year} {2011})}\BibitemShut {NoStop}%
\bibitem [{\citenamefont {Afzal}\ \emph {et~al.}(2020)\citenamefont {Afzal},
  \citenamefont {Zimmerling}, \citenamefont {Ren}, \citenamefont {Perron},\
  and\ \citenamefont {Van}}]{Afzal2020}%
  \BibitemOpen
  \bibfield  {author} {\bibinfo {author} {\bibfnamefont {Shirin}\ \bibnamefont
  {Afzal}}, \bibinfo {author} {\bibfnamefont {Tyler~J.}\ \bibnamefont
  {Zimmerling}}, \bibinfo {author} {\bibfnamefont {Yang}\ \bibnamefont {Ren}},
  \bibinfo {author} {\bibfnamefont {David}\ \bibnamefont {Perron}}, \ and\
  \bibinfo {author} {\bibfnamefont {Vien}\ \bibnamefont {Van}},\ }\bibfield
  {title} {\enquote {\bibinfo {title} {Realization of anomalous {F}loquet
  insulators in strongly coupled nanophotonic lattices},}\ }\href {\doibase
  10.1103/physrevlett.124.253601} {\bibfield  {journal} {\bibinfo  {journal}
  {Phys. Rev. Lett.}\ }\textbf {\bibinfo {volume} {124}},\ \bibinfo {pages}
  {253601} (\bibinfo {year} {2020})}\BibitemShut {NoStop}%
\bibitem [{\citenamefont {Gao}\ \emph {et~al.}(2016)\citenamefont {Gao},
  \citenamefont {Gao}, \citenamefont {Shi}, \citenamefont {Yang}, \citenamefont
  {Lin}, \citenamefont {Xu}, \citenamefont {Joannopoulos}, \citenamefont
  {Solja{\v{c}}i{\'{c}}}, \citenamefont {Chen}, \citenamefont {Lu},
  \citenamefont {Chong},\ and\ \citenamefont {Zhang}}]{Gao2016}%
  \BibitemOpen
  \bibfield  {author} {\bibinfo {author} {\bibfnamefont {Fei}\ \bibnamefont
  {Gao}}, \bibinfo {author} {\bibfnamefont {Zhen}\ \bibnamefont {Gao}},
  \bibinfo {author} {\bibfnamefont {Xihang}\ \bibnamefont {Shi}}, \bibinfo
  {author} {\bibfnamefont {Zhaoju}\ \bibnamefont {Yang}}, \bibinfo {author}
  {\bibfnamefont {Xiao}\ \bibnamefont {Lin}}, \bibinfo {author} {\bibfnamefont
  {Hongyi}\ \bibnamefont {Xu}}, \bibinfo {author} {\bibfnamefont {John~D.}\
  \bibnamefont {Joannopoulos}}, \bibinfo {author} {\bibfnamefont {Marin}\
  \bibnamefont {Solja{\v{c}}i{\'{c}}}}, \bibinfo {author} {\bibfnamefont
  {Hongsheng}\ \bibnamefont {Chen}}, \bibinfo {author} {\bibfnamefont {Ling}\
  \bibnamefont {Lu}}, \bibinfo {author} {\bibfnamefont {Yidong}\ \bibnamefont
  {Chong}}, \ and\ \bibinfo {author} {\bibfnamefont {Baile}\ \bibnamefont
  {Zhang}},\ }\bibfield  {title} {\enquote {\bibinfo {title} {Probing
  topological protection using a designer surface plasmon structure},}\ }\href
  {\doibase 10.1038/ncomms11619} {\bibfield  {journal} {\bibinfo  {journal}
  {Nature Commun.}\ }\textbf {\bibinfo {volume} {7}},\ \bibinfo {pages} {11619}
  (\bibinfo {year} {2016})}\BibitemShut {NoStop}%
\bibitem [{\citenamefont {Gao}\ \emph {et~al.}(2018)\citenamefont {Gao},
  \citenamefont {Gao}, \citenamefont {Zhang}, \citenamefont {Luo},\ and\
  \citenamefont {Zhang}}]{Gao2018}%
  \BibitemOpen
  \bibfield  {author} {\bibinfo {author} {\bibfnamefont {Zhen}\ \bibnamefont
  {Gao}}, \bibinfo {author} {\bibfnamefont {Fei}\ \bibnamefont {Gao}}, \bibinfo
  {author} {\bibfnamefont {Youming}\ \bibnamefont {Zhang}}, \bibinfo {author}
  {\bibfnamefont {Yu}~\bibnamefont {Luo}}, \ and\ \bibinfo {author}
  {\bibfnamefont {Baile}\ \bibnamefont {Zhang}},\ }\bibfield  {title} {\enquote
  {\bibinfo {title} {Flexible photonic topological insulator},}\ }\href
  {\doibase 10.1002/adom.201800532} {\bibfield  {journal} {\bibinfo  {journal}
  {Adv. Opt. Mater.}\ }\textbf {\bibinfo {volume} {6}},\ \bibinfo {pages}
  {1800532} (\bibinfo {year} {2018})}\BibitemShut {NoStop}%
\bibitem [{\citenamefont {Ozawa}\ \emph {et~al.}(2019)\citenamefont {Ozawa},
  \citenamefont {Price}, \citenamefont {Amo}, \citenamefont {Goldman},
  \citenamefont {Hafezi}, \citenamefont {Lu}, \citenamefont {Rechtsman},
  \citenamefont {Schuster}, \citenamefont {Simon}, \citenamefont {Zilberberg},\
  and\ \citenamefont {Carusotto}}]{Ozawa2019}%
  \BibitemOpen
  \bibfield  {author} {\bibinfo {author} {\bibfnamefont {Tomoki}\ \bibnamefont
  {Ozawa}}, \bibinfo {author} {\bibfnamefont {Hannah~M.}\ \bibnamefont
  {Price}}, \bibinfo {author} {\bibfnamefont {Alberto}\ \bibnamefont {Amo}},
  \bibinfo {author} {\bibfnamefont {Nathan}\ \bibnamefont {Goldman}}, \bibinfo
  {author} {\bibfnamefont {Mohammad}\ \bibnamefont {Hafezi}}, \bibinfo {author}
  {\bibfnamefont {Ling}\ \bibnamefont {Lu}}, \bibinfo {author} {\bibfnamefont
  {Mikael~C.}\ \bibnamefont {Rechtsman}}, \bibinfo {author} {\bibfnamefont
  {David}\ \bibnamefont {Schuster}}, \bibinfo {author} {\bibfnamefont
  {Jonathan}\ \bibnamefont {Simon}}, \bibinfo {author} {\bibfnamefont {Oded}\
  \bibnamefont {Zilberberg}}, \ and\ \bibinfo {author} {\bibfnamefont {Iacopo}\
  \bibnamefont {Carusotto}},\ }\bibfield  {title} {\enquote {\bibinfo {title}
  {Topological photonics},}\ }\href {\doibase 10.1103/RevModPhys.91.015006}
  {\bibfield  {journal} {\bibinfo  {journal} {Rev. Mod. Phys.}\ }\textbf
  {\bibinfo {volume} {91}},\ \bibinfo {pages} {015006} (\bibinfo {year}
  {2019})}\BibitemShut {NoStop}%
\bibitem [{\citenamefont {Hafezi}\ \emph {et~al.}(2013)\citenamefont {Hafezi},
  \citenamefont {Mittal}, \citenamefont {Fan}, \citenamefont {Migdall},\ and\
  \citenamefont {Taylor}}]{Hafezi2013}%
  \BibitemOpen
  \bibfield  {author} {\bibinfo {author} {\bibfnamefont {M.}~\bibnamefont
  {Hafezi}}, \bibinfo {author} {\bibfnamefont {S.}~\bibnamefont {Mittal}},
  \bibinfo {author} {\bibfnamefont {J.}~\bibnamefont {Fan}}, \bibinfo {author}
  {\bibfnamefont {A.}~\bibnamefont {Migdall}}, \ and\ \bibinfo {author}
  {\bibfnamefont {J.~M.}\ \bibnamefont {Taylor}},\ }\bibfield  {title}
  {\enquote {\bibinfo {title} {Imaging topological edge states in silicon
  photonics},}\ }\href {https://doi.org/10.1038/nphoton.2013.274} {\bibfield
  {journal} {\bibinfo  {journal} {Nature Photonics}\ }\textbf {\bibinfo
  {volume} {7}},\ \bibinfo {pages} {1001} (\bibinfo {year} {2013})}\BibitemShut
  {NoStop}%
\bibitem [{\citenamefont {Liang}\ and\ \citenamefont
  {Chong}(2013)}]{Liang2013}%
  \BibitemOpen
  \bibfield  {author} {\bibinfo {author} {\bibfnamefont {G.~Q.}\ \bibnamefont
  {Liang}}\ and\ \bibinfo {author} {\bibfnamefont {Y.~D.}\ \bibnamefont
  {Chong}},\ }\bibfield  {title} {\enquote {\bibinfo {title} {Optical resonator
  analog of a two-dimensional topological insulator},}\ }\href {\doibase
  10.1103/PhysRevLett.110.203904} {\bibfield  {journal} {\bibinfo  {journal}
  {Phys. Rev. Lett.}\ }\textbf {\bibinfo {volume} {110}},\ \bibinfo {pages}
  {203904} (\bibinfo {year} {2013})}\BibitemShut {NoStop}%
\bibitem [{\citenamefont {Wang}\ \emph {et~al.}(2017)\citenamefont {Wang},
  \citenamefont {Xiao}, \citenamefont {Liu}, \citenamefont {Zhu},\ and\
  \citenamefont {Chan}}]{Wang2017}%
  \BibitemOpen
  \bibfield  {author} {\bibinfo {author} {\bibfnamefont {Qiang}\ \bibnamefont
  {Wang}}, \bibinfo {author} {\bibfnamefont {Meng}\ \bibnamefont {Xiao}},
  \bibinfo {author} {\bibfnamefont {Hui}\ \bibnamefont {Liu}}, \bibinfo
  {author} {\bibfnamefont {Shining}\ \bibnamefont {Zhu}}, \ and\ \bibinfo
  {author} {\bibfnamefont {C.~T.}\ \bibnamefont {Chan}},\ }\bibfield  {title}
  {\enquote {\bibinfo {title} {Optical interface states protected by synthetic
  {W}eyl points},}\ }\href {\doibase 10.1103/PhysRevX.7.031032} {\bibfield
  {journal} {\bibinfo  {journal} {Phys. Rev. X}\ }\textbf {\bibinfo {volume}
  {7}},\ \bibinfo {pages} {031032} (\bibinfo {year} {2017})}\BibitemShut
  {NoStop}%
\bibitem [{Note4()}]{Note4}%
  \BibitemOpen
  \bibinfo {note} {P. W. Brouwer, Ph.D. thesis, Leiden University,
  1997.}\BibitemShut {Stop}%
\end{thebibliography}%

\end{document}